\documentclass[longauth]{aa} 
\usepackage{graphicx,textcomp}
\usepackage[varg]{txfonts}
\usepackage{subcaption}
\bibpunct{(}{)}{;}{a}{}{,} 
\usepackage[colorlinks = true, linkcolor = blue, citecolor = blue]{hyperref}

\begin{document} 

\title{SPHERE RefPlanets: Search for $\epsilon$~Eridani b and warm dust
}
\author{C.~Tschudi\inst{\ref{instch1}} \and H.~M.~Schmid\inst{\ref{instch1}} \and M.~Nowak\inst{\ref{ioa},\ref{kavli}} \and H.~Le~Coroller\inst{\ref{instf3}} \and S.~Hunziker\inst{\ref{instch1}}  \and R.~G.~van~Holstein\inst{\ref{insteso2}} \and C.~Perrot\inst{\ref{instf4}}  \and D.~Mouillet\inst{\ref{instf1}} 
\and J.-C. Augereau\inst{\ref{instf1}} 
\and A.~Bazzon\inst{\ref{instch1}} \and J.~L.~Beuzit\inst{\ref{instf3}} 
\and A.~Boccaletti\inst{\ref{instf4}} \and M.~J.~Bonse\inst{\ref{instch1}}
\and G.~Chauvin\inst{\ref{instf5}} \and S.~Desidera\inst{\ref{insti1}} 
\and K.~Dohlen\inst{\ref{instf3}} \and C.~Dominik\inst{\ref{instnl2}} 
\and N.~Engler \inst{\ref{instch1}} \and M.~Feldt\inst{\ref{instd1}} 
\and J.~H.~Girard\inst{\ref{STScI}}  \and R.~Gratton\inst{\ref{insti1}} 
\and Th.~Henning\inst{\ref{instd1}} \and M.~Kasper\inst{\ref{insteso1}} 
\and P.~Kervella\inst{\ref{instf4}} \and A.-M.~Lagrange\inst{\ref{instf4}} 
\and M.~Langlois\inst{\ref{instf7}} \and P.~Martinez\inst{\ref{instf5}} 
\and F.~M\'{e}nard\inst{\ref{instf1}} \and M.~R.~Meyer\inst{\ref{MichUSA}} 
\and J.~Milli\inst{\ref{instf1}} \and J.~Pragt\inst{\ref{instnl1}} 
\and P.~Puget\inst{\ref{instf1}} \and S.~P.~Quanz\inst{\ref{instch1}} 
\and R.~Roelfsema\inst{\ref{instnl1}} \and J.-F.~Sauvage\inst{\ref{instf8},\ref{instf3}}
\and J.~Szul\'{a}gyi\inst{\ref{instch1}}  \and Ch.~Thalmann\inst{\ref{instch1}} 
\and A.~Zurlo\inst{\ref{instcl1},\ref{instcl2}}
}
\institute{Institute for Particle Physics and Astrophysics, ETH Zurich, Wolfgang-Pauli-Strasse 27, 8093 Zurich, Switzerland. \email{chtschud@phys.ethz.ch}\label{instch1} 
\and
Institute of Astronomy, University of Cambridge, Madingley Road, Cambridge CB3 0HA, UK \label{ioa} 
\and 
Kavli Institute for Cosmology, University of Cambridge, Madingley Road, Cambridge CB3 0HA, UK \label{kavli} 
\and
Aix Marseille Universit\'{e}, CNRS, CNES, LAM (Laboratoire
d’Astrophysique de Marseille) UMR 7326, 13388, Marseille,
France\label{instf3}
\and
European Southern Observatory, Alonso de Cordova 3107, Casilla
19001 Vitacura, Santiago 19, Chile\label{insteso2}
\and
LESIA, Observatoire de Paris, Universit\'{e} PSL, CNRS, Sorbonne Universit\'{e}, Universit\'{e} Paris-Cit\'{e}, 5 Place Jules Janssen, 92195 Meudon, France\label{instf4}
\and
Univ. Grenoble Alpes, CNRS, IPAG, 38000 Grenoble, France\label{instf1}
\and
Universit\'{e} C\^{o}te d'Azur, Observatoire de la C\^{o}te d'Azur, 
Laboratoire Lagrange, Nice, France\label{instf5}
\and
INAF – Osservatorio Astronomico di Padova, Vicolo
dell’Osservatorio 5, 35122 Padova, Italy\label{insti1}
\and
Anton Pannekoek Astronomical Institute, University of Amsterdam,
PO Box 94249, 1090 GE Amsterdam, The Netherlands\label{instnl2}
\and
Max-Planck-Institut f\"{u}r Astronomie, K\"{o}nigstuhl 17, 69117
Heidelberg, Germany\label{instd1}
\and
Space Telescope Science Institute (STScI), 3700 San Martin Dr, Baltimore, MD 21218, USA
\label{STScI}
\and
European Southern Observatory, Karl Schwarzschild Str. 2, 85748
Garching, Germany\label{insteso1}
\and
Centre de Recherche Astrophysique de Lyon, CNRS, 
Universit\'{e} Claude Bernard Lyon 1, ENS de Lyon, 
France\label{instf7}
\and 
Department of Astronomy, University of Michigan, 1085 S. University, 
Ann Arbor, MI 48109, USA\label{MichUSA}
\and
NOVA Optical Infrared Instrumentation Group at ASTRON, Oude
Hoogeveensedijk 4, 7991 PD Dwingeloo, The Netherlands\label{instnl1}
\and 
DOTA, ONERA, F-13661 Salon cedex Air - France\label{instf8}
\and
Instituto de Estudios Astrof\'isicos, Facultad de Ingenier\'ia y Ciencias, Universidad Diego Portales, Av. Ej\'ercito Libertador 441, Santiago, Chile\label{instcl1}
\and
Millennium Nucleus on Young Exoplanets and their Moons (YEMS)\label{instcl2} 
}
 
\date{Received 31 January 2024 ; accepted 16 April 2024
} 

 \abstract
 {Cold planets, including all habitable planets, produce only scattered light emission in the visual to
near-infrared wavelength range. For this reason it is highly desirable to adapt the technique for the direct imaging of reflected light from extra-solar planets. 
 } 
{For the nearby system $\epsilon$~Eri, we want to set much deeper detection limits for 
the expected scattered radiation from the radial velocity planet candidate ($\approx 0.7~{\rm M_{\rm J}}$) and the warm dust using the VLT/SPHERE adaptive optics (AO) instrument with the ZIMPOL imaging polarimeter.
}
{We carried out very deep imaging polarimetry 
of $\epsilon$~Eri based on 38.5 hours of integration time with
a broad-band filter ($\lambda_{\rm c}=735~{\rm nm}$) for the search of the polarization signal 
from a planet or from circumstellar dust 
using AO, 
coronagraphy, high precision differential polarimetry, and angular differential imaging. The data were collected during 12 nights within four epochs distributed over 14 months and we searched for a signal  
in the individual epochs. We also combined the full data set to achieve an even higher contrast limit  considering the Keplerian motion using the K-Stacker software.
All data were also combined for the search of the scattering signal 
from extended dust clouds. 
We improved various data reduction and post-processing procedures and also developed new ones to enhance the sensitivity of SPHERE/ZIMPOL further. The final detection limits were quantified and 
 we investigated the potential of SPHERE/ZIMPOL for deeper observations. 
}
%
{The data of $\epsilon$~Eridani provide unprecedented contrast limits but no significant detection of a point source or an extended signal from circumstellar dust. For each observing epoch, we achieved a $5\,\sigma_{\mathcal{N}}$ point source contrast for the polarized intensity $C_{\rm P}=Q_{\rm \phi}/I_\star$ between $2\cdot 10^{-8}$ and $4\cdot 10^{-8}$ at a separation of $\rho\approx1\arcsec$, which is as expected for the proposed radial velocity planet at a quadrature phase. The polarimetric contrast limits are close to the photon noise limits for $\rho>0.6\arcsec$ or about six times to 50 times better than the intensity limits because 
polarimetric imaging is much more efficient for speckle suppression.

Combining the data for the search of a planet moving on a Keplerian orbit with the K-Stacker technique improves the contrast limits further by about a factor of two, when compared to an epoch, to 
 about $C_{\rm P}=0.8 \cdot 10^{-8}$ at $\rho=1\arcsec$. This would allow the detection of a planet with a
radius of about 2.5~${\rm R_{\rm J}}$. Should future astrometry provide strong constraints on the position of the planet, then a $3\,\sigma_{\mathcal{N}}$ detection at $1\arcsec$ with $C_{\rm P}\approx 5\cdot 10^{-9}$ would be within reach of our data. 
The surface brightness contrast limits achieved for the polarized intensity from an extended scattering region is about $15~{\rm mag}\,{\rm arcsec}^{-2}$ at $1\arcsec$ or up to $3~{\rm mag}\,{\rm arcsec}^{-2}$ deeper than previous limits. For $\epsilon$~Eri, these limits exclude the presence 
of a narrow dust ring and they constrain the dust properties.
The photon statistics would allow deeper limits but we 
find a very weak systematic noise pattern probably introduced by 
polarimetric calibration errors. 
}
{This $\epsilon$~Eri study shows that the polarimetric contrast limits
for reflecting planets with SPHERE/ZIMPOL can be
improved to a level below $C_{\rm p}<10^{-8}$ by just collecting more data 
during many nights using software such as K-Stacker, which can combine all data considering the expected planet orbit. 
Contrast limits of $C_{\rm p}\approx 10^{-9}$ are within reach for $\epsilon$~Eri
if the search can be optimized for a planet with a well-known orbit. This limit
is also attainable for other bright nearby stars, such as $\alpha$~Cen or Sirius~A. Such data also provide unprecedented sensitivity for the search
of extended polarized emission from warm circumstellar dust.}


\maketitle
%

\section{Introduction}\label{sec:intro}

In the last two decades, many exoplanets have been detected using radial velocity (RV) measurements, transit photometry and spectroscopy, direct imaging, and other techniques and a lot has been learned about exoplanet frequency, orbital parameters, masses, radii, composition, and surface structures. However, almost no progress has been made in the observational characterization of the surfaces and atmospheres of cold exoplanets, including potentially habitable planets. Cold planets are very faint sources in the mid-infrared, and they produce only scattered light in the optical to near-infrared range. Therefore, their investigation with direct imaging requires very deep contrast limits and, as of yet, there exists no successful detection of a cold exoplanet with direct imaging. 

High-contrast imaging is a very successful technique for the investigation of young, self-contracting, self-luminous giant planets that are hot ($T_{\rm eff}\approx 1000~$K) and rather luminous in the near-infrared range \citep{Nielsen19,Vigan12,Vigan17,Bowler16}. Typical examples for current near-infrared techniques provide contrasts of $I_{\rm p}/I_\star \approx 10^{-3.24}$ for an angular separations of about $\rho \approx 0.5\arcsec$ to $\approx 10^{-4}$ for $\rho>2\arcsec$ for the L' band (3.8~$\rm \mu m$) with the Very Large Telescope (VLT) instrument NACO \citep{Cugno23} or of $I_{\rm p}/I_\star \approx 10^{-5}$ for $\rho \approx 0.5\arcsec$ to $\approx 10^{-6}$ for $\rho>2$ for the H band (1.6~$\rm \mu m$) with VLT/SPHERE/IRDIS \citep{Langlois21}. 

Cold planets are much fainter and harder to detect because they have no strong intrinsic energy source and only reprocess stellar irradiation. Therefore, their luminosity is proportional to the irradiation $L_{\rm p}\propto L_\star \, R_{\rm p}^2/d_{\rm p}^2$, where $R_{\rm p}$ is the planet radius and $d_{\rm p}$ the orbital separation. The planet radiation is partly emitted as scattered light at the same wavelengths as the stellar emission, and partly as thermal radiation peaking for cold planets $T_{\rm eq}\leq 300$~K at 10~$\rm \mu m$ or even longer wavelengths. The factor $R_{\rm p}^2/d_{\rm p}^2$ is very small, only $\approx 2.3 \cdot 10^{-7}$ for a Jupiter-sized planet ($R_{\rm p}={\rm R_{\rm J}}$) at a separation of $d_{\rm p}=1$~au. 
The contrast is less demanding for a smaller physical separation $d_{\rm p}$ where, however, the inner working angle for the high-contrast imaging becomes an issue \citep[e.g.][]{Milli13}. Furthermore, the achievable contrast is better for larger angular separations $\rho$, but for large $d_{\rm p}$ the planet signal is weak.
These two conditions -- small $d_{\rm p}$ and at the same time large $\rho$ -- limit the direct imaging search to planets in nearby systems within 5~pc to 10~pc \citep{Lovis17,Kasper21}. 

This paper presents high-contrast imaging observations of $\epsilon$~Eri taken in
2019 and 2020 with VLT. This is a continuation of the SPHERE RefPlanets guaranteed time observation (GTO) programme which uses the SPHERE instrument \citep{Beuzit19} with the Zurich IMaging POLarimeter (ZIMPOL) subsystem \citep{Schmid18} for the search of polarization signals from the scattering of stellar light by planets.
Thereby, polarimetry serves as a powerful differential imaging technique because the polarization signal can be distinguished in the point-spread-function (PSF) halo from the unpolarized light of the much brighter star \citep{Schmid06a}. The expected fractional polarization of a planet is at the level of about $p_{\rm p}\approx 5~\%$ to $50~\%$, it depends on the orbital phase angle $\alpha$, and it strongly constrains the properties of planets as described in \citet{Seager00}, \citet{Stam04}, \citet{Schmid06a}, and 
\citet{Buenzli09}.

The first results of the RefPlanets programme are presented in \citet{Hunziker20} who observed six targets: $\alpha$~Cen~A, $\alpha$~Cen~B, Sirius~A, Altair ($\alpha$~Aql), $\epsilon$~Eri, and $\tau$~Ceti. This programme was executed similar to a blind search and did not target known giant planets or a potential planet candidate. Typically, integration times between 1.5 and 3.5 hours were obtained and the achieved $5\,\sigma$ signal-to-noise ratio (S/N) contrast limits for the polarized flux contrast is about $C_{\rm P}=(p_{\rm p} \cdot I_{\rm p})/I_\star\approx 10^{-7}$ at an angular separation of $\rho=0.5\arcsec$ and about $\approx 10^{-8}$ at $\rho=1.5\arcsec$. Thus, Jupiter-sized planets were only within reach for the nearest targets $\alpha$~Cen~A and $\alpha$~Cen~B, where 1~au corresponds to $\rho = 0.7\arcsec$. The indicated limits also require that the planet has a high albedo, that it produces a scattering polarization of about 20~\%, and that it is located at the right orbital phase during the observations. Therefore, a blind search can easily miss a detectable planet if the observations are carried out during an unfavourable orbital phase. For other systems, which are further away than $\alpha$~Cen, even a planet with $R_{\rm p}\approx {\rm R_{\rm J}}$ would have been too faint to be detected. An important result of this programme was the demonstration that the contrast limits at separations larger than $0.5\arcsec$ reach the photon noise limit and will therefore improve with the square root of the integration time \citep{Hunziker20}. 

The $\epsilon$~Eri observations presented in this work were collected to achieve a point source contrast limit of $C_{\rm P}< 10^{-8}$, which is significantly deeper when compared to the study of \citet{Hunziker20}. This requires the combination of more than 30\,000 integrations of 3~s or 5~s for a total exposure time of more than 38~hours. For this, one has to combine data from different runs and consider that a possible planet moves on its orbit around the star by several pixels per week or many pixels per month. Therefore, one needs to combine the whole time series with a prediction for the Keplerian orbit of possible targets in the post-processing as described in \citet{Nowak18}, \citet{Lecoroller20} and \citet{Dallant23}. 

The system $\epsilon$~Eri is very interesting for such a deep search and performance test with SPHERE/ZIMPOL, because of strong indications for the presence of a  RV planet with a semi-major axis of about 3.5~au or a separation of about $\rho\approx 1\arcsec$ \citep[e.g.][]{Llop21}. Further, $\epsilon$~Eri shows strong thermal dust emission in the infrared including a component peaking at 20~$\rm \mu m$ from warm dust \citep{Backman09}, which should produce an extended polarization signal from light scattering by dust within the ZIMPOL field of view. As $\epsilon$~Eri is the nearest and brightest single solar-type star, it is important for the investigation of extra-solar planetary systems \citep{Backman09}.

This paper is organized as follows. Section~\ref{2EpsEri} summarizes the parameters for the planet candidate $\epsilon$~Eri~b and the dust near the star from the literature and provides predictions for the possible polarization signal from the planet and dust. In Section~\ref{3ObsMeth} we describe the observations, the data reduction including the specific post-processing procedures for the search of a faint point source, and signatures from circumstellar dust. In Section~\ref{4Planet} and \ref{5Dust}, we present the final detection maps and explain how we derived the sensitivity limits of our planet search and the search for dust scattering. In Section~\ref{5Dis}, we discuss our findings and give our conclusions in Section~\ref{6Conc}.
\begin{figure*}[t]
    \includegraphics[trim=10 6 10 4, clip=true,width=1\textwidth]{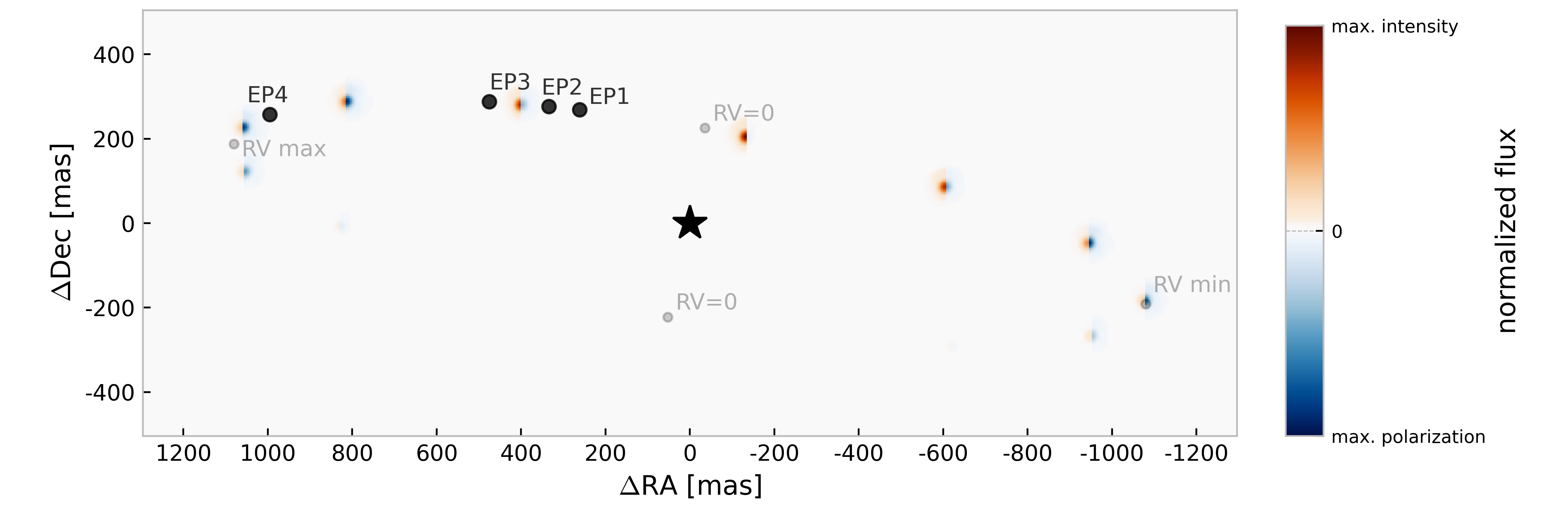}
    \centering
\caption{Illustration for a possible orbit ($i=78^\circ$) and signal strength of $\epsilon$~Eri~b. The blue colour indicates the normalized strength of the polarized intensity and the red colour is analogously for the intensity. The four observing epochs are marked with black dots and the points in the orbit with a minimal, zero, and maximal RV are shown in grey.} \label{1_orbit}
\end{figure*}

\section{$\epsilon$~Eridani}\label{2EpsEri}

The star $\epsilon$~Eri is a single star at a distance of 3.2~pc \citep{Gaia20}, with a spectral type of K2V, $T_{\rm eff}\approx 5040$~K, a mass of 0.82~M$_\odot$, a luminosity of $L_{\rm s}=0.32$~L$_\odot$ \citep[e.g.][]{Baines12}, and an apparent brightness of $m_{\rm V}$=$3.7$~mag, $m_{\rm R}$=$3.0$~mag, $m_{\rm I}$=$2.5$~mag. It is a very active solar type star with a high level of chromospheric activity, with activity cycles of about 13~years and 3~years \citep{Metcalfe13} and a rotation period of 11.67~days \citep{Donahue96}. The system is young and might belong to the 500~Myr old Ursa Major association \citep{Fuhrmann04}, but its spatial motion is one of the most deviant compared to the mean motion of the group. Therefore \citet{Janson08} assume a large age range of 200-800~Myr.

Significant RV variations $\pm 20~$m/s were found by \citet{Campbell88} . They are at least partly caused by chromospheric activity, but there seems to exist also a periodic RV signal from the reflex motion introduced by a giant planet. Initial estimates for the RV period were 10~yr \citep{Walker95}, and then 6.9~yr \citep{Cumming99, Hatzes00}. Longer time series established quite firmly an orbital period of about 7.3~yr \citep{Anglada12, Mawet19, Llop21} while \citet{Zechmeister13} did not find this periodicity. In \citet{Mawet19} and \citet{Llop21} it is strongly suggested that this signal is caused by a planet on an orbit with low eccentricity $e < 0.1$ producing a RV semi-amplitude of about 11~m/s \citep{Anglada10,Mawet19,Llop21}. However, with the available data it is hard to rule out the possibility of another stellar activity cycle introducing such a RV signal. 

We adopt for this work the interpretation of a RV-planet with a period of 7.3~yr (2671 days) and base much of our signal predictions and parts of the data interpretation on this assumption. This solution predicts $a_{\rm p}=3.53$~au for the semi-major axis of the planet orbit and that the planet was further away from us than the star around $T=2019.5$ (JD 2\,458\,666), at RV-phase $\varphi=0.5$ according to \citet{Llop21} while quadrature phase was around 2021.3. Our data were taken between these two dates which correspond roughly to an orbital phase of $\phi=0^\circ$, when we expect maximum intensity for a reflecting planet, and $\phi=90^\circ$ when the fractional polarization of the planet should be highest (see Section~\ref{SectPlanetModel}). 
 These orbital phase estimates could be affected significantly by RV uncertainties introduced by the chromospheric activity. 

The presence of a planet in $\epsilon$~Eri is supported by astrometric measurements taken with the Hipparcos satellite \citep{Reffert11}, the HST Fine Guidance Sensor \citep{Benedict06},
and by the search of proper motion anomalies combining older astrometry with results from the Gaia early data release 3 \citep[e.g.][]{Kervella22,Benedict22,Makarov21}. All these
studies indicate for $\epsilon$~Eri systematic deviations from a constant proper motion 
vector but the remaining uncertainties are still quite large. The results seem to be compatible with the presence of a planet with mass of $\approx 1~$M$_{\rm J}$ as 
measured by radial velocity, on a prograde orbit (N over E), and that it should be located about north-east of the star for our observations from 2019 and 2020 \citep{Benedict22}.
It is expected that accurate astrometric data for $\epsilon$~Eri, for example from the Gaia mission, will provide in the near future, strong, new constraints on the orbit and the mass of the planet. 

Radial velocity and astrometric measurements have been combined to obtain more detailed parameters for $\epsilon$~Eri~b \citep{Benedict06, Reffert11,Llop21,Benedict22}.
However, these studies are based on extra assumptions, in particular it is not considered that additional planets in the system could contribute to the measured reflex motion. Therefore, we use
the derived orbital parameters only as possible values.

If the $\epsilon$~Eri~b planet is on a circular orbit, then the measured
RV semi-amplitude $K = 10.34$~m/s and period $P=2671$~days from \citet[][Table~3]{Llop21}
constrain the radius of the stellar orbit $a_{\rm s}=(K\cdot P)/(2\pi \cdot \sin(i))=$ $2.54\cdot 10^{-3}~{\rm au}/\sin(i) =$ $0.78~{\rm mas}/\sin(i)$ and this defines the minimum mass for the planet $m_{\rm p}\, \sin(i)=(a_{\rm s}/a_{\rm p})\cdot M_{\rm s} = 0.651~$M$_{\rm J}$. For small $i$ the planet mass and therefore the stellar reflex motion measurable by astrometry would be significant larger than the minimum values.

An estimate for the equilibrium temperature for $\epsilon$~Eri~b, using $a_{\rm p}=3.53$~au, $L_{\rm s}=0.32$~L$_\odot$, and Bond albedo $A_{\rm B}=0.3$ gives $T_{\rm eq} \approx (L/L_\odot)^{1/4}\,(a/au)^{-1/2}\,T_{\rm Earth} = 112$~K similar to Jupiter. If there is no strong internal energy source, then the scattered intensity will dominate in the visual and near-infrared range up to at least 3~$\mu$m and the expected planet to star intensity contrast is to first order wavelength independent at a level of about $\Delta m\approx 20$~mag to $21$~mag. At 10~$\mu$m and perhaps also 5~$\mu$m the planet brightness could be dominated by thermal radiation 
, from Kelvin-Helmholtz contraction in case $\epsilon$~Eri is younger than estimated. 

Several high-contrast imaging searches for faint point sources around $\epsilon$~Eri have been carried out and we give here an incomplete selection of reported intensity contrast limits from the 
literature for a separation of $1\arcsec$: 
about $\Delta m \approx 10$~mag in the N-band at 11~$\mu$m \citep{Pathak21},
$\Delta m \approx 13$~mag in Ms at 4.7~$\mu$m \citep{Mawet19},
$\Delta m \approx 13.5$~mag in Lp at 3.8~$\mu$m \citep{Mizuki16},
$\Delta m \approx 14$~mag in H at 1.6~$\mu$m \citep{Janson07},
$\Delta m \approx 14$~mag in the RI-band at 0.75~$\mu$m \citep{Hunziker20}.
These contrast limits are more than $6$~mag away from the detection of the reflected intensity of a planet. For polarimetric imaging \citet{Hunziker20} report in the RI-band a polarimetric contrast of $\Delta m_{\rm p}\approx 18$~mag which is only about $4$~mag from the expected planet signal considering that the polarized intensity is about $1.5$~mag fainter than the intensity (see Section~\ref{SectPlanetModel}).

The $\epsilon$~Eri system is also well known for its infrared excess of $L_{\rm dust}/L_\star\approx 1.0\cdot 10^{-4}$
caused by the thermal emission of circumstellar dust
\citep{Gillett86,Decin03,Backman09}. This infrared-emission is composed of emission from an outer circular ring of cold dust with an inclination of about $30^\circ$ and $T_{\rm dust}\approx 50$~K located at $r\approx 65$~au or $\rho\approx 20\arcsec$ \citep[e.g.][]{Greaves14,Chavez16}.
 
Also a region of warm dust has been resolved and it extends to about $r=14$~au or $\rho\approx 4\arcsec$ and a temperature of about 100~K \citep{Greaves14}. According to the SED analysis of \citet{Backman09} and \citet{Su17},
the warm dust has two components, one at 20~au and one at a separation of around 3~au producing an infrared-bump at 20~$\mu$m from dust with $T\approx 120$~K. This latter 'warm' dust component with an infrared excess of $L_{\rm warm}/L_\star = 3.3\cdot 10^{-5}$ \citep{Backman09} is most interesting for this study, because a signal of scattered light is expected in the field of view of our SPHERE/ZIMPOL observations. Therefore we also search for extended emission of polarized light in our data. Even finding only upper limits could be of interest as $\epsilon$~Eri is the closest debris disk system to the Sun and therefore a prototype and important test case for the theoretical modelling \citep[e.g.][]{Reidemeister11,Su17}. With HST intensity data \citet{Wolff23} could not find dust scattering outside of the inner working angle of 1$\arcsec$.

\subsection{Predictions for the signal of $\epsilon$~Eri~b}
\label{SectPlanetModel}

The results from the RV and astrometry data are very helpful for predicting the expected signal of the light reflection from $\epsilon$~Eri~b. The intensity contrast $C_{\rm I} = I_{\rm p}/I_\star$ and the closely linked polarized intensity (or polarization) contrast $C_{\rm P}$ can be calculated according to
\begin{equation}
C_{\rm P} = p(\alpha) C_{\rm I} = p(\alpha) f(\alpha) \frac{R_{\rm p}^2}{d_{\rm p}^2}\,,
\label{eq:Cpol}
\end{equation}
where $R_{\rm p}$ is the planet radius and $d_{\rm p}$ the star to planet separation \citep{Stam04}. The reflectivity $f(\alpha)$ and the fractional polarization $p(\alpha)$ of the scattered light are both functions on the planet scattering angle $\alpha$ and depend on the wavelength. 

For our simple estimate we consider circular orbits because then the factor ${R_{\rm p}^2}/{d_{\rm p}^2}$ does not change with orbital phase $\phi$. Its value is $1.85\cdot 10^{-8}$ for $R_{\rm p}={\rm R_{\rm J}}$ and $d_{\rm p}=3.53~$au. The relation between $\alpha$ and $\phi$ depends on the inclination~$i$ of the orbital plane  
\begin{equation}
\alpha = {\rm arccos}(\sin(i) \cos(\phi))\,,
\end{equation}
where $\phi=0^\circ$ is the phase when the planet illumination as seen by the observer is maximal. For a face-on orbit ($i=0^\circ$) the scattering angle is constant at $\alpha =90^\circ$. For an edge-on orbit ($i=90^\circ$), there is $\alpha=\phi$ and
we define $\alpha=0^\circ$ for the back-scattering configuration when $f(\alpha)$ is maximal and $f(0^\circ)$ identical to the geometric albedo of the planet, while $f(180^\circ)=0$. The $p(\alpha)$ phase curve has typically its maximum close to right-angle scattering $p_{\rm max} \approx p(90^\circ)$. 

\begin{figure}[t!]
    \includegraphics[trim=30 10 45 2, clip=true,width=0.46\textwidth]{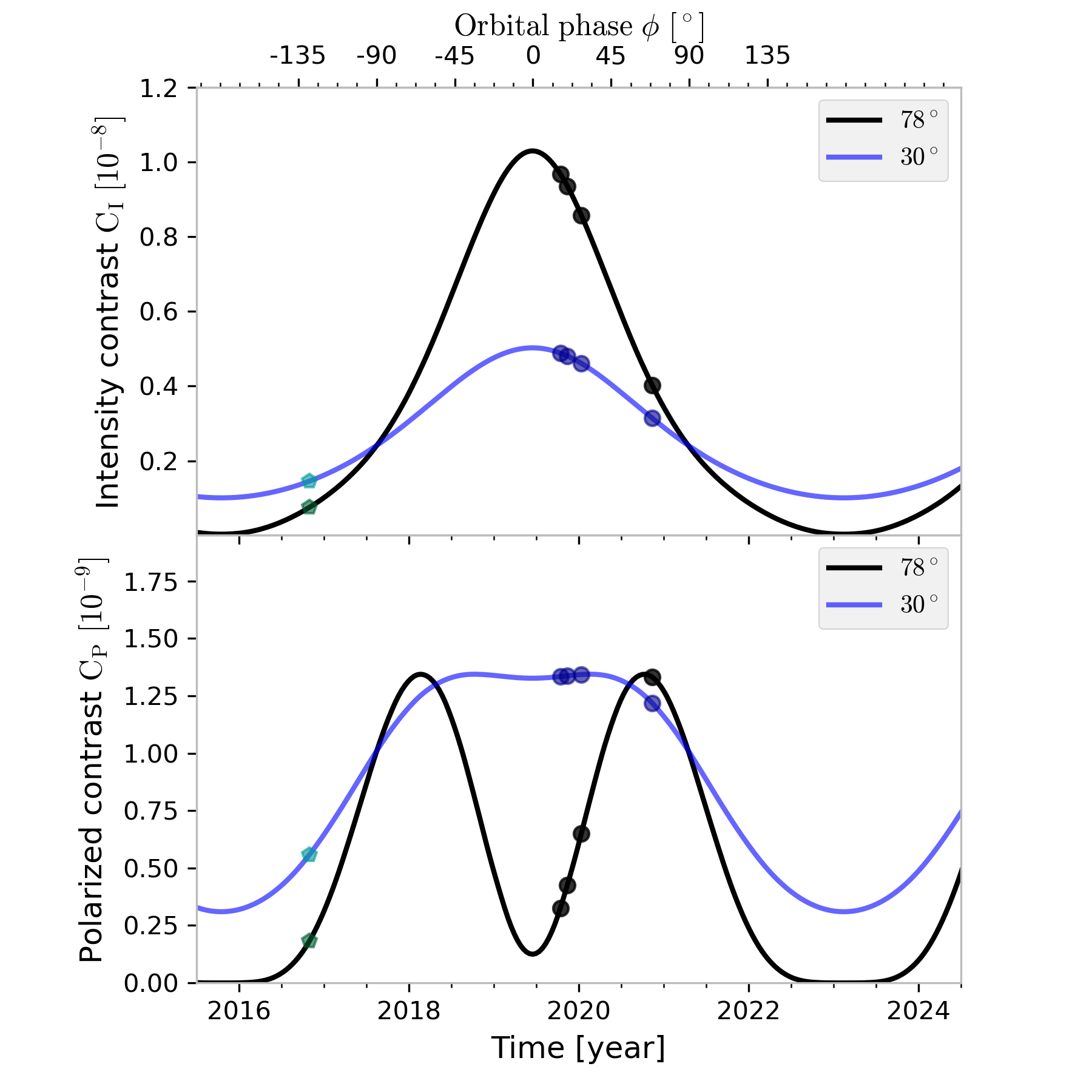}
    \centering
\caption{Planet model contrast as a function of time for the intensity (upper panel) and the polarized intensity (lower panel) for the orbit with an inclination of $78^\circ$ (black curves) as in Figure~\ref{1_orbit} and for an inclination of $30^\circ$ (blue curves). The dots show our four observing epochs, while the octagon stands for earlier ZIMPOL observations in \citet{Hunziker20}.} \label{1_orbit_b}
\end{figure}

Figure~\ref{1_orbit} illustrates the expected prograde orbital motion of $\epsilon$~Eri~b and the time of our four observing epochs. The inserted PSFs are splitted left/right and show the relative signal strengths along an orbit for the intensity in red and the polarized light in blue. The orientation and inclination of the plotted orbit is not well
known and therefore tentative.

The movement of the planet is small enough to allow the combination of data from
consecutive nights of an epoch. The maximum separation of the planet is 1.1$\arcsec$ and an orbit takes 2671~d \citep{Llop21}. If the orbital inclination is $0^\circ$ and perfectly circular, then the orbital speed is constant: $(2\pi r)/P = 2 \pi\cdot 1.1\arcsec/2671 = 2.585$~mas/d $ = 0.718$~px/d. The projected speed is equal or slower for inclined orbits 
and particularly small near quadrature phase for high
$i$.

The reflectivity $f(\alpha)$ and the fractional polarization $p(\alpha)$ of a planetary atmosphere depends on the atmospheric structure, which is hard to predict. 
There exist detailed model calculations for the reflected intensity and polarized intensity of giant planets \citep[e.g.][]{Stam04,Bailey18} 
and simple parameteric models \citep[e.g.][]{Seager00,Buenzli09,Madhusudhan11}.
For our estimate we just pick a model for a Rayleigh scattering planet from \citet{Buenzli09}, with an optical depth $\tau_{\rm sc} = 2$ for a Rayleigh scattering layer, with a single scattering albedo of $\omega_R = 0.95$, above a cloud layer with a Lambertian surface with an albedo of $A_{\rm S} = 1$. This is the same model as shown in Figure~1 of \citet{Hunziker20}. This model produces a polarized reflectivity 
$p(\alpha)f(\alpha)$ of about $\approx 0.055$ for quadrature phase $\phi=\alpha=90^\circ$. 

In Figure~\ref{1_orbit_b} we show the expected intensity contrast $C_{\rm I}$ and polarized intensity contrast $C_{\rm P}$ (Equation~\ref{eq:Cpol}) for $i=78^\circ$ and $30^\circ$.
For inclined orbits $p(\alpha)f(\alpha)$ and therefore also $C_{\rm P}$ is small or very small for orbital phases $\phi\approx 120^\circ-240^\circ$ when the planet is closer to us than the star so that we can hardly see the illuminated hemisphere. For $i\ge 30^\circ$ the favourable orbital phases are around $\phi\approx \pm 70^\circ$, when the planet is further away from us than the star so that we see a substantial fraction of the illuminated hemisphere, but still near a scattering angle which is strongly polarizing. The polarized reflectivity is then at ${\rm max}(p(\alpha)f(\alpha))\approx 0.07$ or about $25~\%$ higher than at quadrature phase. For conjunction $\phi=0^\circ$ or maximum illumination the polarized reflectivity is still high for low inclination $(i\approx 30^\circ)$, but has a dip for $i\approx 40^\circ-70^\circ$, while there is only a weak signal for $i>70^\circ$, because then the back-reflection produces only a very small scattering polarization. 
We show the expected contrast as a function of time: in the upper panel for the intensity contrast $C_{\rm I} = f(\alpha) \cdot {R_{\rm p}^2}/{d_{\rm p}^2}$ and in the lower image for the polarized intensity contrast $C_{\rm P}$ (Equation~\ref{eq:Cpol}). 
The black curve illustrates the signal strength for the orbit as shown in Figure~\ref{1_orbit} with an inclination of $78^\circ$ and the blue curve for an alternative inclination of $30^\circ$ closer to the inclination of the outer dust. The alternative curve for $30^\circ$ only shows the effect of a different illumination and scattering of the same hypothetical planet. This means that this curve does not account for the fact that the planet with an orbital inclination of $30^\circ$ has, according to RV measurements and the resulting factor $m \sin(i)$ a higher mass. This higher mass could lead to a different planetary radius which ultimately would change the signal strength overall. The dots in the colour of the curves show our four observing epochs in 2019 and 2020, while the octagons stand for earlier SPHERE/ZIMPOL observations in \citet{Hunziker20}. Trusting the available RV data, our fourth observing epoch is at the ideal timing with a large angular separation and a scattering angle close to 90$^\circ$ for a strong polarization signal.

The curves in Figure~\ref{1_orbit_b} are only rough estimates, because the reflectivities $f(\alpha)$ and $p(\alpha) f(\alpha)$ depend on atmospheric parameters \citep{Buenzli09}). For orbits with an eccentricity of $e \approx 0.2$ the separation $d_{\rm p}(\phi)$ can be 20~\% larger or smaller during the orbit. Also the adopted planet radius $R_{\rm p}={\rm R_{\rm J}}$ could be larger or smaller by 10~\% to 20~\% as follows from the distribution of measured giant planet radii \citep{Thorngren19}. We conclude that the expected planet signal could be up to a factor two larger or a factor of a few smaller than the estimates given above. Circumplanetary rings similar to Saturn could also boost the signals of the reflected light by a factor of two or even more \citep{Arnold04,Dyudina05}.

\subsection{Predictions for the scattered light from warm dust}
\label{SectDustModel}

The spectral energy distribution of $\epsilon$~Eri shows a multi-component infrared excess with a 'cold' dust component in the far-infrared and a warm dust component peaking around 20~$\mu$m. The warm component has a relative luminosity of $L_{\rm warm}/L_\star\approx 3.3 \cdot 10^{-5}$, and it is estimated that this dust is located at a separation of roughly 3~au from the star according to \citet{Backman09} or
between 2.5 and 6 au according to \citet{Su17}, depending a lot
on the adopted grain properties.
This dust should also produce a scattered light signal within the field of view of SPHERE/ZIMPOL which covers about $r=5$~au.

\begin{figure}[t!]
    \includegraphics[trim=1 6 4 6, clip=true,width=0.4\textwidth]{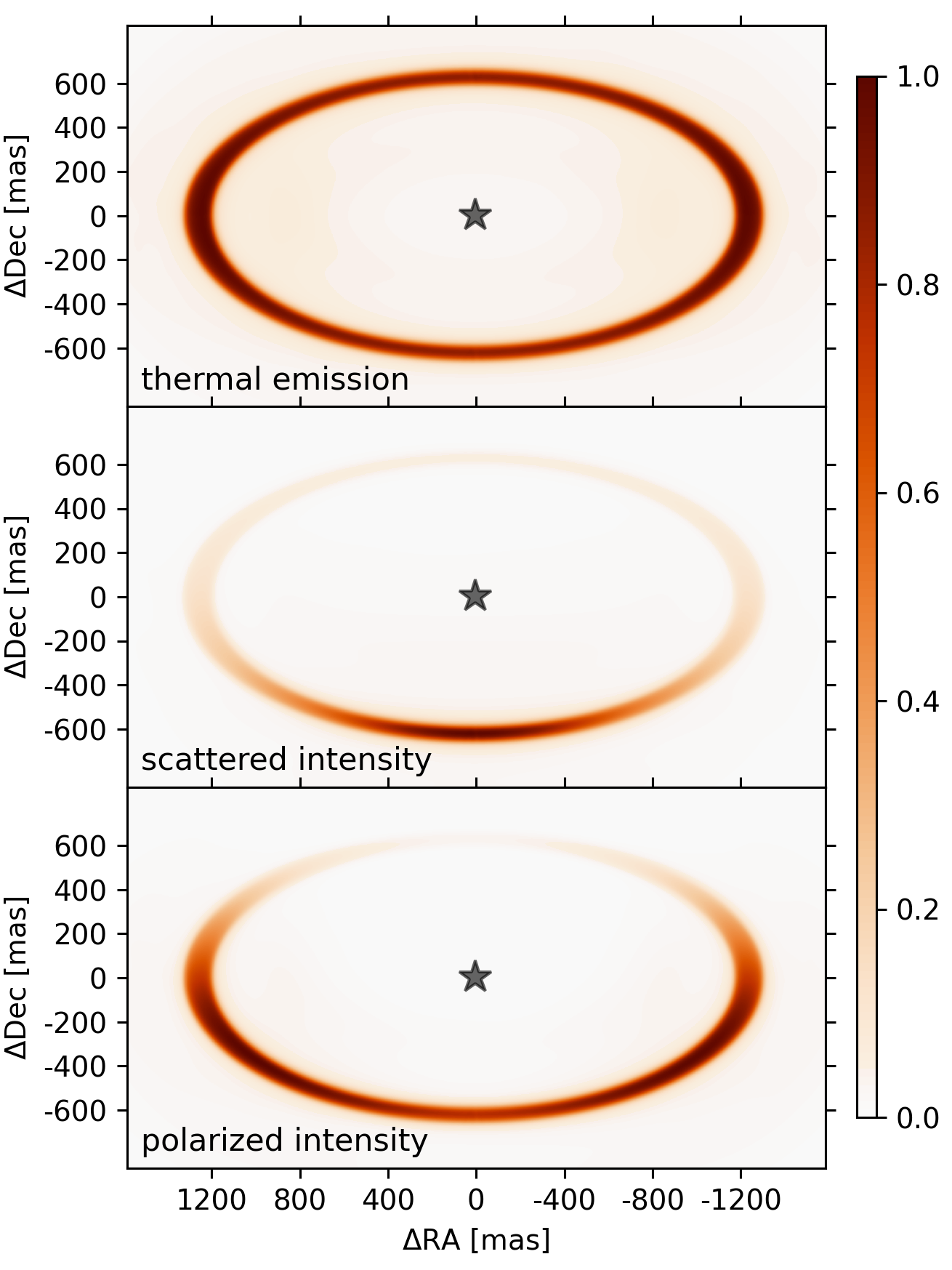}
    \centering
\caption{Relative flux distribution for a narrow dust ring model in $\epsilon$~Eri with $r=4$~au based on the warm component
in the infrared excess. Top: thermal emission. Middle: expected
scattered intensity $I_{\rm disk}$ for $g=0.6$. Bottom: polarized 
intensity $Q_{\rm \phi}$ for $g=0.6$. The model images are
convolved with the PSF of SPHERE/ZIMPOL, normalized to their flux maximum and
aligned with the sky coordinate axes.} \label{2d_dustmodels_b}
\end{figure}

Based on this infrared-emission of the warm dust, we can make estimates about the polarization $Q_{\rm \phi}$, which depend strongly on the adopted dust scattering properties and the disk geometry. We use for a rough estimate of $Q_{\rm \phi}$ simple axisymmetric and optically thin models as illustrated by a disk ring in Figure~\ref{2d_dustmodels_b}. 

The thermal emission of the dust is radiated isotropically, while the disk integrated polarized emission $Q_{\rm \phi}/I_\star$ depends strongly on the disk inclination and the polarized scattering phase function $f_{\rm \varphi}(\theta)$ of the dust, where $\theta$ is the scattering angle measured as angle of deflection. 
The function $f_{\rm \varphi}(\theta)= f_{\rm I}(\theta) \times p(\theta)$ is described by a Henyey-Greenstein function $f_{\rm I}=f_{\rm HG}(\theta,g)$ with asymmetry parameter $g$ for the distribution of the scattered intensity, while the fractional polarization has the same angle dependence as for Rayleigh scattering $p(\theta,p_{\rm max})=p_{\rm max}(\sin^2(\theta))/(1+\cos^2(\theta))$, but with a scaling factor of $p(90^\circ)=p_{\rm max}$ which accounts for the reduced scattering polarization produced by large, compact dust particles. The adopted parameters are 
roughly representative for the zodiacal dust in the solar system \citep{Leinert75}.

The integrated polarization $Q_{\rm \phi}/I_\star$ of an optically thin,
axisymmetric disk depends on the total cross section of
the scattering dust $\sigma$, and on a disk averaged polarized scattering phase function $\langle f_{\rm \varphi}(i,g,p_{\rm max})\rangle$ \citep{Schmid21}. One can relate $\langle f_{\rm \varphi}(i,g,p_{\rm max})\rangle$ to the intensity phase function for isotropic scattering $\langle f_{\rm I}(g=0)\rangle$ and with 
the total dust absorption cross section~$\kappa$ also to the isotropically emitted infrared excess according to
\begin{equation}
\frac{Q_{\rm \phi}}{I_\star} = \frac{\sigma}{\kappa}\, \frac{\langle f_{\rm \varphi}(i,g,p_{\rm max})\rangle}
   {\langle f_{\rm I}(g=0)\rangle}\, \frac{L_{\rm warm}}{L_\star}\,.
\end{equation}
We obtain for $i=60^\circ$, $g=0.6$ and $p_{\rm max}=0.25$ the disk phase function $\langle f_{\rm \varphi}(60^\circ,0.6,0.25)\rangle/\langle f_{\rm I}(g=0)\rangle = 0.085$, and with $\sigma/\kappa=\omega_d/(1-\omega_d)=1.0$ the intrinsic, disk integrated polarization signal of $Q_{\rm \phi}/I_\star = 0.085\cdot {L_{\rm warm}}/{L_\star}=2.8\cdot 10^{-6}$ for the warm dust in $\epsilon$~Eri. 

\begin{figure}[t!]
    \includegraphics[trim=10 18 12 48, clip=true,width=0.48\textwidth]{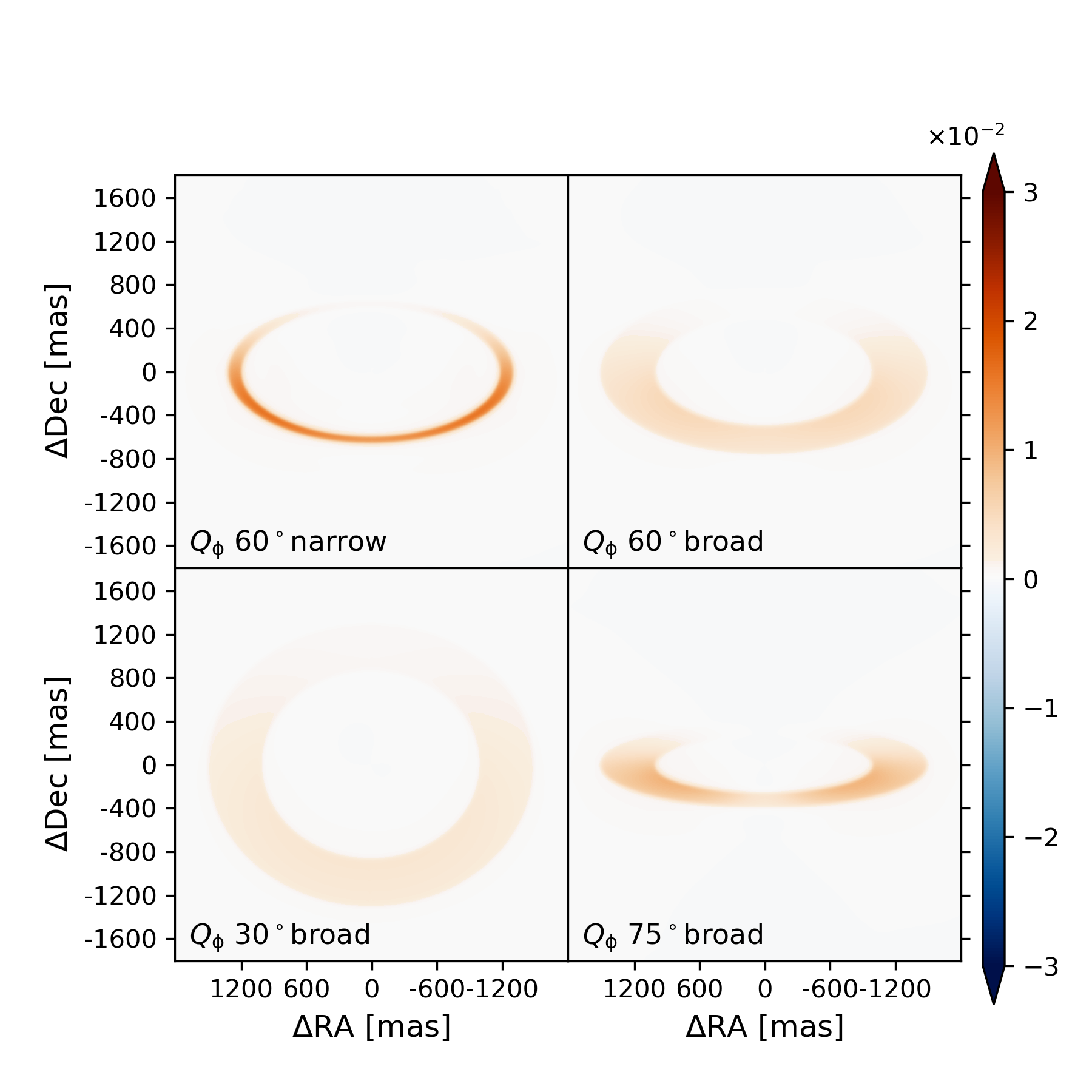}
    \centering
\caption{Predicted polarized flux $Q_{\rm \phi}$ [ct/(s $\cdot$ px)] produced by
scattering in a narrow and a broad dust ring model for $\epsilon$~Eri with an inclination 
of $i=60^\circ$ (top row), and for broad rings with $i=30^\circ$ and $i=75^\circ$ (bottom row). All models would produce the measured infrared excess of 
$L_{\rm warm}/L_\star=3.3\cdot 10^{-5}$.
} \label{2d_dustmodels}
\end{figure}

For the comparison with our observations we can now calculate the
$Q_{\rm \phi}$ signal with the dust parameters given above and 
investigate the expected surface brightness for the polarized
signal $S\!B_{\rm p}$ for different disk inclinations and 
radial distributions of the dust.
Figure~\ref{2d_dustmodels} shows four 
cases with a mean ring radius $r_{\rm c}=1.25 \arcsec$, one with a 
small width of $\Delta r_{\rm narrow}=0.1\cdot r_{\rm c}$ and 
$i=60^\circ$, and three broader disk rings 
$\Delta r_{\rm broad} = 0.4\cdot r_{\rm c}$, with $i=30^\circ$, $i=60^\circ$ and $i=75^\circ$ to investigate the inclination dependence.
All four models produce the same amount of infrared excess $L_{\rm warm}/L_\star=3.3\cdot 10^{-5}$
and they are convolved with the mean PSF of the $\epsilon$~Eri data for a prediction of the expected observational signal $Q_{\rm \phi}/I_\star$ as given in Table~\ref{Modeldisks} (Column~5).

Most important for dust detection is the resulting surface brightness of the polarized intensity $S\!B_{\rm p}$ which depends strongly on the disk geometry. 
It is clearly visible in Figure~\ref{2d_dustmodels}, that a much higher surface brightness $S\!B_{\rm p}$ is obtained for inclined disks and for the narrow dust ring when compared to the wide ring. In our simple models with constant dust emissivities $\epsilon(r)=\epsilon_0$, the intrinsic surface brightness is anti-correlated with the disk widths $S\!B_{\rm p}\propto 1/\Delta r$. After convolution the ratio of the peak $S\!B_{\rm p}$ between the narrow and wide ring models is about a factor of three, as the convolution degrades more the peak surface brightness flux of narrow structures.

The three models for the wide disks illustrate that $S\!B_{\rm p}$ is lower for low $i$ because only a small amount of light is scattered towards the observer by dust with strong forward scattering. For high $i$ the intensity surface brightness $S\!B_{\rm I}$ of the front side increases strongly 
because of the strong forward scattering. The polarization $S\!B_{\rm p}$ has 
a dip on the disk front for $i=75^\circ$ because small
scattering angles $\theta < 30^\circ$ 
produce only a weak 
polarization $p(\theta)$. The locations with the maximum $S\!B_{\rm p}$-signal shifts for higher $i$ towards the apparent major axis of the projected disk, where the scattering angle is close to $\theta\approx 90^\circ$ and therefore 
$p(\theta)\approx p_{\rm max}$.

\begin{table}[t]
\caption{Integrated disk polarization, intensity, 
and peak surface brightness contrasts for the selected 
scattering models.}
\label{Modeldisks}
\begin{tabular}{lcccccc}
\hline
\hline
\noalign{\smallskip}
\noindent $i$     & geom. & $Q_{\rm \phi}/I_\star$ 
                          & $I_{\rm disk}/I_\star$ 
                                  & $Q_{\rm \phi}/I_\star$
                                      & $\Delta{S\!B}_{\rm p} $ 
                                             & $\Delta{S\!B}_{\rm I}$ \cr
\noindent [$^\circ$]  &  &   intrin. & intrin. & conv.   & \multicolumn{2}{c}{mag/arcsec$^2$}  \cr  
\noalign{\smallskip}
\hline
\noalign{\smallskip}
\noindent 60    &  narrow &  2.80e-6  & 30.7e-6   &  2.29e-6   &  13.9  &  10.5   \cr 
\noalign{\smallskip}
\noindent 30    &  broad  &  3.14e-6  & 16.3e-6   &  2.65e-6   &  15.6  &  13.5   \cr
\noindent 60    &  broad  &  2.80e-6  & 30.7e-6   &  2.28e-6   &  15.1  &  11.7   \cr
\noindent 75    &  broad  &  2.51e-6  & 45.9e-6   &  1.79e-6   &  14.4  &  10.3   \cr
\noalign{\smallskip}
\hline
\end{tabular}
\tablefoot{The columns give the disk inclination $i$, the geometry of the disk, the intrinsic, disk integrated polarization $Q_{\rm \phi}$ and intensity
$I_{\rm disk}$ relative to the stellar intensity $I_\star$, 
the $Q_{\rm \phi}/I_\star$-value after PSF convolution, and the peak surface 
brightness contrasts in mag/arcsec$^2$ for the polarization $\Delta{S\!B}_{\rm p}$ and the intensity $\Delta{S\!B}_{\rm I}$ derived after PSF convolution.}
\end{table}

The models provide expected maximum surface brightness contrasts $\Delta {S\!B}_{\rm p}$ which is $\Delta {S\!B}_{\rm p}={S\!B}_{\rm p}-m_\star$ 
for the brightest disk section ${S\!B}_{\rm p}$ measured relative to the central star $m_\star$. For the narrow disk model ($i=60^\circ$) the intrinsic contrast is $\Delta {S\!B}_{\rm p}= 13.3~{\rm mag/arcsec}^2$, or $\approx 13.9~{\rm mag/arcsec}^2$ if we also consider the signal degradation by the PSF convolution. Compared to the stellar PSF peak ${S\!B}(0)$, the brightest disk region is about ${S\!B}_{\rm p}-{S\!B}(0)\approx 20.3~{\rm mag/arcsec}^2$ fainter because there is 
${S\!B}(0)-m_\star \approx -6.4~{\rm mag/arcsec}^2$ between the peak and the total brightness of the central star. 
This disk polarization to PSF peak contrast is about seven times higher ($\approx 7\cdot 10^{-9}$) than the point source contrast calculated for the RV planet in Section~\ref{SectPlanetModel}. In addition, the two brightest regions of the disk on opposite sides of the ring are roughly $0.125\arcsec$ wide and $0.4\arcsec$ long and cover each a surface area of about $50\cdot 10^{-3}~{\rm arcsec}^2$, while a point source has only a size of about $1\cdot 10^{-3}~{\rm arcsec}^2$. Thus, the signal-to-noise ratio for the predicted disk polarization signal is about a factor of $\sim 50$ higher than for the RV planet, if the data would only be limited by photon noise.

The situation is less favourable for wider disks because the peak contrasts are only $\Delta{S\!B}_{\rm p}\approx 15.6$, $15.1$, and 14.4~${\rm mag/arcsec}^2$ (Table~\ref{Modeldisks}) for $i=30^\circ$, $60^\circ$ and $75^\circ$, respectively. The signal is distributed over a larger area and might therefore become visible with strong pixel binning.

The intensity signal of the scattered radiation has for inclined disks a strong maximum
on the disk front side because of the strong forward scattering (Figure~\ref{2d_dustmodels_b}). The highest surface brightness contrast for the narrow disk with $i=60^\circ$ is about
$\Delta {S\!B}_{\rm I}= 10.5~{\rm mag/arcsec}^2$ or a factor of 22 brighter than for the peak polarization signals. 
Despite this, we expect for SPHERE/ZIMPOL observations a higher disk detection sensitivity with polarimetry because the polarized disk signal can be distinguished from the strong steller intensity halo. Nonetheless, a bright disk detected with imaging polarimetry could also produce a detectable intensity signal in the data. Figure~\ref{2d_dustmodels_b} shows also the expected surface brightness distribution for the thermal radiation of the dust in the mid-infrared. All these model results depend strongly on the adopted disk parameters. Therefore, a search should also consider signals which could be a factor of a few higher or lower than the predictions for the selected disk models shown in Figure~\ref{2d_dustmodels}.

\begin{table*}[t]
\caption{\label{avdataSPHERE}Parameters for the used SPHERE/ZIMPOL
polarimetric observation cycles of $\epsilon$~Eridani.} 
\label{obsSHOW}
\begin{tabular}{lcccccccl}
\hline
\hline
\noalign{\smallskip}
N &  date &  DIT  & $n_{\rm cyc}$ (usable) &  $t_{\rm exp}$ (usable) & seeing  & $\tau_{\rm 0}$ & field rotation &  note \\
      &       & [s]   &               &                &  [\arcsec]   &   [ms]       &  [°]   &\\
\noalign{\smallskip}
\hline
\noalign{\smallskip}
$1$  & 2019-10-10  & 3 & 48 (39) & 3h 12min (2h 36min)& 0.71 [ 0.52, 1.45 ]    &  5.6 [3.2,11.1]  & 118  & $\rm rd^{1}$ \\
$2$  & 2019-10-11  & 3 & 48 (35) & 3h 13min (2h 20min)& 0.63 [0.43, 1.54]    &  7.4 [3.6,13.6]  & 92  & $\rm rd^{1}$, $\rm lw^{2}$ \\
\noalign{\smallskip}
$3$  & 2019-11-30  & 5 & 44 (41) & 2h 56min (2h 44min)& 0.70 [0.41, 1.0]    &  4.5 [2,8.8]  & 127  & $\rm rd^{1}$ \\
$4$  & 2019-12-01  & 5 & 46 (43) & 3h 4min (2h 52min)& 0.76 [0.52, 1.2]    &  5.5 [3.4, 8.1]  & 131  & $\rm rd^{1}$ \\
$5$  & 2019-12-02  & 5 & 42 (37) & 2h 50min (2h 28min)& 1.12 [0.59, 2.02]    &  2.5 [1.5, 5.9]  & 137  & $\rm rd^{1}$ \\
$6$  & 2019-12-03  & 5 & 48 (47) & 3h 12min (3h 8min)& 0.66 [0.45, 0.95]    &  6.1 [3.6,9.2]  & 136  & $\rm rd^{1}$ \\
\noalign{\smallskip}
$7$  & 2020-01-12  & 5 & 37 (36) & 2h 28min (2h 24min)& 0.92 [0.59, 2.41]    &  4.6 [2.2, 6.8]  & 34  & $\rm rd^{1}$ \\
$8$  & 2020-01-13  & 5 & 39 (39) & 2h 36min & 0.86 [0.62, 1.35]    &  4.3 [2.1, 8.5]  & 44  & $\rm rd^{1}$ \\
$9$  & 2020-01-14  & 5 & 39 (39) & 2h 36min & 0.65 [0.42, 0.86]    &  7.7 [4.1, 11.8]  & 48  & \\
\noalign{\smallskip}
$10$  & 2020-11-23  & 5 & 89 (82) & 5h 58min (5h 28min) & 0.49 [0.32, 1.3]    &  5.8 [2.9, 11.5]  & 148  &  \\
$11$  & 2020-11-24  & 5 & 79 (72) & 5h 19min (4h 48min) & 0.65 [0.37, 1.09]    &  6.5 [2.8, 11.4]  & 142  &  \\
$12$  & 2020-11-25  & 5 & 73 (68) & 4h 53min (4h 32min) & 0.99 [0.71, 1.6]    &  4.1 [2, 6.8]  & 145  &  \\
\noalign{\smallskip}
\hline
\end{tabular}
\tablefoot{Seeing and atmospheric coherence time $\tau_0$ are median and [min, max] values for the $\epsilon$~Eridani data of that night (N); all observations taken in fast polarization detector mode and VBB filter for both ZIMPOL cameras; $t_{\rm exp}$ is
$n_{\rm cyc}\times 4 \times {\rm nDIT}\times {\rm DIT}$; $\rm rd^{1}$: readout electronic issue camera 1; $\rm lw^{2}$: low wind effect. 38~h~32~min is total usable time of the coronagraphic cycles. Epochs of consecutive days are grouped with small skips.}
\end{table*}

\section{Observations and data analysis}\label{3ObsMeth}
\subsection{SPHERE/ZIMPOL instrument}\label{13}
The planet $\epsilon$~Eri~b was searched with direct imaging using the SPHERE instrument \citep{Beuzit19} at the Nasmyth focus of the VLT unit telescope UT3 of the European Southern Observatory (ESO). This instrument consists of an extreme adaptive optics (AO) system, an image derotator, stellar coronagraphs and three focal plane instruments \citep{Fusco06,Petit14,Sauvage14,Fusco14} including the Zurich Imaging Polarimeter (ZIMPOL) used for this programme. ZIMPOL works in the visual spectral regime 500-900~nm and is tuned for the search of reflecting planet and circumstellar disks using fast-modulation imaging polarimetry with a modulation frequency of about 1~kHz to 'freeze' the speckle variations \citep{Schmid18}. The SPHERE AO system achieves under good observing conditions regularly a Strehl ratio of about 40~\% in 
the I-band \citep{Fusco15} and a resolution of about 25~mas full width at half maximum (FWHM). ZIMPOL has a detector field of view of $3.6\arcsec \times 3.6\arcsec$ and a pixel scale of 3.6~mas~$\times$~3.6~mas and is equipped with different coronagraphs, filters, and instrument calibration components. For polarimetry, the telescope and instrument polarization are compensated and calibrated with rotatable half-wave plates and further calibration components \citep{Bazzon12}. 
ZIMPOL has two arms, each of them with one camera and its own filter wheel, and data are taken simultaneously in both arms. The two CCD detectors are operated in frame transfer mode and provide a high gain, fast read-out mode with small detector overheads for high flux applications. 

The deep planet search with SPHERE/ZIMPOL is based on the high-contrast imaging, which provides at a separation of about $1\arcsec$ a raw contrast at the level of $10^{-4}$ using AO and coronagraphy. Combining this with polarimetric differential imaging (PDI) and angular differential imaging (ADI) gives an additional polarimetric contrast improvement of about $10^{-4}$. This provides a total contrast at the level of about $C_{\rm P} \approx 10^{-8}$ and this can be improved further by reducing the photon noise with sufficiently long integrations for bright targets \citep{Schmid06a,Thalmann08,Hunziker20}.

\subsection{Observations}
The observations of $\epsilon$~Eridani were taken in visitor mode with an observing strategy similar to the observations of \citet{Hunziker20}. Three runs were
executed between Oct. 2019 and Jan. 2020 and one run in Nov.~2020. This planning considered the planet motion from run to run, and the combination of the data 
with a Keplerian motion prediction.
The observation log and basic information on exposure times and observing conditions are listed in Table~\ref{obsSHOW}.

The ZIMPOL instrument setup was optimized for the highest possible throughput, using the Very Broad Band (VBB) filter ($\lambda_{\rm c}=735.4~{\rm nm},\,\Delta\lambda=290.5~{\rm nm}$) for both camera arms, polarimetry in fast modulation mode and a classical Lyot coronagraph (V$\_$CLC$\_$MT$\_$WF) with a spot radius of $\rho = 77.5$~mas. The mask has a transmission of around 0.1~$\%$ so that, under good conditions the stellar spot is visible behind the Lyot mask and can be used for centring. Approximately once per hour we were offsetting the star from the coronagraph and added the ND2 neutral density filter (reduces flux by a factor of about 100) to measure the flux throughput and the PSF shape for an improved beam-shift correction (see appendix~\ref{A3_bs}).

Observations are taken in polarimetric cycles which consist of a sequence of $Q^+$, $Q^-$, $U^+$ and $U^-$ images taken with four different half wave plate (HWP) orientation each with nDIT subintegrations. The $Q^+$ and $Q^-$ data provide the total intensity for the $Q$-measurement $I_{\rm Q}=I_0+I_{90}$ and Stokes $Q=I_0-I_{90}$, which is already corrected for the instrument polarization thanks to the switch between the $Q^+$ and $Q^-$ measurement. This does not correct for the telescope polarization introduced before the HWP switch. The equivalent parameters are obtained for Stokes $U=I_{45}-I_{135}$ and $I_{\rm U}=I_{45}+I_{135}$ with $I_{\rm Q} = I_{\rm U} = I$. The detector integration time DIT = 3~s was initially used for the first two runs, but then extended to DIT = 5~s for a higher count level with typically $>100$~counts, or photo electron numbers of $N_{\rm e}>1000$~e$^-$ (gain factor of 10.5~e$^-$/ct) for $\rho \gtrsim 1.2\arcsec$ (see intensity PSF in Figure~\ref{PolarimetricCycle}).
This level is required to achieve photon noise limited observations, with $(N_{\rm e})^{1/2}$ larger than the read-out noise level of $N_{\rm ron}\approx 20$~e$^-$ (2~counts). 

The ZIMPOL P1 derotator mode was used which is optimized for polarimetry because the derotator and all other components after the rotating and switching HWP2 with their corresponding instrument polarization effects are constant during the night. Therefore the sky rotates in the image and very importantly this allows to use ADI together with PDI to correct better for quasi-static speckles and other instrumental effects of SPHERE and ZIMPOL. However, in P1 mode also the telescope pupil rotates, most notably the M2 spider pattern, but with a different rotation law than the sky image.

\subsection{Data reduction}\label{datared}
The data were mainly reduced with the IDL-based sz (SPHERE-ZIMPOL) software developed at the ETH Zurich. Basic steps include signal extraction for the two simultaneously measured polarization modes, bad pixel cleaning, bias subtraction, flat-field correction, calibration of the polarimetric modulation efficiency, and the polarimetric combination of the four frames of each cycle. Additionally the frame transfer smearing was corrected in the intensity frames by subtracting the average row level multiplied by 56~ms/DIT to account for the illumination during the frame transfer for the detector mode used for fast modulation polarimetry. Important for our analysis are the corrections for the differential polarimetric beam shifts \citep{Schmid18,Hunziker20}, which can be tricky to define from coronagraphic data. Therefore we also use the PSF images as described in Appendix~\ref{A3_bs}. It is important for polarimetry to consider the telescope polarization which adds a fractional polarization $p_{\rm tel}$ with a position angle $\delta_{\rm tel}$ to the signal. This is rotating with the parallactic angle of the telescope $\theta_{\rm para}$ so that the uncorrected polarization of $\epsilon$~Eridani rotates along a circle in the $Q/I$-$U/I$-plane. The correction for the telescope polarization and the second order radial dependence thereof are described in Appendix~\ref{A2_telpol}, while particular detector corrections are discussed in \ref{A1_cam1rd} and \ref{A4_ct}.

For the search of scattered light from a planet and from an extended cloud of dust, we use the azimuthal Stokes parameters $Q_{\rm \phi}$ and $U_{\rm \phi}$ with respect to the central star. The following formulas relate the Stokes $Q$ and $U$ to the azimuthal polarization $Q_{\phi} = - Q \cos(2 \phi) - U \sin(2 \phi)$ and $U_{\phi} = - Q \sin(2 \phi) + U \cos(2 \phi)$, where $\phi$ is the position angle with respect to the central star measured from north over east \citep[][]{Schmid06,Monnier19}. The scattered light from a planet or a pole-on disk is mostly azimuthally polarized and therefore should be visible in the $Q_{\rm \phi}$ image only, while $U_{\rm \phi}$ is zero.

In Figure~\ref{PolarimetricCycle} we show one complete polarimetric cycle from night 9 after all the data reduction steps mentioned above. The plot is splitted to emphasize the large difference in residual signal between the total intensity $I$ in the top half and Stokes $Q_{\rm \phi}$ for the lower half. In the intensity half, one can clearly see the bright speckle ring at around 0.45$\arcsec$ up to which the AO system corrects well the atmospheric seeing. Outside of $0.5\arcsec$ the residual intensity halo of the star decreases steadily with separation. Additional features are the four cross shaped stripes from the secondary mirror mount (telescope spiders) which are partially visible, dominant PSF features left and right of the star in the speckle ring, and small black astrometric dots from the coronagraph mask, for example at $(y=+1\arcsec,x=-1\arcsec,0\arcsec,+1\arcsec)$. In \citet{Schmid18} full plots of many PSFs for SPHERE/ZIMPOL are shown and described. The differential polarization signal $Q_{\rm \phi}$ is much smoother and with significantly lower count numbers than the intensity image (note colour scales in Figure~\ref{PolarimetricCycle}). The exposure time of this polarimetric cycle is $\rm NDIT~\times~DIT~\times$~four half wave plate positions $= 12 \cdot 5 $~s~$ \cdot\ 4 = 240$~s.

\section{Planet search}\label{4Planet}
\subsection{Pushing the limits}\label{postproc}

Additional steps are necessary for the search of a reflecting planet because the expected signal is much weaker than the stellar halo in the coronagraphic intensity image or the noise in the differential $Q_{\rm \phi}$ image. The aim is essentially to reduce the relative noise in the halo by averaging a large amount of data such as in Figure~\ref{raw_mean_09} where all observations of night 9 are rotationally aligned, so that north is up and east to the left, and combined to enhance the S/N for the detection of a faint companion. We have injected fake planets at different separations and azimuthal locations where the points for a given angle have all the same contrast value as indicated. A source with a given contrast value is easier to detect at larger separations and much easier for a polarized contrast value $C_{\rm P}$ in the $Q_{\rm \phi}$ image than the corresponding intensity contrast $C_{\rm I}$ in the intensity image. The PSF peak flux of the star is roughly $6.5 \cdot 10^6~{\rm ct/(px \cdot DIT)}$ and therefore a planet with a contrast of $10^{-6}$ has $6.5~{\rm ct/(px \cdot DIT)}$. 

\begin{figure}[t!]
    \includegraphics[trim=1 1 1 1, clip=true,width=0.46\textwidth]{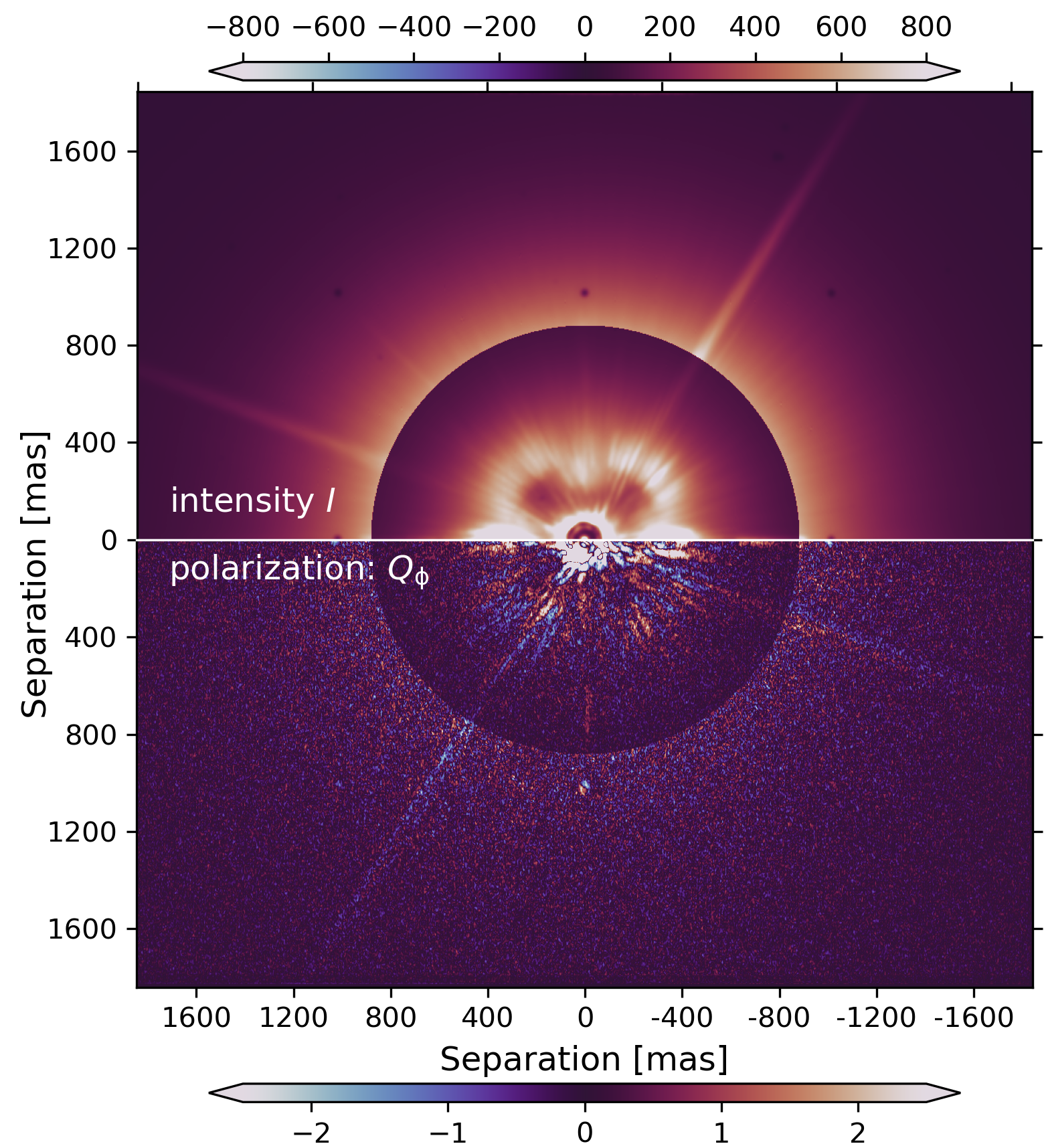}
    \centering
\caption{One polarimetric cycle $t_{\rm exp} = 240$~s 
corrected for the telescope polarization: The top half shows the intensity 
$I$ and bottom half the polarized intensity $Q_{\rm \phi}$ with corresponding colour 
scales in ct/(px$\cdot$ DIT). 
The counts inside 0.9$\arcsec$ are scaled down for a better visibility by a factor of six for the intensity and a factor of three
for $Q_{\rm \phi}$. 
} \label{PolarimetricCycle}
\end{figure}

\begin{figure}[t!]
    \includegraphics[trim=1 1 1 1, clip=true,width=0.46\textwidth]{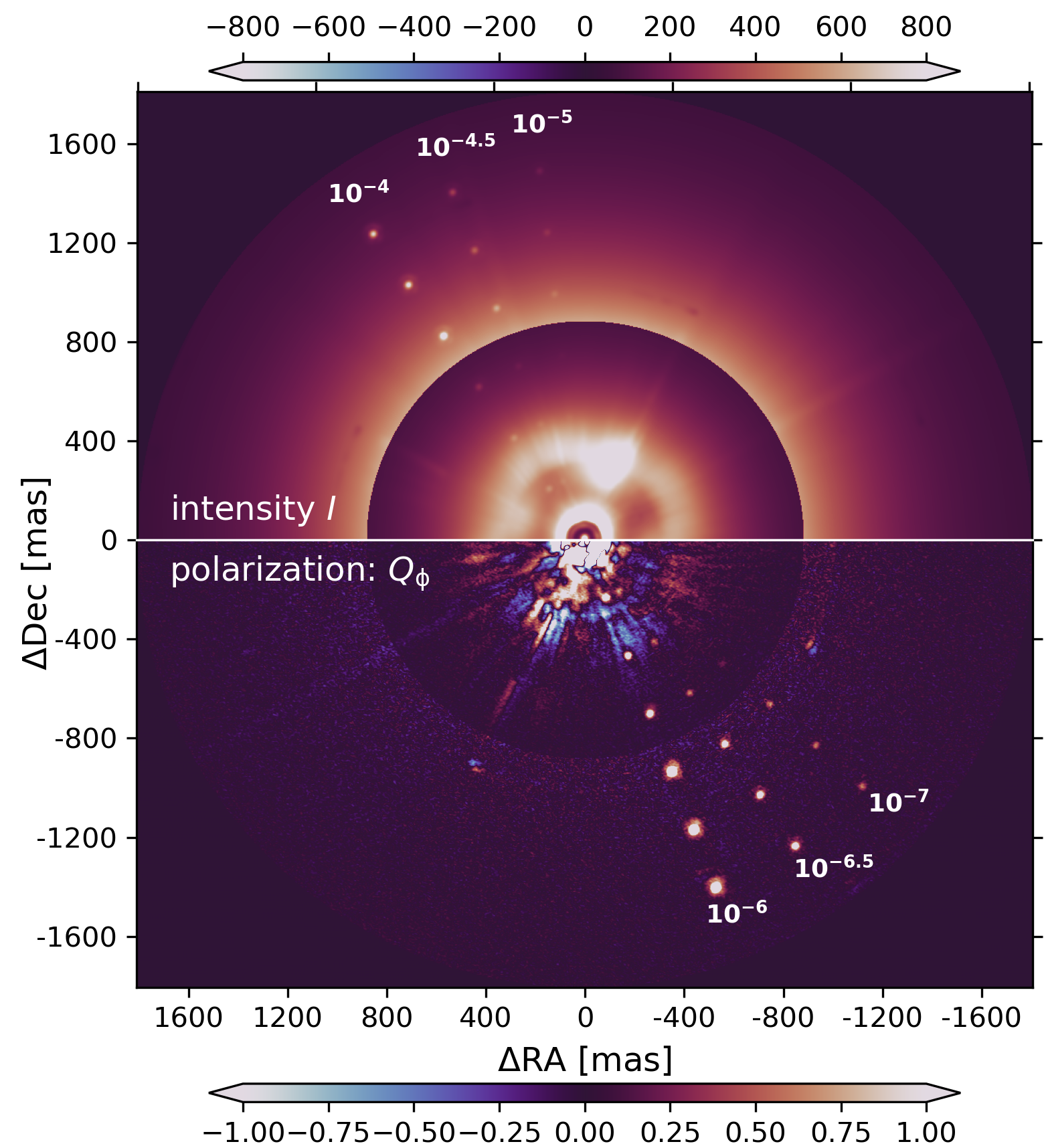}
    \centering
\caption{All data of night 9 ($t_{\rm exp}= 9360$~s), derotated, averaged,
and corrected for telescope polarization: The top half shows $I$ and the bottom half $Q_{\rm \phi}$ with corresponding colour scales in ct/(px $\cdot$ DIT) as in Figure~\ref{PolarimetricCycle}. Artificial point sources are inserted at different separations and contrasts $C_{\rm I}$ and $C_{\rm P}$ are changed with position angle as indicated. Counts inside 0.9$\arcsec$ are scaled down for better visibility with a factor of six for $I$ and a factor three for $Q_{\rm \phi}$.}
\label{raw_mean_09}
\end{figure}

The averaging reduces the noise for $Q_{\rm \phi}$ to a level of about $\pm 0.1$ ct/(px $\cdot$ DIT) depending on separation and this would allow a detection of a point source with a contrast of about $C_{\rm P}\approx 10^{-7}$ outside a separation $>0.9\arcsec$ as demonstrated by the artificial point sources inserted in the image. The rotational alignment smooths the stellar halo in the intensity image and averages out localised features fixed to the instrument such as the telescope spiders or the dark points from the coronagraphic mask. However, higher contrast limits require the subtraction of the strong PSF halo in the intensity data and a few improvements for the polarimetric data as described below. 

The data from $\epsilon$~Eri, such as those shown in Figures~\ref{PolarimetricCycle} and \ref{raw_mean_09}, allow rough estimates of the photon noise in the images compared to a planet signal with a contrast of $C_{\rm I}=4\cdot 10^{-9}$. Such a planet would produce an intensity signal with a PSF peak flux of $0.004~{\rm ct/(px \cdot s)}$ because the corresponding $\epsilon$~Eri peak flux is about $10^6~{\rm ct/(px \cdot s)}$. This competes at $\rho \approx 1\arcsec$ with a count intensity for the stellar halo of about $I_{\rm ct}\approx 40~{\rm ct/(px \cdot s)}$ or the total number of photons collected during 38.5 hours of $n_{\rm \gamma}\approx 6\cdot 10^{7}~{\rm px}^{-1}$. This considers the detector gain of 10.5~e$^-$/ct. The corresponding relative photon noise limit $(n_{\rm \gamma})^{-1/2}=1.3\cdot 10^{-4}$ can be expressed as statistical noise limit per pixel $I_{\rm ct}\cdot(n_{\rm \gamma})^{-1/2}\approx 0.005~{\rm ct/(px \cdot s)}$ for the entire data set around $1\arcsec$. 

This is comparable to the expected PSF peak of the planet with $C_{\rm I}=4\cdot 10^{-9}$ or a ${\rm S/N}\approx 1$ for a pixel near the PSF peak of the planet. Because the PSF has a FWHM of about 6 pixels the S/N for the whole planet PSF would be roughly at a level $\approx 5$ with respect to the photon noise of the stellar halo. 

The estimated polarimetric signal of the planet is lower, about $C_{\rm P}\approx 0.25\,C_{\rm I}=1\cdot 10^{-9}$ and one needs to take $Q$ and $U$ measurements, if the position angle of the polarization is not known. Therefore, twice the measuring time is required to collect $2\, n_{\rm \gamma}$ to reach a required relative photon noise limit of $(n_{\rm \gamma})^{-1/2}$ for the signal in the $Q_{\rm \phi}$ image. However, only in polarimeteric imaging the detection limit is close to the photon noise, while in intensity imaging the detection limits are clearly above the photon noise because of the strong speckle noise (Section~\ref{polintcomp}). 

\subsection{Post-processing} 

To push the noise levels further we combine in our post-processing ADI with fitting and subtraction of the PSF halo in the intensity imaging or of the large scale pattern in the residual differential polarization for polarimetric imaging. We also apply additional steps to suppress special instrument features in the data. 

The subtraction of the structure of the stellar PSF halo is rather simple for the calibrated polarimetric imaging data of $\epsilon$~Eri because the residual signal of a polarimetric cycle is very weak (see Figure~\ref{PolarimetricCycle}). For $\rho>0.6\arcsec$ the systematic structures are smaller than the photon noise except for residuals at the position of the eight astrometric spots of the coronagraph mask and some along the telescope spider. Residuals from the telescope spider remain because they rotate in our data with another rotation law than the sky field and this introduces small systematic alignment errors in the combination of the polarimetric data. 
The features from the coronagraphic spots in the polarized intensity are introduced by the beam shift calibration which corrects the sky image for the differential polarimetric shifts introduced by the inclined mirrors of the VLT and the SPHERE instrument. This correction produces artifical polarimetric beam shifts for all image features introduced by components located after the inclined mirrors, such as the astrometric spots of the used coronagraph or dust on optical components of the science cameras. The position of these residual spot features vary during the night after centring the star because of drifts in the alignment between the star and the focal plane coronagraph. Therefore a simple subtraction procedure gives unsatisfactory results in the residual images. Consequently, we mask these localized features in all $Q_{\rm \phi}$ images with pixels having 'not a number' (nan) values. This masking leads to a loss of around 6~$\%$ of the photons in return for a much smoother residual image. We then used for the subtraction of the residual halo structure in all $Q_{\rm \phi}$-images of a night the median of the masked, non-derotated data of that night. 

\begin{figure}[t]
    \includegraphics[trim=1 1 1 1, clip=true,width=0.46\textwidth]{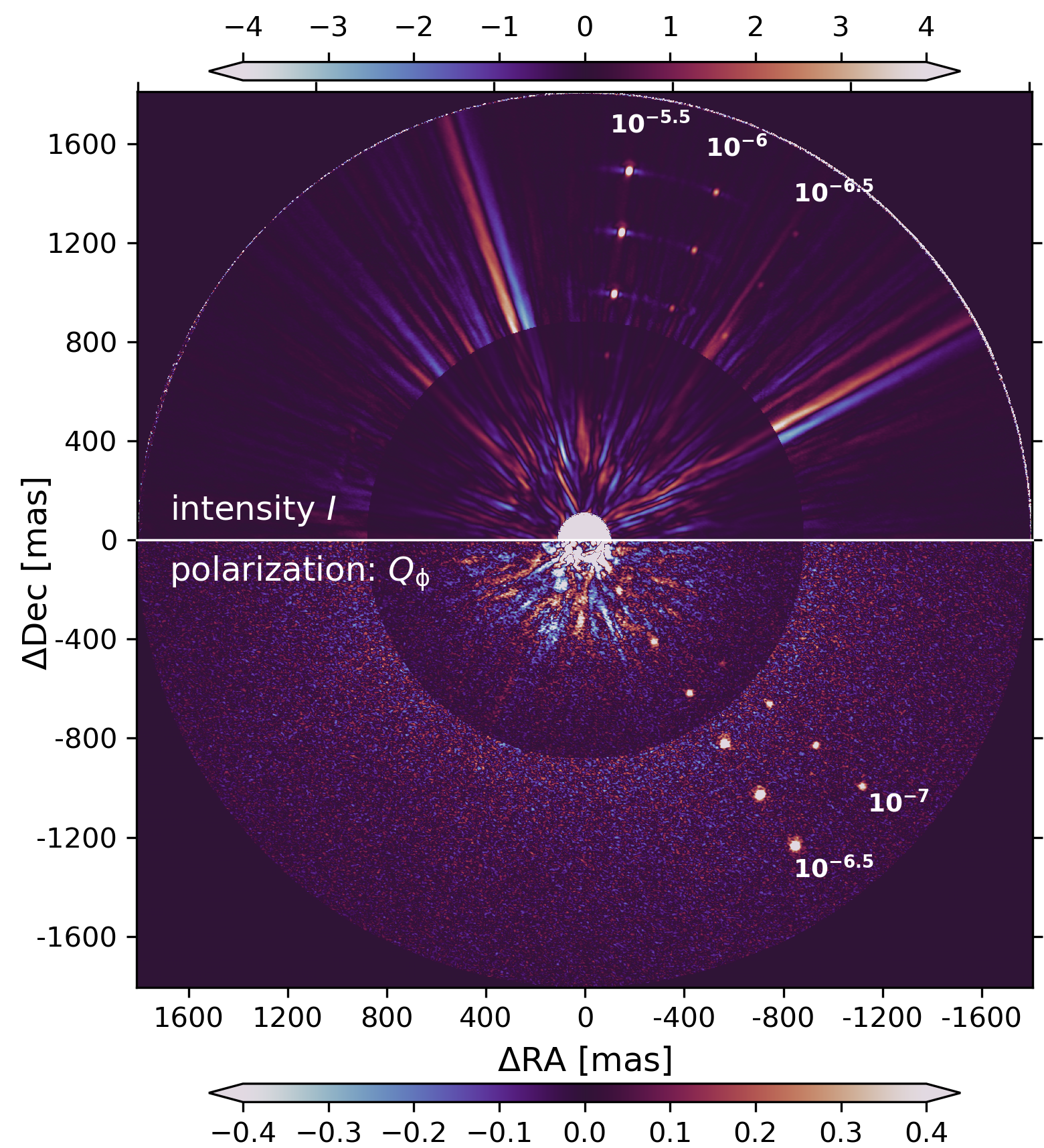}
    \centering
\caption{All data of night 9 ($t_{\rm exp}= 9360$~s), derotated, PSF subtracted, masked regions for the $Q_{\rm \phi}$, averaged
and corrected for telescope polarization: The top half shows $I$ and the bottom half $Q_{\rm \phi}$ with corresponding colour scales in ct/(px $\cdot$ DIT) as in Figure~\ref{PolarimetricCycle}. Artificial point sources are inserted at different separations and
contrasts $C_{\rm I}$ and $C_{\rm P}$ are changed with position angle as indicated. 
Counts inside 0.9$\arcsec$ are scaled down for better visibility with a factor of six for $I$ and a factor three for $Q_{\rm \phi}$.}
\label{psfsub_09}
\end{figure}

The situation is more delicate for the intensity images, because the stellar intensity halo is strong and quite variable due to AO performance variations with strong short-lived speckles and the overall PSF halo structure changes. For this reason, the stellar halo is fitted for each image using a principal component analysis (PCA) \citep{Amara12} and then these fits are subtracted from the data. This procedure was carefully investigated to avoid the introduction of spurious point-like features or possible self-subtraction of real point sources in the resulting data residuals, which are used for the search of a planet. Best results are obtained using about 20 principal components in the PCA analysis. The astrometric spots from the coronagraph (one such spot is visible in Figure~\ref{PolarimetricCycle}, 1$\arcsec$ above the centre), and the telescope spiders were not masked because the use of nan - values introduces strong spurious effects in the averaging of data with variable flux levels. 

Finally, we compared different combination methods, mean, median, or noise-weighted mean \citep{Bottom17}, for the derotated, differential polarimetric and intensity images of each night. Best results for both types of data are obtained with noise-weighted means (in case of the polarized intensity with masked spiders and astrometric spots). 

The resulting data for night 9 after all these post-processing steps are shown in Figure~\ref{psfsub_09}. The PSF halo subtraction for the intensity image yields as final data sets difference images with a mean value of zero and deviation at the level of a few counts for separation $>0.9\arcsec$ and deviations of the order 10 counts inside $<0.9\arcsec$. This allows to spot artificial point sources which are ten to 30 times fainter than what could be seen in the averaged imaging data of Figure~\ref{raw_mean_09}. Dominant noise sources in the halo subtracted intensity image are the speckle noise at small $\rho$ and the residuals from the diffraction pattern of the telescope spider for larger $\rho$. The spider effect is a result of the ZIMPOL P1 mode because ZIMPOL offers no polarimetry with pupil stabilization. The spider effect is small, less than 1~\% of the halo intensity but it might still be beneficial to correct for this in future investigations. 

The post-processing of the $Q_{\rm \phi}$ image removed most instrument features and shows for night 9 a very clean pattern of Gaussian noise on top of which the artificial sources with a contrast of $C_{\rm P}=10^{-6.5}$ and $C_{\rm P}=10^{-7}$ are clearly visible. This illustrates that the differential polarization data reach a contrast very close to the photon noise limit for the search of point sources. 

\subsection{Contrast curves and detection maps}\label{sensitivity_meth}
To characterize the sensitivity of our search, we inject fake planets in the images before the post-processing and then evaluate how well we can retrieve the planet. The template for the fake planets for a given night is the median PSF from unsaturated images of $\epsilon$~Eri taken with the ND2 filter. The counts are scaled to the flux of the coronagraphic images used for the planet search considering the wavelength dependent telescope and instrument transmission for the broad VBB filter with and without ND2, the atmospheric transition, and the $\epsilon$~Eri spectrum. The derived scale factors are 174 for nights 1 and 2 when the data were taken with 3~s DITs, and 290 for the remaining nights taken with 5~s DITs, respectively. 

We then injected many such template PSFs multiplied with a small contrast factor (e.g. $10^{-6}$) as fake planets in the science image. The fake planets were distributed in a spiral pattern with sufficient separation to avoid cross talks in the signal extraction. We then changed the contrast factor until the planets could be retrieved with a $5\,\sigma_{\mathcal{N}}$ Gaussian significance detection. This procedure was repeated six times with the planet spiral pattern rotated each time by 60$^\circ$ so that for each radial separation the planet was injected at six different angle positions. Finally the mean of the six measurements was taken. This whole process was repeated with planets at radii between the previous radii to get a denser radial sampling. For the planet apertures a radius of 5~pixel (or 18~mas) was used as this is the optimal size to extract as much point source signal as possible while keeping at the same time the background noise low. Around each aperture an annulus of 4~px width and 2~px separation to the aperture was used to subtract the local background. This background subtraction does not improve the contrast, but reduces the standard deviation between the six injected planets for a given radius. 

We use for the detection metric the false positive fraction (FPF) as described in 
\citet{Mawet14}. As explained in \citet{Bonse23}, \citet{Gonzalez17} and \citet{Christiaens23} the S/N for the student's t-distribution should not be confused with the Gaussian sigma significance. In short, we first calculate the S/N as defined in \citet{Mawet14} (based on two sample t-test): $T = (P-B)/(s_{\rm B} \sqrt{1+{1}/{n}})$, where $P$ is the planet signal measured in the aperture, $B$ the mean and $s_{\rm B}$ the standard deviation for the background apertures, and $n$ the number of background apertures. The same $T$ value corresponds to different FPF for different separations ${\lambda}/{D}$. $T$ follows a student t-distribution with $n$~-~1 degrees of freedom and one can calculate for each radius the ${\rm FPF} = \int_{T}^{\infty} p(T=t|H_0)dx$ \citep{Bonse23} for a found value of $T$. For better readability we express the FPF in terms of the quantiles of the standard normal distribution, so that the $5\,\sigma_{\mathcal{N}}$ Gaussian significance corresponds to a FPF of $2.87 \cdot 10^{-7}$ and $3\,\sigma_{\mathcal{N}}$ to a FPF of $1.35 \cdot 10^{-3}$. Because n increases with separation, a measured S/N value of $T = 5$ corresponds to a 4.76$\,\sigma_{\mathcal{N}}$ at a separation of 500~mas (4.87$\,\sigma_{\mathcal{N}}$ at 1000~mas and 4.92$\,\sigma_{\mathcal{N}}$ at 1500~mas).
It was checked that the behaviour of the noise was close to Gaussian to fulfil the statistical assumptions, as expected for planet signals located at separations of $\rho >6 ({\lambda}/{D})$.

\begin{figure}[t!]
    \includegraphics[trim=2 20 4 20, clip=true,width=0.46\textwidth]{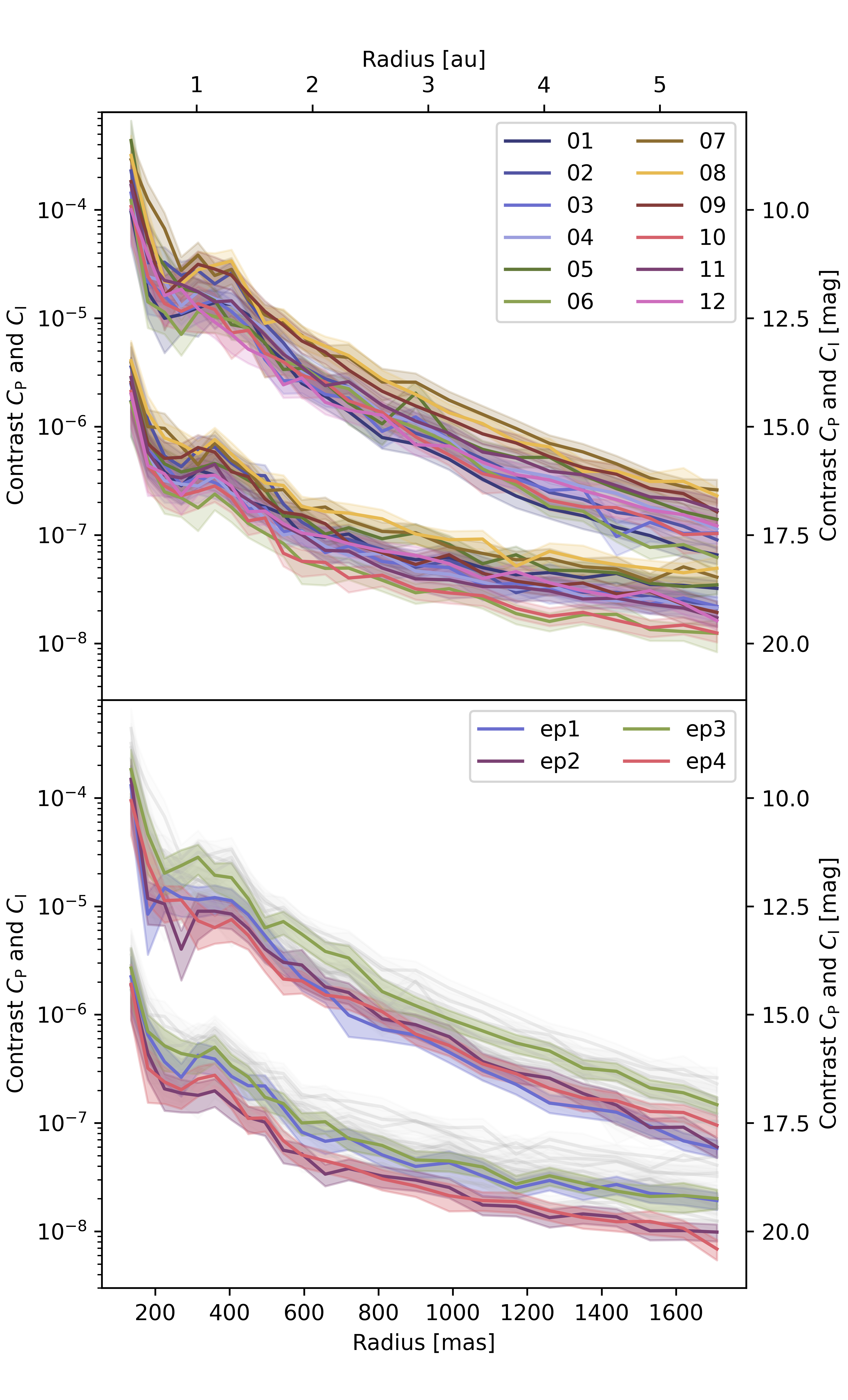}
    \centering
\caption{Sensitivity expressed as contrast curves with $5\,\sigma_{\mathcal{N}}$ Gaussian significance. Top: Individual nights for the intensity $ C_{\rm I} = I_{\rm p} / I_\star$ (upper curves), the polarized light $C_{\rm P} = p_{\rm p} \cdot I_{\rm p} / I_\star$ (lower curves). Bottom: The four epochs in colour in comparison to the individual nights in grey.} \label{contrcurve}
\end{figure}

We derived with this method for $\epsilon$~Eri azimuthally averaged radial contrast curves shown in Figure~\ref{contrcurve} for $5\,\sigma_{\mathcal{N}}$ Gaussian significance for the intensity and the polarization for each individual night and for each epoch. The twelve nights are spread over four epochs of a few consecutive nights as indicated in Table~\ref{obsSHOW}. The PSF changes significantly from one night to another night and therefore also the contrast curves. 
Because the expected motion of a planet around $\epsilon$~Eri is less than 0.7~pixels per night (Section~\ref{SectPlanetModel}) we can combine the nightly results from one epoch to a time weighted 'mean epoch' data set. 

We also produce detection maps for the $Q_{\rm \phi}$-data to search for a significant signal of a polarized point source in the entire field of view (Figure~\ref{det_maps}). For this we use for an individual night the averaged, residual $Q_{\rm \phi}$ frame and treat each pixel as central pixel of an aperture and calculate the Gaussian sigma detection values. Detection maps for the epochs are based on time weighted $Q_{\rm \phi}$ averages of the individual nights. A source with a significant azimuthal polarization $Q_\phi$ should then show up in these maps as bright spot with a significance of $> 5\,\sigma_{\mathcal{N}}$.

\subsection{Results: Individual nights}\label{indnights}

The sensitivity of our survey is illustrated with $5\,\sigma_{\mathcal{N}}$ contrast curves in Figure~\ref{contrcurve} for all the individual nights for the intensity (upper curves) and the polarized light $Q_{\rm \phi}$ (lower curves) as a function of separation $\rho$ from the star.
\paragraph{Polarized intensity.}
For $\rho=1\arcsec$ the mean polarized contrast of the twelve nights is $(5.57 \pm 1.85 ) \cdot 10^{-8}$. Between the best night~10 ($(2.90 \pm 0.37 ) \cdot 10^{-8}$) and the least sensitive night~8 ($(9.0 \pm 1.42 ) \cdot 10^{-8}$) is roughly a factor of three independent of the separation. The uncertainties reflect the standard deviation of the contrast values derived for the six fake planets with the same separation but inserted at different position angles. At $\rho=0.6\arcsec$ the mean contrast is $(11.7 \pm 3.81 ) \cdot 10^{-8}$ and at 1.6$\arcsec$ it is $(2.79 \pm 1.04 ) \cdot 10^{-8}$. The contrast is less good at small $\rho$ because of strong, short lived speckles, larger halo flux, and less field rotation in absolute pixel values. For $\rho > 1\arcsec$ no contrast curve improvement could be obtained after PDI and ADI with different post-processing methods such as median or PCA halo subtraction probably because PDI and ADI already achieved the photon noise limit.

Comparing the nights 5 and 6 exemplifies well the effect of the observing conditions on the achieved limits. The integration time was similar, but the conditions and the PSF were much better in night~6 with better seeing and longer coherence time $\tau_0= 6.1$~ms instead of only $2.5$~ms in night 5 (Table~\ref{avdataSPHERE}). The achieved contrast limits are approximately 2.5 times better for night 6 than for night~5, also because 20~minutes of integration time was unusable in night~5 because the AO system was not stable.

\begin{table}[t]
\caption{Points with $\sigma_{\mathcal{N}}>5$ found in the $Q_\phi$ detection maps of single nights 
or averaged data sets for the four epochs.}
\label{TabPeakcounts}
\begin{center}
\begin{tabular}{lcccccl}
\hline
\hline
\noalign{\smallskip}
data & $\rho$  & $\theta$ & $\sigma_{\mathcal{N}}$    & contrast  & npx  & notes\\
          & [mas]  & [deg] &           &        &      &  \\ 
\noalign{\smallskip}
\hline
\noalign{\smallskip}
\multicolumn{5}{c}{nights}\\
N2  & 903 & 261 & 5.1  & $5.3 \cdot 10^{-8}$ & 3 &  \\
N4  & 178 & 238 & 5.8  & $7.8 \cdot 10^{-7}$ & 9 &  r1\\
N6  & 865 & 65  & 6.3  & $4.3 \cdot 10^{-8}$ & 12 &  \\
N8  & 318 & 23  & 5.8  & $6.8 \cdot 10^{-7}$ & 3 &  r1\\
N9  & 1504 & 222 & 5.2 & $3.1 \cdot 10^{-8}$ & 1 &  r3 \\
N9  & 702  & 331 & 5.1 & $9.9 \cdot 10^{-8}$ & 1 &   \\
N10 & 1521 & 323 & 5.4 & $1.6 \cdot 10^{-8}$ & 4 &  r3 \\
N10 & 947 & 30 & 5.4  & $3.3 \cdot 10^{-8}$  & 1 &  \\
N10 & 703 & 120 & 5.1  & $4.6 \cdot 10^{-8}$ & 1 &   \\
N10 & 371 & 150 & 6.5  & $3.4 \cdot 10^{-7}$ & 2 &  r1, r2 \\[1.3ex]
\multicolumn{5}{c}{epochs}\\
E3     & 472 & 101 & 5.04  & $2.2 \cdot 10^{-7}$ & 1 &  r1 \\
E4     & 686 & 337 & 5.1  & $4.3 \cdot 10^{-8}$ & 3 &   \\
\noalign{\smallskip}
\hline
\end{tabular}
\end{center}
\tablefoot{The columns give for each high $\sigma_{\mathcal{N}}$ point the night (N) or epoch (E) number, the angular separation $\rho$ in milli-arcsec, the position angle $\theta$ measured 
from N over E, the $\sigma_{\mathcal{N}}$, the point source contrast $Q_{\rm \phi}/I_\star$, number of additional neighbouring pixels with $\sigma_{\mathcal{N}}$>5 (npx) and notes: 'r1': too bright, 'r2': too low separation, 'r3': too large separation.}
\end{table}

The Gaussian significance detection maps for the polarized light $Q_{\rm \phi}$ of the individual nights are shown in the top panel of Figure~\ref{det_maps}. Out of the twelve images, each consisting of 723\,736 pixels, there are 37~pixels with a significance larger than 5 and these 37~pixels belong to 10~points of neighbouring pixels (see Table~\ref{TabPeakcounts}). As the Gaussian significance corresponds to a FPF of $2.87 \cdot 10^{-7}$ one would expect roughly 2.5~pixels with a $\sigma_{\mathcal{N}} > 5 $ in the twelve images assuming perfect Gaussian noise. 

One should note that most points with more than $5\,\sigma_{\mathcal{N}}$ signal have a contrast significantly brighter than our expectation of about $10^{-9}$ for $\epsilon$~Eri~b (see Section~\ref{SectPlanetModel}). Furthermore some points are at a separation $\rho$ larger than expected from the RV orbit. In particular, the predicted separation for the fourth epoch should be close to 1.16$\arcsec$, making detections at much smaller or larger separations unlikely. The noise properties of the $U_{\phi}$ residual images, in which we expect no signal from a planet, look indistinguishable from the $Q_{\phi}$ images. 

\paragraph{Total intensity.}
We get for the intensity contrast curves for the 12 nights in Figure~\ref{contrcurve}
 a $5\,\sigma_{\mathcal{N}}$ mean contrast of 
$(6.15 \pm 2.90 ) \cdot 10^{-6}$ at $\rho = 0.6\arcsec$,
$(1.58 \pm 0.83 ) \cdot 10^{-6}$ at $\rho = 1\arcsec$, and $(2.55 \pm 1.15 ) \cdot 10^{-7}$ at $\rho = 1.6\arcsec$. In the intensity residual images there is more systematic noise left from short-lived speckles and the artefacts related to the telescope spiders. We expect that some contrast improvement might be possible with future, more sophisticated post processing methods.

\subsection{Results: Epochs}\label{res_ep}
\paragraph{Polarized intensity.} The $5\,\sigma_{\mathcal{N}}$ Gaussian significance contrast curves for the epochs consisting of 2-4 consecutive nights are illustrated in Figure~\ref{contrcurve}. The mean contrast for the $Q_{\phi}$ polarization for the four epochs is $(7.04 \pm 2.14 ) \cdot 10^{-8}$ for $\rho=0.6\arcsec$, $(3.29 \pm 1.01 ) \cdot 10^{-8}$ for 1$\arcsec$, and $(1.60 \pm 0.54) \cdot 10^{-8}$ for 1.6$\arcsec$. Epoch~4 is most interesting because the planet separation is at a maximum, the scattering angle is close to 90$^\circ$ and ideal for a strong $Q_{\rm \phi}$ signal from the planet, and epoch~4 is the longest with the best contrast: $(5.02 \pm 0.76 ) \cdot 10^{-8}$ for 0.6$\arcsec$, $(2.11 \pm 0.43) \cdot 10^{-8}$ for 1$\arcsec$, and $(1.10 \pm 0.24 ) \cdot 10^{-8}$ for 1.6$\arcsec$. The three nights of epoch~4 have an individual mean contrast of approximately $4 \cdot 10^{-8}$ for $\rho=1\arcsec$ and each night is about 5~hours long. The combined epoch sensitivity is approximately $\sqrt{3}$ times higher than for one night or an improvement similar to $\sqrt{t_{\rm exp}}$. 
Most notably, the achieved $5\,\sigma_{\mathcal{N}}$ contrast is $(1.22 \pm 0.32 ) \cdot 10^{-8}$ for the expected maximum planet separation of 1.16$\arcsec$ occurring around epoch~4.

In lower panel of Figure~\ref{det_maps} we show the Gaussian significance epoch maps. For perfect Gaussian noise we would expect 0.8~pixels with a significance greater than 5 and we find 4~pixels with a $\sigma_{\mathcal{N}} >5$ attributed to two locations (see Table~\ref{TabPeakcounts}): In epoch~3 with $\sigma_{\mathcal{N}} = 5.04$ at $\rho=472$~mas at an angle (north over east) of 101$^\circ$. This would correspond to a contrast of $(2.21 \pm 0.46 ) \cdot 10^{-7}$ which is at least one order of magnitude higher than the expected planet signal. The other spot is in epoch 4 and consists of three neighbouring pixels with a $\sigma_{\mathcal{N}} >5$ with the central pixel at $\rho=686$~mas, 337$^\circ$ and significance of $5.1$, which corresponds to a contrast of $(4.32 \pm 0.66 ) \cdot 10^{-8}$. The separation is smaller than expected at that time and the contrast is a rather high value. Another point in epoch 4 which might be worth mentioning is at $\rho=994$~mas, 30$^\circ$ with $\sigma_{\mathcal{N}}=4.33$ and contrast $1.96 \pm 0.46 \cdot 10^{-8}$.

\paragraph{Total intensity.}
For the search of the intensity signal of a companion the mean $5\,\sigma_{\mathcal{N}}$ contrast for the four epochs is $(3.06 \pm 1.36 ) \cdot 10^{-6}$ for 0.6$\arcsec$, $(6.05 \pm 1.73 ) \cdot 10^{-7}$ for 1$\arcsec$ and $(1.21 \pm 0.46 ) \cdot 10^{-7}$ for 1.6$\arcsec$. For most separations epoch~1 is the most sensitive although for example epoch 4 contains more than twice more observing time. Probably collecting more individual images (DIT=3~s instead of DIT=5~s) offers an advantage in post-processing and PSF halo subtraction. This consideration does not take into account the read noise at larger separation as the fake planet injection uses a high signal PSF template multiplied by the small contrast number. 
For the PSF subtraction technique in the post-processing, the used PCA with 20 principal components gives approximately a factor two contrast improvement when compared to a median PSF subtraction. For small separations this factor is slightly larger and for larger separations a bit smaller. 
Going through the full parameter space and combining the best PCA contrast curves with different number of components \citep[see e.g. contrast curve documentation of \texttt{applefy} package][]{Bonse23}, using different settings of aperture sizes depending on the night or using more advanced PSF subtraction techniques, for example, \citep[see e.g.][]{Cantalloube21, Gebhard22} could lead to an additional improvement in contrast. 

\subsection{Advantage of using PDI }\label{polintcomp}

The $\epsilon$~Eri data were taken with angular differential imaging (ADI) and polarimetric differential imaging (PDI) with the use of the ZIMPOL P1 mode. This offers an ideal opportunity to compare the performances for the high-contrast searches of the intensity signal and the polarization signal of point sources because the analysis can be based on data taken with the same instrument and under the same observing conditions. 

Thus, we compare observing and data analysis methods for the total intensity planet search using ADI, plus PSF fitting and subtraction with a PCA method \citep{Amara12} with the polarimetric planet search using PDI, ADI and residual pattern subtraction. This should provide a very reliable assessment because we are using state of the art procedures for the observations, the data reduction and post-processing. However, it should be noted, that the Strehl ratio provided by the SPHERE AO system is about 40~\% for the short wavelength range of ZIMPOL, which is significantly lower than for the near-infrared \citep{Fusco15}. 

Polarimetry is a very efficient high-contrast technique because it provides the differential signal of the opposite polarization modes simultaneously and this minimizes very strongly the temporal variability effects of the speckle noise \citep[see e.g.][]{Schmid22}. 

\begin{figure}[t]
    \includegraphics[trim=1 1 26 1, clip=true,width=0.46\textwidth]{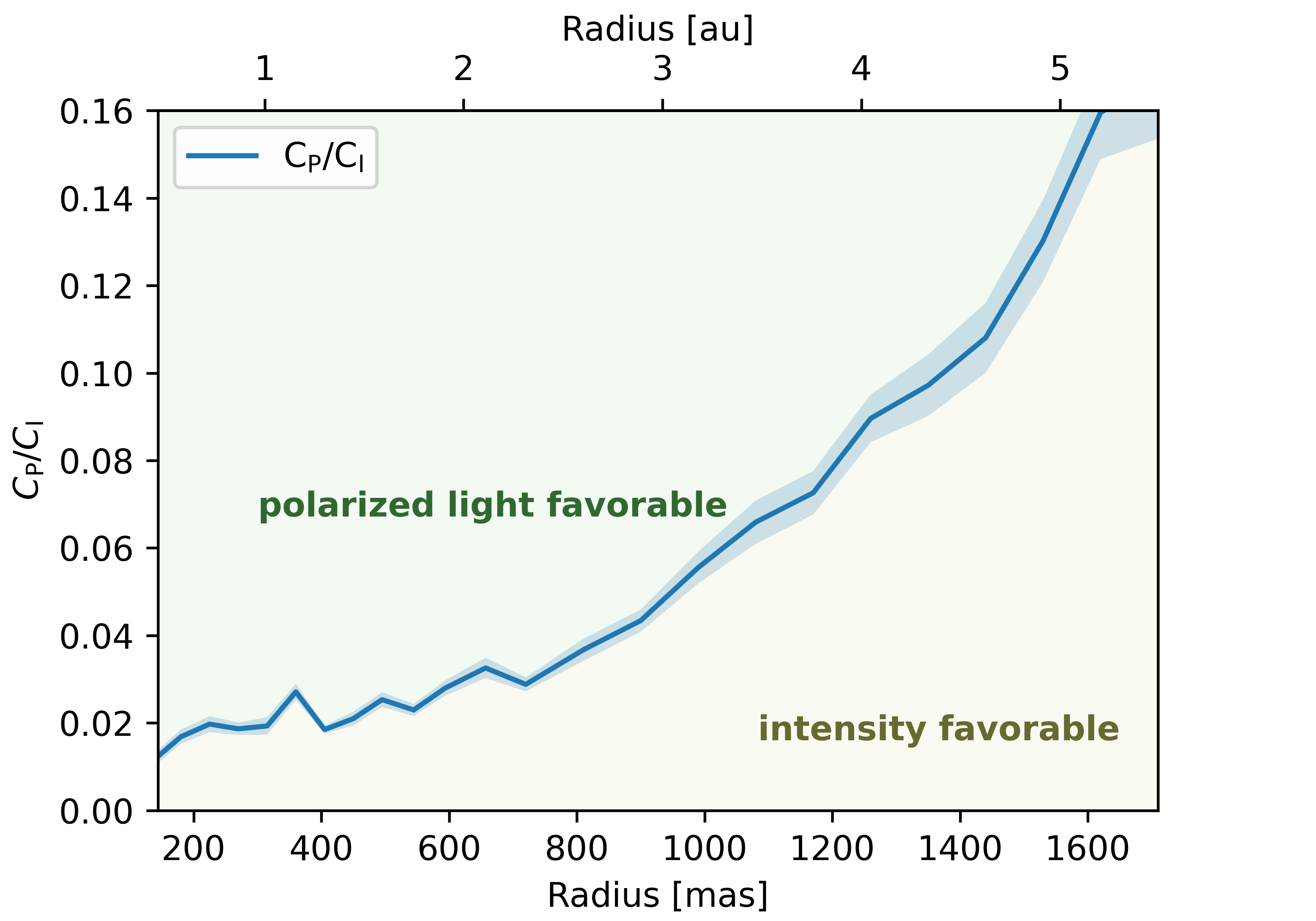}
    \centering
\caption{Ratio between polarization
$C_{\rm P}$ and intensity $C_{\rm I}$ contrast 
limits for the full data set, illustrating the advantage of PDI for the speckle suppression. Planets with a fractional polarization $p>C_{\rm P}/C_{\rm I}$ should be easier to detect in the polarimetric data.} \label{contrcurveb}
\end{figure}

The advantage of polarimetry is clearly visible from the much better polarized contrast curves $C_{\rm P}$ when compared to the intensity contrast curves $C_{\rm I}$ in Figure~\ref{contrcurve}. The ratio $C_{\rm P}/C_{\rm I}$ based on the curves for the mean contrast limits derived from the full data set as function of separation is shown in Figure~\ref{contrcurveb}. 

A planet detection is easier with polarimetic imaging for planets with a fractional polarization $p>C_{\rm P}/C_{\rm I}$, and easier with intensity imaging for planets with $p<C_{\rm P}/C_{\rm I}$. Especially for small $\rho$ , searches in polarized light offers a strong advantage because of the very efficient suppression of unpolarized speckles. 

Of course, the polarization signal $C_{\rm P}$ from a planet is strictly smaller than the intensity signal $C_{\rm I}$, typically by a factor of 3 to 20 for a scattering angle of about $\alpha\approx 90^\circ$. The measured fractional polarization $p_{\rm p}$ of solar system gas planets in the R-band are for the Rayleigh scattering planets Uranus and Neptune about $p_{\rm p}\approx 20$~\% \citep{Schmid06a,Buenzli09}.
Reflections by clouds produce less $p_{\rm p}$ while atmospheric haze can produce a very high polarization $p_{\rm p}$ of up to 50~\%. Therefore the integrated polarization of Jupiter is about 10~\% \citep{Smith84} for equatorial sight lines and about $p_{\rm p}\approx 15$~\% for polar sight lines when the reflection from the polar haze is well visible. For Saturn $p_{\rm p}$ is only about $5~\%$ (without disk) because of the predominant atmospheric clouds \cite{Tomasko84}. 

For the selection of the best observing strategy for future planet searches with SPHERE/ZIMPOL one should carefully consider the ratio curve $C_{\rm P}/C_{\rm I}$ in Figure~\ref{contrcurveb} together with the expected fractional polarization of the planet. This curve is based on a large and representative data set and it varies only slightly between observations taken under good or bad atmospheric conditions. This can be inferred from the $C_{\rm I}$ and polarization $C_{\rm P}$ contrast curves in Figure~\ref{contrcurve}, which are for given nights both going up or down in step with the observing conditions. The $C_{\rm P}/C_{\rm I}$ curve derived for SPHERE/ZIMPOL gives a useful benchmark for the design of future instruments, but one should also consider that the overall performance depends on many instrument parameters for the AO-system and the differential imaging concept.

\subsection{Results of K-Stacker orbital search}\label{sec:reskstacker}

Combining consecutive individual nights into a set of four different epochs is straightforward because the expected orbital motion from day to day is less than 3~mas or one ZIMPOL detector pixel. Improving the detection limits by the combination of the data from different epochs requires much more care. Adopting for $\epsilon$~Eri an orbital semi-major axis of $a\sim{} 3.5$~au around a $0.82\,\mathrm{M}_\odot{}$ star gives an orbital period of about 7~yr, or an astrometric motion of 3~au/yr, or $1\arcsec/$yr for the distance of 3.2~pc. This converts to about 80~mas per month for a face-on, circular orbit, which is more than 3 times larger than the width (FWHM) of the PSF of SPHERE/ZIMPOL. Thus, the epochs must be properly combined considering the orbital motion of any putative planet. To do so, we used the K-Stacker algorithm \citep{Nowak18, Lecoroller20}, which searches for potential companions in series of images along a grid of pre-determined orbital parameters. For this we use the $Q_{\rm \phi}$-maps of all nights, such as the one shown for night 9 in the lower half of Figure~\ref{psfsub_09}. 

\subsubsection{Potential solutions}
 
The overall process can be summarized in a few steps according to the detailed description of the algorithm in the reference publications 
\citep{Nowak18, Lecoroller20}:
\begin{enumerate}
    \item The algorithm first calculates the planet position for each night for a set of orbital parameters $p$.
    \item For each night $t$, it extracts a 'signal' value $s_{\rm t}(p)$ which is the integrated photometry in a circle of 6~px diameter, which corresponds to the size of the instrumental PSF, as well as a 'noise' value $n_t(p)$ calculated from the distribution of signal values extracted along a circle whose radius corresponds to the separation at night $t$ for orbit $p$.
    \item The algorithm calculates the total estimated S/N for all points of an orbit $p$ using: $\mathrm{(S/N)}(p) = \sum_{\rm t} s_{\rm t}(p)/\sqrt{\sum_{\rm t} n_{\rm t}(p)^2}$. It then ranks all the orbits of the grid by order of decreasing S/N. 
    \item A subset of the best orbits (in our case, the best 70 orbits) are further optimized using a gradient-descent algorithm to allow for parameter values between the initial grid-points.
    \item The calculation ends with a report of all the values calculated along the grid of orbital parameters, and the optimized results for the best 70 orbits.
\end{enumerate}

To setup the grid of orbital parameters, we followed \cite{Nowak18} and \cite{Lecoroller20}, and first determined the typical width of a maximum in the (S/N)$(p)$ function along the different parameters. We then define the step sizes for each parameter as typically one fifth of the corresponding (S/N)$(p)$ peak width, to ensure that K-Stacker would not miss any potential S/N maximum. The explored range of orbital parameters in our grid was defined based on previous studies of the planet $\epsilon$~Eri~b, mainly on \citet{Llop21}. The used parameter grid is given in Table~\ref{tab:KS_search_parameters}. It restricts orbital periods according to $(P[{\rm yr}])^2=(a[{\rm au}])^3/M_{\rm star}[{\rm M}_\odot]$ to the range between about $P\approx 5.5$~yr and $10.9$~yr and the orbits' eccentricity to $e\leq 0.6$. The grid search does not constrain the orbital inclination $i$ and the orientation of line of nodes with respect to sky plane $\Omega$, nor the argument of the periapsis $\omega$. Thus, the search allows for prograde and retrograde planetary orbits.

\begin{table}[t]
\begin{center}  
\caption{Grid of orbital parameters used for the K-Stacker search.}
  \begin{tabular}{llll}
    \hline
    \hline
    \noalign{\smallskip}
    \multicolumn{2}{l}{Parameter} & Range & Number of steps\\
    \noalign{\smallskip}
    \hline
    \noalign{\smallskip}
    $d_\mathrm{star}$ & [pc] & [3.22] & fixed value \\       
    $M_{\text{star}}$ & $[\mathrm{M}_\odot{}]$ & [0.76, 0.9] & 6  \\
    $a$ & [au] & [3.0, 4.5] & 21 \\
    $e$ & - & [0, 0.6] & 48  \\
    $t_0$ & [yr] & [0, 11] & 220 \\
    $\Omega$ & [deg]  & [-180, 180] & 110 \\
    $i$ & [deg] & [0, 180] & 40  \\
    $\omega$ & [deg] & [-180, 180] & 110 \\
    \noalign{\smallskip}
    \hline
  \end{tabular}
  \tablefoot{
    Time of passage at periapsis $t_0$, given in decimal year elapsed since a reference epoch of MJD = 58766 (which corresponds to the $\rm 10^{\rm th}$ of October 2019)
  }
  \label{tab:KS_search_parameters}
\end{center}  
\end{table}

\begin{figure}[t]
  \begin{center}
    \includegraphics[width=0.9\linewidth, clip=True, trim=0.4cm 0.4cm 0cm 0.4cm]{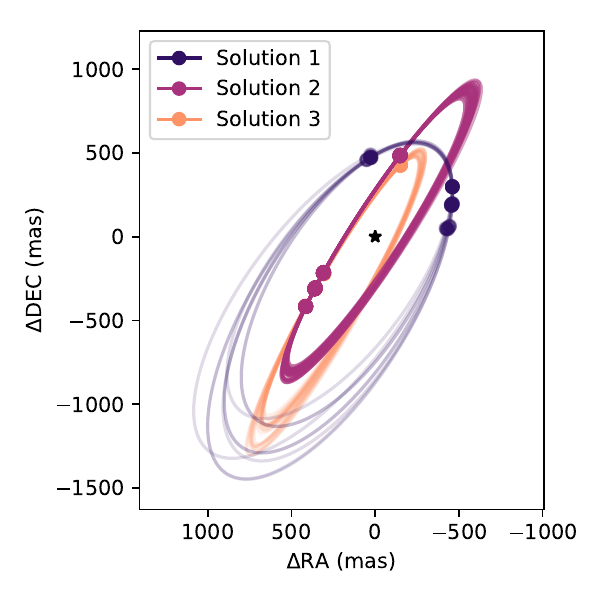}
    \caption{Illustration of the 70 best orbits found by K-Stacker, after the re-optimization step. The markers correspond to the positions along each orbit at the epoch of the individual SPHERE/ZIMPOL observations. These 100 orbits are actually distributed along 3 main orbits, among which two (orbit 1 and 2) correspond to similar positions in the image.}
  \label{fig:kstacker_solutions}    
  \end{center}
\end{figure}

\begin{table*}[t]
\begin{center}  
\caption{Comparison of the three solutions found by K-Stacker with the solution of \cite{Llop21}.}
    \begin{tabular}{llllll}
      \hline
      \hline
      \multicolumn{2}{c}{Parameters} & KS Solution 1 & KS Solution 2 & KS Solution 3 & \cite{Llop21} \\
      \hline
      $M_{\text{star}}$ & $[\mathrm{M}_\odot{}]$ & $0.82 \pm 0.04$ & $0.83 \pm 0.03$ & $0.82 \pm 0.04$ & $0.82 \pm 0.02$ \\
      $a$ & [au] & $3.43 \pm 0.23$ & $3.24 \pm 0.08$ & $3.13 \pm 0.13$ & $3.52 \pm 0.04$ \\
      $e$ & - & $0.40 \pm 0.04$ & $0.05 \pm 0.02$ & $0.43 \pm 0.02$ & $0.07 \pm 0.07$\\
      $t_0$ & [yr] & $7.69 \pm 0.81$ & $5.83 \pm 0.94$ & $7.72 \pm 0.49$ & $4.02 \pm 1.89$ \\
      $\Omega$ & [deg] & $146.52 \pm 1.56$ & $147.15 \pm 0.21$ & $150.75 \pm 0.38$ & $190.06^{+109}_{-152}$ \\
      $i$ & [deg] & $116.91 \pm 0.14$ & $81.00 \pm 0.15$ & $80.72 \pm 0.28$ & $89.7 \pm 25$\\
      $\omega$ & [deg] & $181.45 \pm 3.93$ & $17.49 \pm 39.95 $ & $196.70 \pm 4.74$ & $-29.84^{+105}_{-116}$\\
      \hline
  \end{tabular}
  \label{tab:kstacker_solutions}
\end{center}  
\end{table*}

\begin{figure*}[t]
  \begin{center}
    \includegraphics[width = 0.9\linewidth]{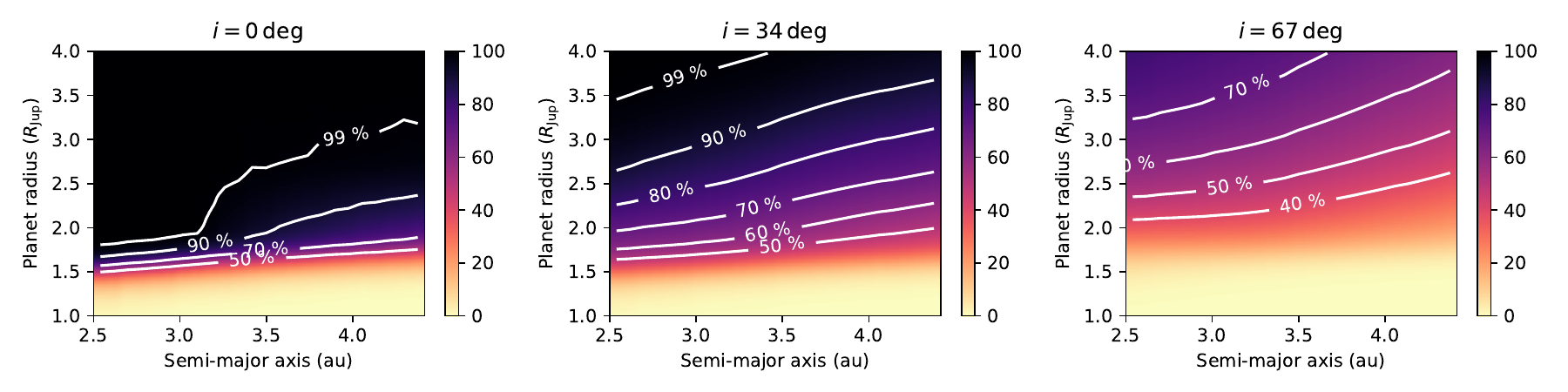}
    \caption{Detectability of planets 
as a function of the planet radius and semi-major axis for three different values of the inclination. The inclination has a significant impact on the detectability of planets, due to the influence of the phase angle on the polarization contrast.}
    \label{fig:kstacker_limits}
  \end{center}
\end{figure*}

Among the $6\times{}10^{11}$ planetary orbits explored by the algorithm, the best solution found reached a S/N$(p)$ of up to 7.4. The re-optimization of the best 70 solutions with a gradient-descent algorithm leads to typical improvements of about 0.4 in the S/N, and it yields a distribution of best results along 3 potential orbits, represented in Figure~\ref{fig:kstacker_solutions}.

Solutions 2 and 3 obtained by K-Stacker (see Figure~\ref{fig:kstacker_solutions}) are actually very similar in terms of position for the individual epochs, with a difference of at most 50 mas for the last epoch, and about 15~mas for the other epochs. This suggests that the algorithm has actually caught some feature within the images. The emergence of two solutions arise from degeneracies in the orbital parameters, a consequence of the limited number of independent epochs available for our data. 
Interestingly, solution 2 shows orbital parameters which are largely compatible with the solution presented by \cite{Llop21}, as shown in Table~\ref{tab:kstacker_solutions}. It should be noted, though, that despite their use of astrometric data in the analysis, their orbital parameters are poorly constraining planet positions, with large error bars on $i$ and $\Omega$, which increases the chance of having ``compatible solutions''. We note, however, that solution 1 found by K-Stacker corresponds to a completely different set of orbital parameters, and to different positions at the ZIMPOL epochs. This demonstrates that K-Stacker converges towards multiple potential solutions for this data set.

The fact that K-Stacker reports a solution compatible with the orbit of \citep{Llop21} is interesting, and could suggest that this is indeed a true signal from a point-like source. However, the existence of solution 1 with a very similar (S/N)$(p)$ tells us that solutions 2 and 3 cannot be taken as evidence of a detection: at most, they are only marginally better than the noise floor in K-Stacker. 

We also notice, that all solutions find point signals  around $\rho\approx 0.5\arcsec$, which corresponds to the separation of the strong speckle ring in the ZIMPOL PSF (see Figures~\ref{PolarimetricCycle} and \ref{raw_mean_09}), where the contrast performance for an individual night is only about $C_{\rm P}\approx 4\cdot 10^{-7}$. A $1\,R_\mathrm{J}$ is expected to have a typical contrast of $C_{\rm P}\approx 1.4\times{}10^{-9}$ for low eccentricity orbit with $a\approx 3.5$~au (Equation~\ref{eq:Cpol}), and according to \cite{Llop21}, $\epsilon$~Eri~b is expected to have a mass of $\sim{}0.8\,M_\mathrm{J}$. Therefore, even taking into account a gain of factor $\sqrt{12}$ by combining all the images, it seems unlikely that K-Stacker would be able to detect such a small planet in this data set. This is confirmed by the estimated detection limits shown in Figure~\ref{fig:kstacker_limits}.
Probably, the strong speckles at this separation are responsible for a non-Gaussian noise distribution which could result in the detection of high S/N noise features by K-Stacker.

A less likely explanations could be, that an unresolved dust cloud with a diameter of $<0.1$~au could produce a polarization signal with a detectable contrast at the level of $C_{\rm P}\approx 2\cdot 10^{-7}$. This signal would be ten times fainter than the expected integrated signal of a dust responsible for the so-called warm 20~$\mu$m infrared excess. Thus, if the narrow dust ring model from Figure~\ref{2d_dustmodels} has a clumpy structure made of a few dozen components, then it seems quite likely, that the brightest one has a detectable contrast of $\approx 2\cdot 10^{-7}$. Small dust clouds were observed in scattered light for the debris disk in AU Mic \citep{Boccaletti15}, and a body with a dust cloud could also explain the point-like scattering object Fomalhaut b \citep{Kalas13,Gaspar20}.

\subsubsection{K-Stacker detection limits}

Determining the detection limits of K-Stacker is difficult for at least two main reasons. Firstly, as revealed by the two truly different solutions found with the grid of orbital parameters, the algorithm can converge to relatively high (S/N)$(p)$ values (higher than the commonly used threshold of 5) without the presence of a true companion. Secondly, the algorithm takes into account the orbital motion of the planet, and therefore combines different separations and phase angle together, which makes the usual presentation of contrast as function of separation irrelevant. The true detection limits of K-Stacker can only be understood in orbital parameter space.

Nonetheless, to provide at least a rough estimate of the sensitivity achieved by combining all the available data, we created a set of ``detection maps'' as follows:

\begin{enumerate}
\item For each set of orbital parameters $p$ of the K-Stacker grid, we calculate the amount of ``missing signal'' $\Delta{}S$ which would be required to reach a threshold of $\mathrm{(S/N)}_{\mathrm{KS}} = 8$, defined as $\Delta{}S(p) = [8 - \mathrm{(S/N)}_{\mathrm{KS}}(p)]\times{}\sqrt{\sum_t n(p)^2}$.
This corresponds to the missing signal along orbit $p$ to reach an $\mathrm{(S/N)}_\mathrm{KS} = 8$. This arbitrary threshold is chosen to be slightly higher than the K-Stacker 'noise-floor', which corresponds to the maximum $\mathrm{(S/N)}_\mathrm{KS} = 7.5$ found by K-Stacker.

\item Injecting a companion at a known contrast $C = 2\times{}10^{-7}$ 
in the SPHERE/ZIMPOL data, and extracting the corresponding K-Stacker signal, we determine a ``contrast-to-signal'' conversion factor $\gamma_{\rm t}$ for each night.
\item From the set of orbital parameter $p$, we calculate the 3-dimensional positions at each SPHERE/ZIMPOL epoch $t$. From this, we extract both the distance to the central star $d_{\rm t}(p)$ and the phase angle $\alpha_{\rm t}(p)$.
\item From these distances and phase angles, and using the same Rayleigh scattering as discussed in Section~\ref{SectPlanetModel}, we calculate the polarization contrasts $C_{\mathrm{P}, \mathrm{t}}(p)$ of a $1\,R_\mathrm{J}$ planet using Equation~\ref{eq:Cpol}, and the corresponding total reference K-Stacker signal $S_{R_\mathrm{J}(p)} = \sum_\mathrm{t} \gamma_\mathrm{t} C_{\mathrm{P}, \mathrm{t}}(p)$.
\item The minimum detectable radius on orbit $p$ is finally taken as $R_\mathrm{min}(p) = \sqrt{\Delta{}S(p)/S_{R_\mathrm{J}}}.$

\end{enumerate}

In Figure~\ref{fig:kstacker_limits}, we show, as a function of planetary radius and semi-major axis, the fraction of the orbits from the initial grid on which the planet could have been detected. Since the minimum detectable radius is strongly dependent on the inclination (through the phase angle), this map is presented for 3 different values of the inclination.

For $i=0^\circ$ orbits the scattering angle is always $\alpha_\mathrm{t}(p)=90^\circ$ and the contrast of reflecting planets is constant for $e=0$ or varies with the separation $1/d_{\rm p}^2$. The fact that the planet radius for a given detectability line decreases slightly with decreasing semi-major axis indicates that the planet brightness increases a bit faster for smaller separation $d_{\rm p}$ than the detection limits. The probability for the detection of a planet changes very rapidly for $i=0^\circ$, because once the radius of a planet is large enough to be detectable then it is close to the detection limit along the whole orbit. Thus, for $a=3.5$~au a planet needs to have a radius of about 2.5~R$_{\rm J}$ or a contrast of $C_{\rm P}\approx 8\times{}10^{-9}$ to be detected with a high probability of $> 95$~\% with K-Stacker. For planets on inclined orbits, there are orbital phases where the planet is bright or faint (Figure~\ref{1_orbit_b}): for $i\approx 30^\circ$ the planet is only bright during about 40~\% of its orbit and for $i\approx 70^\circ$ only about 20~\% of its orbit and this leads to the strongly reduced detectability rates in Figure~\ref{fig:kstacker_limits} for a given planet radius and high $i$.

\section{Search for dust}\label{5Dust}
\subsection{Method}\label{dustmeth}

The infrared SED of $\epsilon$~Eri indicates the presence of warm dust emitting thermal radiation within the field of view of ZIMPOL as described in Section~\ref{SectDustModel}, and therefore we search for the scattered, polarized radiation from this dust. We assume that the dust distribution around $\epsilon$~Eri did not change significantly during our four epochs of observations and we expect an azimuthal polarization signal with a positive signal for $Q_{\rm \phi}(x,y)$ and zero signal in $U_{\rm \phi}(x,y)$.

We process first the polarimetry of each night individually and then combine the $Q_{\rm \phi}$ and $U_{\rm \phi}$ maps of all twelve nights to achieve the highest possible signal to noise. Important is the correction for the radial dependence of the instrument polarization as described in Section~\ref{A2_telpol}, because a residual instrumental signal could introduce a faint, extended $Q_{\rm \phi}$ feature. Further, the processing of the individual nights does not include a subtraction of an extended residual $Q_{\rm \phi}$-pattern before derotating the individual images. Such a subtraction was applied for the search of a point source, but for an extended source this could cause significant self-subtraction, or introduce a spurious extended signal. 
However, the residual artefacts from the telescope spiders and the eight astrometry spots of the coronagraph are again masked as for the search of a point source. Further we can increase the sensitivity by pixel binning or image smoothing. A binning of $20\times20$~pixels ($72~{\rm mas}\times 72~{\rm mas}$) would still resolve the dust structures of a narrow ring with a width of about 0.4~au.

\subsection{Derived signal}\label{dust_res}

\begin{figure}[t!]
    \includegraphics[trim=10 18 12 48, clip=true,width=0.46\textwidth]{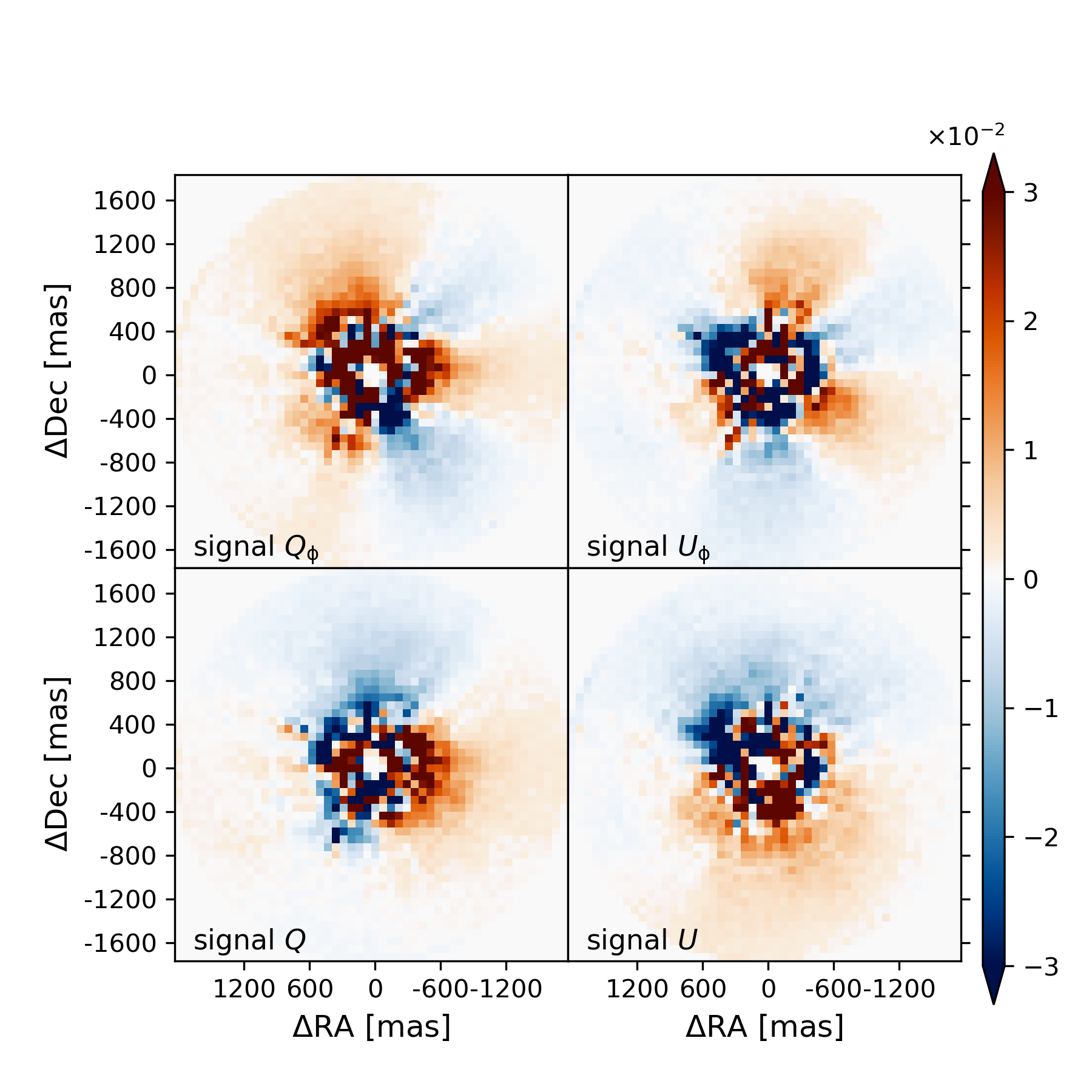}
    \centering
\caption{ 
Final polarimetric maps $Q_{\rm \phi}$, $U_{\rm \phi}$, $Q$, and $U$ derived from the 38.5~hours of the SPHERE/ZIMPOL integration of $\epsilon$~Eri. The colour scale gives
the surface brightness signal in $\rm counts/(s\cdot px)$, where one pixel 
is 3.6~mas $\times$ 3.6~mas.}
\label{2d_dustsearch_b}
\end{figure}

\begin{figure}[t!]
    \includegraphics[trim=10 18 12 48, clip=true,width=0.46\textwidth]{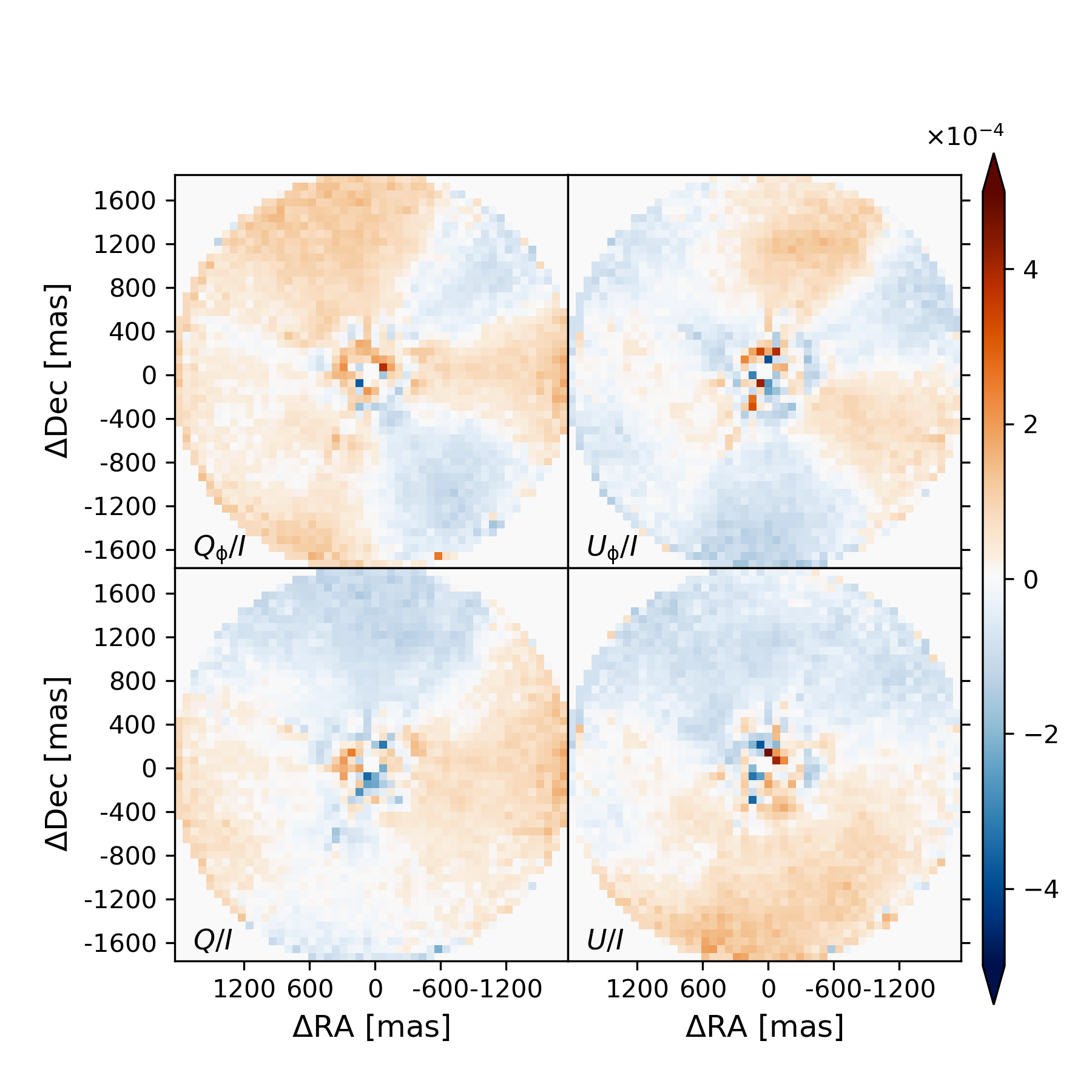}
    \centering
\caption{
Final fractional polarization maps
$Q_{\rm \phi}/I$, $U_{\rm \phi}/I$, $Q/I$, and $U/I$ derived from the 38.5~hours of the SPHERE/ZIMPOL integration of $\epsilon$~Eri. 
} \label{2d_dustsearch_c}
\end{figure}

The final $Q_{\rm \phi}(x,y)$ and $U_{\rm \phi}(x,y)$ maps for $\epsilon$~Eri based on 38.5~hours of SPHERE/ZIMPOL integration are shown in Figure~\ref{2d_dustsearch_b}. Both maps, $Q_{\rm \phi}$ and $U_{\rm \phi}$, show for $\rho>0.6\arcsec$ a quite smooth wedge pattern, with alternating positive and negative signals. The morphology of the patterns is quite similar for $Q_{\rm \phi}$ and $U_{\rm \phi}$, with wedges with a width of roughly $\approx 60^\circ$ but the location of the positive and negative wedges are at different position angles for $Q_{\rm \phi}$ and $U_{\rm \phi}$. The Stokes $Q$ and $U$ maps show smooth structures with one dominant positive and one dominant negative wedge region in the field. 

The patterns in Figure~\ref{2d_dustsearch_b} decrease in strength for larger separations while there is a noisy central region $\rho<0.6\arcsec$ without clear structure. The maps for the fractional polarization $Q_{\rm \phi}/I$, $U_{\rm \phi}/I$, $Q/I$ and $U/I$ plotted in Figure~\ref{2d_dustsearch_c} show for $\rho>0.6\arcsec$ to first order a wedge pattern which does not depend strongly on separation. This indicates that the polarization signals are roughly proportional to the intensity $I(\rho,\theta)$ of the 
stellar halo.

The wedge pattern is very weak, despite the fact that it dominates the final image. The differential polarization $Q_{\rm \phi}$ and $U_{\rm \phi}$ signals at $\rho\approx 1\arcsec$ are of the order of $\pm 0.01~{\rm ct/(s \cdot px)}$ or a surface brightness contrast of $\Delta{S\!B}_{\rm p}=13.6~{\rm mag}/{\rm arcsec}^2$. The pixel count rates are about five times lower at the border of the field of view. The fractional polarization $Q_{\rm \phi}(x,y)/I(x,y)$, $U_{\rm \phi}(x,y)/I(x,y)$ relative to the stellar halo is for $\rho>0.6\arcsec$ at the level of $\pm 0.01~\%$, while the noise in the central region has an amplitude of $\pm 0.03~\%$.

The rather constant signal in fractional polarization could indicate that a large part of the observed pattern is introduced by a cross-talk $I(x,y)\rightarrow Q(x,y),U(x,y)$ from the intensity halo of the stellar PSF. We find in each individual night roughly the same $Q_{\rm \phi}(x,y)/I(x,y)$ and $U_{\rm \phi}(x,y)/I(x,y)$ pattern for $\rho>0.6\arcsec$. This is also true for the first half and the second half of a given night, where parallactic angles and altitudes for the telescope are different. This polarization pattern must therefore rotate together with the sky field on the detector. We expect, that dust scattering in the $\epsilon$~Eri system would only produce a ring-like or disk-like signal in the $Q_{\rm \phi}$ image and none in $U_{\rm \phi}$ and not a wedge pattern as in our data. Therefore, we must consider other potential sources for the obtained signal.
Interstellar polarization could produce in our data a field independent polarization offset. However, for nearby stars the interstellar polarization is small, and indeed for $\epsilon$~Eri only a polarization of $Q/I=+0.0028~\%$ and $U/I=-0.0012~\%$ was measured by \cite{Cotton17} and we obtained very similar values from our data for the integrated polarization signal (Section~\ref{A2_telpol}). This signal is about ten times weaker than the amplitude of the measured polarization patterns shown in Figure~\ref{2d_dustsearch_c} and therefore we can exclude interstellar polarization as significant contributor to the residual pattern.

We think, that the obtained pattern results possibly from an instrumental effect, most likely related to the half wave plate HWP2, which is inserted in the beam for polarimetric observation. 
This component is rotating the polarization position angle from the sky synchronously with the field rotation into the sky coordinate system in the detector plane. A systematic effect introduced by this component would be rotating with the field, and not be averaged out by the image derotation applied in the reduction, and therefore also be constantly present in all data sets. 

One might hope, that the origin of the unexplained 
polarization pattern can be understood and perhaps calibrated if more very deep polarimetric imaging data with SPHERE/ZIMPOL are taken and analysed. 

We would like to note, that we obtain a positive net $Q_{\rm \phi}$ signal at the level of $Q_{\rm \phi}/I_\star = (4.6 \pm 1.0) \cdot 10^{-6}$ when we integrate $Q_{\rm \phi}$ from $0.6\arcsec$ to $1.6\arcsec$ and normalize this to the total flux of the star. This value is about 3.5 times higher than $U_{\rm \phi}/I_\star = (1.3 \pm 0.6) \cdot 10^{-6}$ obtained for the same integration region. 
This net $Q_{\rm \phi}/I_\star$ signal could originate from circumstellar scattering and it has the strength expected from the models presented in Section~\ref{SectDustModel}. The unexpected and strong wedge pattern seen for $Q_{\rm \phi}(x,y)$ and $U_{\rm \phi}(x,y)$ casts some doubts on this interpretation because the weak integrated $Q_{\rm \phi}/I_\star$-signal could simply be a net effect of systematic errors. 

A polarization offset, as introduced by interstellar or instrumental polarization would create constant offsets for the fractional Stokes maps $Q/I$ and $U/I$ shown in Figure~\ref{2d_dustsearch_c} and quadrant patterns in the $Q_{\rm \phi}(x,y)$ and $U_{\rm \phi}(x,y)$ images but only a very small net contributions for 
the integrated $Q_{\rm \phi}$ and $U_{\rm \phi}$. Similar arguments hold for an uncorrected beamshift effect, which produces a gradient in the Stokes $Q(x,y)$ and $U(x,y)$ and for $Q_{\rm \phi}(x,y)$ and $U_{\rm \phi}(x,y)$, but without a significant impact on the integrated $Q_{\rm \phi}$ and $U_{\rm \phi}$. Higher order effects are required to create offsets in the integrated $Q_{\rm \phi}$ and $U_{\rm \phi}$ signals but this needs still to be investigated. 

One may suspect that unidentified instrumental effects may introduce a negative $Q_{\rm \phi}$-signal as likely as a positive $Q_{\rm \phi}$ signal, and a strong $U_{\rm \phi}$ signal ($\lvert U_{\rm \phi} \rvert > \lvert Q_{\rm \phi} \rvert$) as likely as a weak $U_{\rm \phi}$ signal ($\lvert U_{\rm \phi} \rvert < \lvert Q_{\rm \phi} \rvert$). The fact that we obtain a positive $Q_{\rm \phi}$ signal which is a few times stronger than the absolute value of $\lvert U_{\rm \phi} \rvert$ as expected for a real circumstellar scattering signal should therefore attract our attention. This could be a real signal, but also an instrumental effect. The latter case would be an unfortunate coincidence, at the level of one out of four possibilities, that the introduced spurious signals $Q_{\rm \phi}$ and $U_{\rm \phi}$ behave as expected for a weak signal from circumstellar scattering. 

\subsection{Disk detection limits}
\begin{figure}[b!]
    \includegraphics[trim=10 18 12 48, clip=true,width=0.46\textwidth]{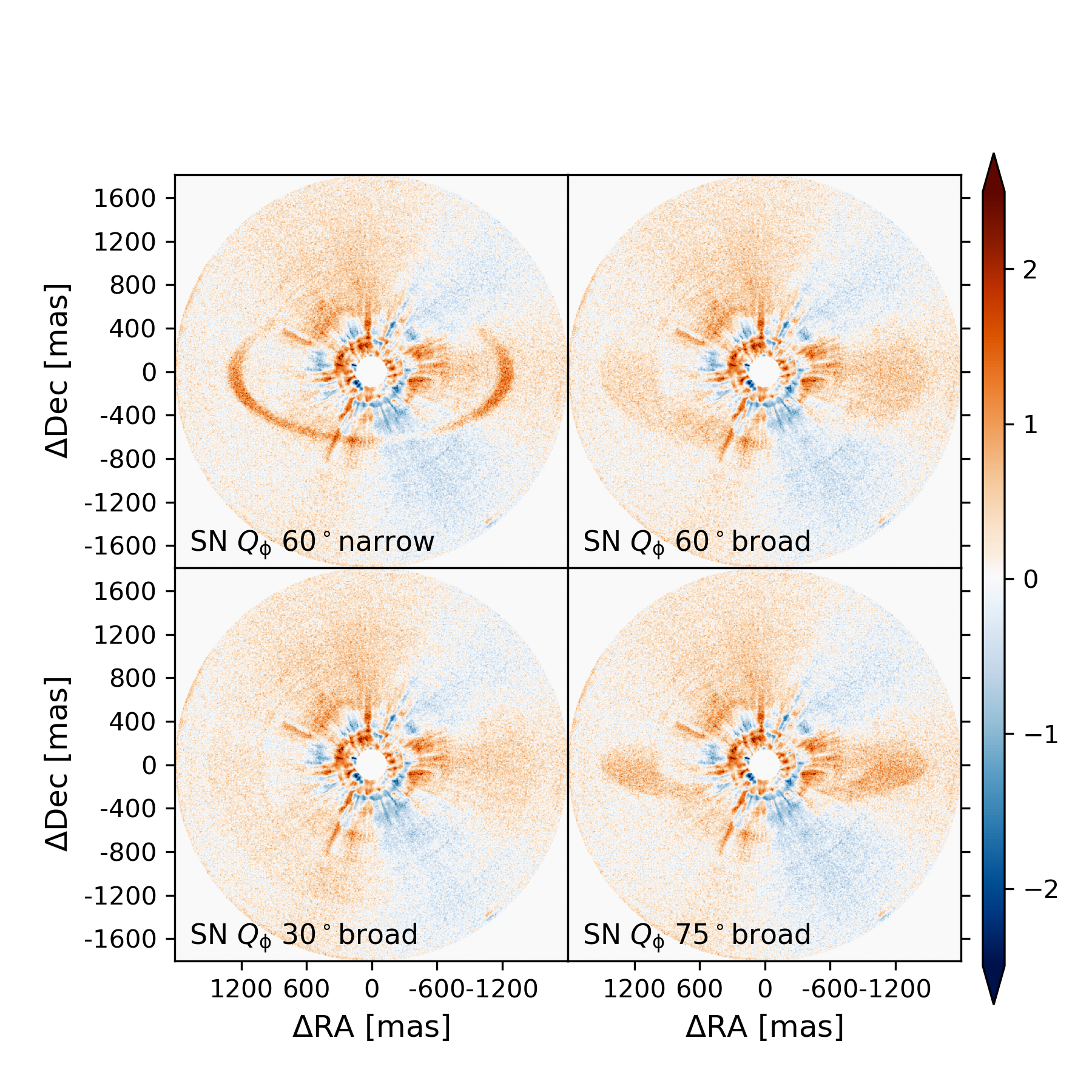}
    \centering
\caption{Convolved disk models from Section~\ref{SectDustModel} for $\epsilon$~Eri injected in the S/N map for the $Q_{\rm \phi}$ observations. The noise per pixel is defined by the standard deviation along the 12 nights in the image cube.}
\label{2d_dustsearch_d_sn}
\end{figure}

\begin{figure*}[t]
    \includegraphics[trim=50 8 70 30, clip=true,width=1\textwidth]{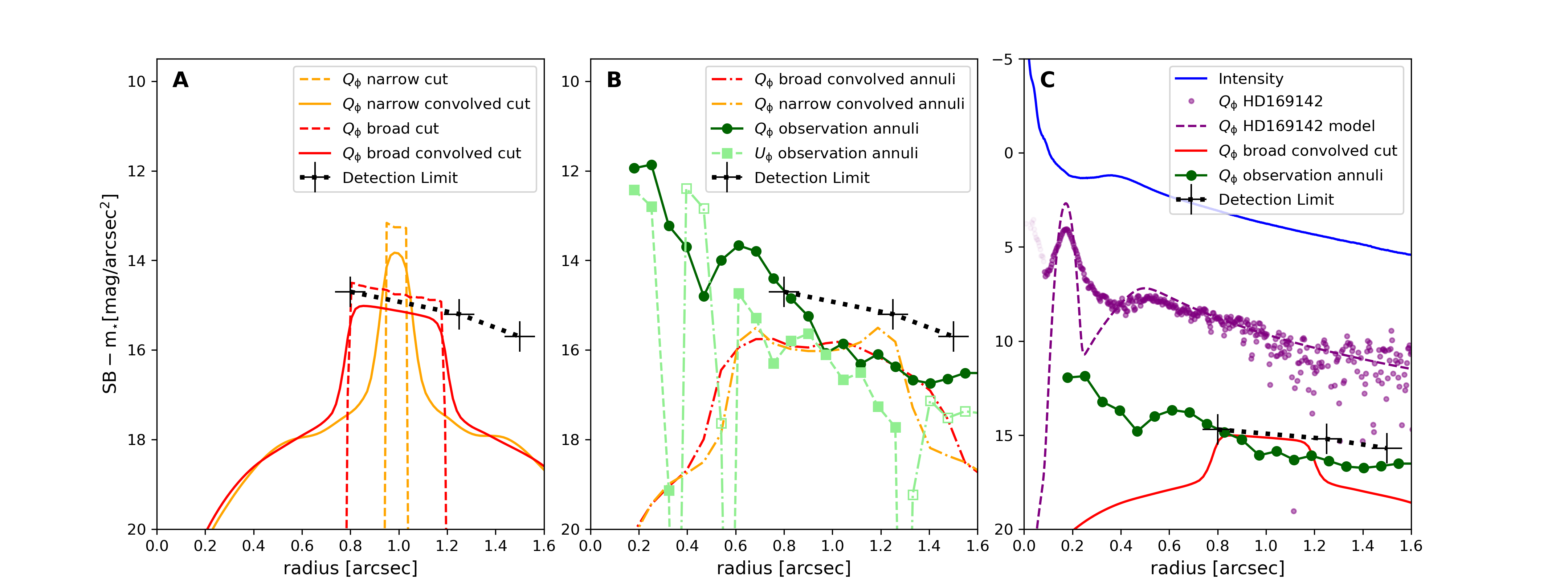}
    \centering
\caption{Comparison of the observational contrast limits for the polarized surface brightness (black crosses and dotted line) with the disk model calculations. 
(A): Comparison with cuts through the dust disk with $i=60^\circ$ for the intrinsic (dashed line) and convolved (full line) models for the narrow ring (orange) and broad ring (red) models. 
(B): Annuli comparison of the convolved dust models (dash dotted line) with respect to the averaged $Q_{\rm \phi}$ and $U_{\rm \phi}$ observations.
(C): Surface brightness comparison between $\epsilon$~Eri and a prominent protoplanetary disk such as HD~169\,142 together with the intensity PSF profile of $\epsilon$~Eri.} \label{an_dustsearch}
\end{figure*}

 We can search in the final $Q_{\rm \phi}$ or $Q_{\rm \phi}/I$ maps of $\epsilon$~Eri in Figures~\ref{2d_dustsearch_b} and \ref{2d_dustsearch_c} for a ring-like or disk-like structure from circumstellar dust scattering on top of the instrumental pattern. 
 A careful inspection reveals no obvious such structure in these maps. However, we can estimate rough sensitivity limits by inserting the calculated polarized flux $Q_{\rm \phi}(x,y)$ for the four disk models described in Section~\ref{SectDustModel} into the data. Figure~\ref{2d_dustsearch_d_sn} shows S/N maps, where S is the average $Q_{\rm \phi}(x,y)$ signal with the inserted disk model and N is the standard deviation between the 12 individual nights' $Q_{\rm \phi}(x,y)$ images. The standard deviation is large for separations close to the star and small for large angular separations. This type of map allows to inspect the full field of view with the same scaling. 

For the narrow ring model the polarized intensity is easily visible in our data, despite the systematic $Q_{\rm \phi}$ pattern. The presence of the disk is more difficult to recognize for the broad ring models. Only the high inclination systems with $i=60^\circ$ and $i=75^\circ$ have disk regions with sufficient surface brightness to be recognized. The disk model with $i=30^\circ$ is hardly visible despite the fact that the disk integrated $Q_{\rm \phi}$ signal is 16~\% higher than for $i=60^\circ$, or 32~\% higher than for the disk with $i=75^\circ$. Important for a detection is the peak surface brightness, which is higher for more inclined disks, and the presence of sharp disk boundaries which are helpful for distinguishing between disk structure and the systematic wedge pattern. We consider the wide disk model with $i=60^\circ$ as rough disk detection limit. In the fractional polarization map the signal needs to be of the order of $Q_{\rm \phi}(x,y)/I_\star(x,y) \approx +0.007~\%$ (Figure~\ref{2d_dustsearch_c}) to be visible as distinct positive signal on top of the smooth 'background' pattern. 

The detection limits can be quantified as a polarized surface brightness contrast $\Delta{S\!B}_{\rm p} ={S\!B}_{\rm p}-m_\star$ This is obtained from the count rates per pixel ${\rm ct/(s \cdot px)}$ using the total count rates $(2.09\pm 0.07)\cdot {10^{8}}~{\rm ct/s}$ for $\epsilon$~Eri for an aperture with a diameter of 3$\arcsec$ and the pixel size of $1/77\,160~{\rm arcsec}^2$. 
This yields contrast limits for the $\epsilon$~Eri data of ${\Delta S\!B}_{\rm p}\approx 15.2~{\rm mag/arcsec}^2$ at $\rho=1.25\arcsec$, and $\approx 14.7~{\rm mag/arcsec}^2$ at smaller ($0.8\arcsec$) or $\approx 15.7~{\rm mag/arcsec}^2$ at larger ($1.5\arcsec$) separation, respectively, as plotted in all panels of Figure~\ref{an_dustsearch} in black. The comparison with the disk model predictions excludes the existence of the narrow ring model, because a radial cut through the brightest ring section plotted with an orange line in Figure~\ref{an_dustsearch}(A), would introduce a detectable polarization signal. Also, any other compact, and therefore high ${S\!B}_{\rm p}$ dust cloud structure responsible for the measured $20~\mu$m infrared emission, such as circumplanetary dust, can be excluded within the covered separation range. The non-detection is compatible with a low inclination disk with a large width $\Delta r>1$~au, similar to the red lines in Figure~\ref{an_dustsearch}(A), or dust located partly outside of the SPHERE/ZIMPOL field of view $r>5~$au. 

We noticed in Section~\ref{dust_res} a weak positive net signal for $Q_{\rm \phi}$, if we integrate the $Q_{\rm \phi}$ from $0.6\arcsec$ to $1.6\arcsec$. We can also derive an azimuthally averaged surface brightness contrast profile $\Delta{S\!B}_{\rm p}(r)$ for $Q_{\rm \phi}$ based on the mean value for annuli with $\Delta r=20~{\rm px}$, which is shown with filled dark green circles in Figure~\ref{an_dustsearch}(B). This signal would be compatible with the two $i=60^\circ$ models, if their convolved surface brightness is also averaged in these concentric annuli. This might indicate that an extended disk with an integrated disk signal of $Q_{\rm \phi}/I_\star \approx 4.6\cdot 10^{-6}$ but without sharp edges could be present. The caveat is the unexplained, strong systematic pattern in the data which causes significant doubts about the nature of the integrated $Q_{\rm \phi}$-signal. The surface brightness contrast profile $\Delta{S\!B}_{\rm p}(r)$ for the $U_{\rm \phi}$ signal illustrates this uncertainty, as it should be zero for optically thin circumstellar scattering. Clearly, there is a lot of noise for the innermost region $<0.6\arcsec$. For $r>0.6\arcsec$ the $Q_{\rm \phi}(r)$ signal is typically a factor 2-3 larger than the absolute value for the average $U_{\rm \phi}(r)$ signal, which is often used as noise indicator in imaging polarimetry of circumstellar disks. 

More studies are required on the instrument polarization of SPHERE/ZIMPOL to clarify and improve the measurements for $\epsilon$~Eri. This is not easy, because the achieved limit of the presented data is already very deep. This is illustrated in panel (C) of Figure~\ref{an_dustsearch}, which compares the surface brightness contrast limits $\Delta{S\!B}_{\rm p}(r)$ of the $\epsilon$~Eri observations with $\Delta{S\!B}_{\rm I}(r)$ and the 'typical' $\Delta{S\!B}_{\rm p}(r)$ of the bright circumstellar disk HD~169\,142 \citep{Tschudi21} also observed with SPHERE/ZIMPOL. The contrast limit $\Delta{S\!B}_{\rm p}(r)$ for $\epsilon$~Eri is more than $5~{\rm mag/arcsec}^2$ deeper than for the proto-planetary disk and therefore we are faced with previously not recognized systematic noise effects.

\section{Discussion}\label{5Dis}

This work presents for the $\epsilon$~Eri system a very deep search for scattered light from a planet or from circumstellar dust using high-contrast imaging polarimetry in the visual spectral range with SPHERE/ZIMPOL. We achieve for $\epsilon$~Eri in two out of the four epochs $5\,\sigma_{\mathcal{N}}$ polarimetric contrast limits for point sources at levels of $C_{\rm P}\approx 5 \cdot 10^{-8}$ at a separation of 0.6$\arcsec$, $2 \cdot 10^{-8}$ at 1.0$\arcsec$, and $1 \cdot 10^{-8}$ at 1.6$\arcsec$ (Figure~\ref{contrcurve}). The achieved detection limits for the surface brightness contrast for polarized light from an extended source of dust scattering is at a level of about ${S\!B}_{\rm p}\approx 15 {\rm mag/arcsec}^2$ at 1.25$\arcsec$. These limits are discussed in this section with respect to the expected signal for the $\epsilon$~Eri system considering also possible observational improvements and requirements towards a successful detection. 

\subsection{Search for a point source}

The system $\epsilon$~Eri is an attractive target for pushing the detection limits for the search of reflecting planets because several studies postulate the presence of the planet candidate $\epsilon$~Eri~b with an orbital period of about 7.3-7.6~yr based on RV and astrometric data \citep{Mawet19,Llop21,Benedict22}. The RV measurements predicted for Nov. 2020 an orbital quadrature phase which is the best phase for a polarimetric search of reflecting planets because we can expect a larger orbital separation $\rho\approx 1.1\arcsec$ and possibly the maximal polarization signal of $C_{\rm P}\approx 1.2\cdot 10^{-9}$ assuming a giant planet with an atmosphere producing a lot of scattering polarization. For the 2019 observations the planet separation and polarization could be similar, if the orbit inclination $i$ is low, but it could also be substantially less favourable ($\rho\approx 0.5\arcsec$ and $C_{\rm P}\approx 0.5 \cdot 10^{-9}$) for a high $i\approx 80^\circ$ (Section~\ref{SectPlanetModel}). Unfortunately the inclination of the $\epsilon$~Eri~b orbit is not well known.

Our contrast limits of about $C_{\rm P}\approx 1 \cdot 10^{-8}$ can exclude for $\epsilon$~Eri~b ``exotic'' models, such as a gas planet with a giant ring system $R_{\rm ring}\approx 10 \cdot {\rm R_{\rm J}}$ \citep[e.g.][]{Arnold04}, which would increase strongly the reflecting surface and therefore enhance the scattering polarization by a factor of ten with respect to the reflection from the planetary atmosphere. A strong signal from a giant circumplanetary disk requires also the right values for the disk inclination and dust scattering properties. Therefore, such a system is unlikely for the nearest, single, solar type star.

Our observations show first and foremost, that the contrast limits for the search of a faint point source get deeper by increasing the integration time by roughly $C_{\rm P}\approx 1/\sqrt{t_{\rm exp}}$. This was already shown in \citet{Hunziker20} for $t_{\rm exp}$ up to 100~min. This trend continues when data from two to four nights from one observing epoch with $t_{\rm exp}$ of up to 15~hours are combined, as shown in Figure~\ref{contrcurve} and numbers given in Section~\ref{res_ep}.

We also investigated the possible improvement by combining data from different epochs using the K-Stacker software \citep{Lecoroller20}, which combines the data based on an Keplerian orbit prediction. This method allows the combination of data from different epochs for an object with substantial orbital motion and further increases the contrast roughly according to the $C_{\rm P}\approx 1/\sqrt{t_{\rm exp}}$ law. 

This study clarifies the possible improvements for future deep searches with SPHERE/ZIMPOL. Significant deeper contrast limits are achievable within a given observing time, if the observations are only taken under very good seeing conditions. 
For example the contrast is about a factor 2.5 better for night~6 with an average seeing of $0.66\arcsec$ when compared to night~5 with very similar $t_{\rm exp}$ but a seeing of $1.12\arcsec$. The typical seeing for our $t_{\rm exp}=38.5$~hours of $\epsilon$~Eri was about 0.75$\arcsec$. It can be expected that the same contrast as in this work would be achievable within about half the exposure time if the average seeing of the observations would be 0.5$\arcsec$. Also favourable for a deep contrast limit is the coverage of a large field rotation during one night, which helps to improve the averaging of the residual speckle noise in the data with angular differential imaging. 

The required significance for claiming a detection is substantially reduced for follow-up observations if the position of the planet $\epsilon$~Eri~b is known from accurate stellar astrometry of the reflex motion or from the direct detection of the planet with imaging. If the planet position is known to a precision of about 25~mas, then a $3\,\sigma_{\mathcal{N}}$ detection is sufficient to claim a significant polarimetric signal. A $3\,\sigma_{\mathcal{N}}$ detection limit with a polarization contrast of $C_{\rm P} \approx 5\cdot 10^{-9}$ could be possible with the combination of all our data, if $\epsilon$~Eri~b turns out to have a favourable orbit with a low inclination, low eccentricity, and the same high brightness in polarized light for all our epochs.

Knowing the astrometric orbit would also allow to obtain for each night only Stokes $Q$ measurements with a position angle aligned with the $Q_{\rm \phi}$ of the planet. This is possible with ZIMPOL and would save 50~\% of the measuring time as the corresponding Stokes $U$ signal from the planet is expected to be zero. 

Taking all these steps into account, a $3\,\sigma_{\mathcal{N}}$ detection for the planet 
$\epsilon$~Eri~b with a contrast of about $C_{\rm P} \approx 1 \cdot 10^{-9}$ 
would be possible with a well known planet orbit, using only observations
taken under best seeing conditions during the best orbital phases of the planet, 
and measuring only Stokes $Q \parallel Q_{\rm \phi}$ with an integration time of about 200~hours with ZIMPOL. This seems to be technically feasible with VLT/SPHERE within the framework of an ESO large programme.

\subsection{Search for extended emission from circumstellar dust}

We searched for an extended polarization signal from circumstellar dust in $\epsilon$~Eri and achieve a contrast limit of about $\Delta {S\!B}_{\rm p}\approx 15~{\rm mag/arcsec}^2$ at a separation of $1.25\arcsec$ (4 au). This limit is set by an unknown systematic noise effect. 
This is very unfortunate, because the statistical noise limit for our data is much lower. For a single pixel at a separation of about 1$\arcsec$ the photon noise of the whole data set is roughly $\approx 14~{\rm mag/arcsec}^2$ (0.005~${\rm ct/(s \cdot px)}$) and this could be strongly pushed by pixel binning, for example by a factor 30 to $\approx 17.7~{\rm mag/arcsec}^2$ for an area of $30\times 30$~pixels. This binning provides still a very good spatial sampling of $0.11\arcsec\times 0.11\arcsec$ or $0.35~{\rm au} \times 0.35~{\rm au}$ for the detection of extended dust scattering around $\epsilon$~Eri. At this contrast level the predicted polarization signal should clearly show up, even for very unfavourable spatial distribution and scattering properties of the inner dust in $\epsilon$~Eri. 

Actually, we measure an integrated azimuthal polarization signal of about $Q_{\rm \phi}/I_\star \approx 4.6\cdot 10^{-6}$ for an annulus covering the separation range from 2~au to 5~au. This is roughly at the predicted level of the simple scattering models for the warm dust in $\epsilon$~Eri. However, we hesitate to claim a detection without a better understanding of the dominant systematic $Q_{\rm \phi}$ noise pattern. Clearly, a better correction or calibration of the systematic pattern shown in Figures~\ref{2d_dustsearch_b} and \ref{2d_dustsearch_c} would allow a major progress for the investigation of the inner dust in $\epsilon$~Eri. Also other bright targets with warm dust, such as Fomalhaut \citep[e.g.][]{Gaspar23} could be investigated with high sensitivity in scattered light. 

For this reason we have put quite some efforts to understand the systematic effects by using various types of alternative post-processing procedures. We also analysed deep polarimetric imaging data for $\alpha$~Cen~A available in the ESO archive (ESO programme ID 2107.C-5008), which were taken with the same instrument mode as $\epsilon$~Eri. We find also for $\alpha$~Cen~A a wedge pattern in the residual polarization image similar to Figures~\ref{2d_dustsearch_b} and \ref{2d_dustsearch_c} with roughly the same contrast $\Delta{S\!B}_{\rm p}$ but a different wedge geometry \citep{Tschudi23}. 
The $\alpha$~Cen~A data are too different to be used for a calibration of the systematic pattern in the $\epsilon$~Eri data. More studies are required to improve further the polarimetric sensitivity for extended sources with SPHERE/ZIMPOL. 

\subsection{$\epsilon$~Eri with the Roman Space Telescope}
The Nancy Grace Roman Space Telescope, which is planned to become operational in a few years, includes a Coronagraph Instrument technology demonstrator (hereafter Roman-CGI) for the detection of the reflected visible light from cold planets and circumstellar dust \citep{Kasdin21,Mennesson22,Bailey23,Doelman23}. The $\epsilon$~Eri system will also serve for this instrument as important test case for the instrument performance \citep{Douglas22,Anche23}, because it is one of best planetary systems known for a successful detection \citep{Carrion21}. Therefore, it is interesting to compare briefly the expectations of this space instrument using advanced coronagraphy with our SPHERE/ZIMPOL using a 'basic' stellar Lyot coronagraph and PDI for the suppression of the strong, residual speckle halo of a ground based AO instrument. 

It is expected that Roman-CGI will achieve for $\epsilon$~Eri contrast levels of about $1\cdot 10^{-9}$ for the reflected intensity with an integration time of about 100~hours using the 'wide' field imaging mode for separations $0.45\arcsec - 1.4\arcsec$, in the I-band ($\lambda_{\rm c}=825$~nm) \citep[e.g.][]{Bailey23}. This is ideal for the planet $\epsilon$~Eri~b for which one can expect a contrast between $C_{\rm I}\approx 10^{-9}$ and $10^{-8}$ as illustrated in Figure~\ref{1_orbit_b}. Such data could provide the planet intensity phase curve for the I-band, an accurate orbit, and pin-down for SPHERE/ZIMPOL the best orbital phase and location for a polarimetric detection of $\epsilon$~Eri with follow-up observations as described above.

The Roman-CGI is also expected to be very sensitive for the polarimetric mapping of the scattered light from the warm dust in $\epsilon$~Eri, and it could provide a detection with an integration of about 10~min \citep{Anche23,Douglas22}. Similar to our ZIMPOL study a narrow disk ring would be easy to measure while a smooth, extended cloud will require deeper observations and more accurate calibrations. However, a detection of the dust in $\epsilon$~Eri seems to be straight forward with Roman-CGI \citep{Anche23}, even for difficult circumstances, because of the excellent coronagraphic contrast and the high sensitivity. SPHERE/ZIMPOL could complement the Roman-CGI dust scattering polarimetry with data having significantly higher spatial resolution, provided the polarimetric calibration problem described above can be solved. Additionally, ZIMPOL multi-wavelength polarimetry could constrain dust properties based on the colour of the reflected signal.

\section{Conclusions}\label{6Conc}

This work demonstrates the potential of PDI with SPHERE/ZIMPOL despite the fact that we could not detect the planet $\epsilon$~Eri~b or map the extended polarization signal from circumstellar dust with the combination of $t_{\rm exp} = 38.5$~hours of data from 12 nights spread over more than a year.  For $\epsilon$~Eri, this provides the deepest contrast limits for a point source so far and unprecedented contrast limits for the polarimetric search for extended emission from dust scattering. This pilot study is therefore useful to understand the limits of SPHERE/ZIMPOL imaging polarimetry better and it clarifies strategies to optimize future searches of faint sources around bright stars in polarized light. 

\paragraph{On the search of point sources.} The presented imaging polarimetry of $\epsilon$~Eri reaches $5\,\sigma_{\mathcal{N}}$ point source contrast limits at the level of $C_{\rm P}\approx 10^{-8}$ at a separation of $1\arcsec$. It seems that the limits would just further improve with the square root of $t_{\rm exp}$ or the number of collected photons. Similar point source contrast limits were previously achieved with this instrument for the much brighter stars $\alpha$~Cen~A and $\alpha$~Cen~B by selecting data of a single night with perfect seeing \citep{Hunziker20}. The $\epsilon$~Eri data prove that the combination of data from different nights can be used as a standard procedure to push planet detection limits using the service observation mode offered for the VLT telescopes. However, this observing strategy must consider the substantial orbital motion of potentially observable reflecting planets around nearby stars using a software that searches for the planet with a Keplerian motion prediction as described in this work for the K-Stacker software \citep{Lecoroller20,Nowak18}.

We also compared the chances for a successful detection of a reflecting planet with SPHERE/ZIMPOL in searching for a polarized signal or the intensity signal. We find that a planet is easier to find with polarimetry. At small separations, polarimetry is up to 30 times more efficient in suppressing the strong speckle noise than the search of the corresponding intensity signal. Because Earth \citep{Stam08,Bazzon13}, Jupiter \citep{Smith84}, or Uranus and Neptune \citep{Schmid06,Buenzli09} show a scattering polarization of about 15~\% or higher in the R band, a polarimetric search of a planet with SPHERE/ZIMPOL is attractive. 

The polarimetric contrast limits that were reached of about $C_{\rm P}\approx 1\cdot 10^{-8}$ at $\rho\approx 1\arcsec$ are still a factor of about ten above the expected signal for the RV-planet $\epsilon$~Eri~b. Steps to reduce the gap between the planet signal and detection limit are discussed and we conclude that a $3\,\sigma_{\mathcal{N}}$ contrast limit of $C_{\rm P}\approx 1\cdot 10^{-9}$ would require about 200~hours of VLT integration time under good seeing conditions and a well-known planet orbit to optimize the observing strategy in a follow-up search. This is very demanding but technically feasible. 

Contrast limits of $C_{\rm P}\approx 1\cdot 10^{-9}$ at $\rho\approx 1\arcsec$ are easier to achieve for planets around the nearest bright stars Sirius, $\alpha$~Cen~A and B, Altair, and a few others. For these objects it is possible to measure a photon flux that is ten times higher  with SPHERE/ZIMPOL without harmful detector saturation effects and to reach contrasts of $C_{\rm P}\approx 1\cdot 10^{-9}$ within 20~hours. Moreover, because $\alpha$~Cen is so close, a contrast of $\approx 1\cdot 10^{-9}$ would allow the detection of a planet with a radius of roughly 0.4~${\rm R_{\rm J}}$, because $1\arcsec$ corresponds to a physical separation of only $d_{\rm p}=1.3$~au.

Such a programme with SPHERE/ZIMPOL (ESO programme ID 2107.C-5008) was approved following the announcement of a planet candidate in $\alpha$~Cen~A by \citet{Wagner21}; however, because of scheduling issues during the corona pandemic, only observations were executed for 4~hours. We analysed these $\alpha$~Cen~A observations very similar to the $\epsilon$~Eri data and reached a $5\,\sigma_{\mathcal{N}}$ detection limit of $C_{\rm P}\approx 8\cdot 10^{-9}$ at $1\arcsec$ for these 4~h \citep{Tschudi23}. This is a similar contrast to the 38.5~hours for $\epsilon$~Eri because of the higher photon flux and the better seeing conditions for the $\alpha$ Cen A data. We could not detect a point source in these data. Because the planet candidate around $\alpha$~Cen~A has not been confirmed yet, a future SPHERE/ZIMPOL programme aiming for deeper observations would face the risks of a blind search, where the location and brightness of the planet for a given epoch are unclear and therefore could be very unfavourable if the observations are scheduled at the 'wrong' time. 

\paragraph{On the search of extended emission.} 
SPHERE/ZIMPOL is the only polarimeter regularly used for high-contrast imaging polarimetry in the visual 500-900~nm range \citep[e.g. ][]{Schmid21}. It is therefore very useful for the study of wavelength dependencies of the scattered radiation of circumstellar disks and shells \citep[e.g. ][]{Ma24,Haubois23}. For these applications the ZIMPOL performance is competitive with respect to the state-of-the-art near-infrared polarimetric modes of SPHERE/IRDIS \citep{deBoer20} or of GPI at Gemini \citep{Perrin15}. For dust around bright stars $R<5$~mag, ZIMPOL has the advantage that it was designed for the search of planets around very bright stars and the instrument can therefore fully exploit the photon collecting power of the VLT telescope in broad-band filters without harmful detector saturation effects.

For $\epsilon$~Eri we added 38.5~hours of integration for the search of extended polarized emission of dust. By averaging the signal over areas of $0.11\arcsec \times 0.11\arcsec$, we could further reduce the statistical noise to theoretical contrast limits of $\Delta{S\!B}_{\rm p}\approx 17.7~{\rm mag/arcsec}^2$ at a separation of about $1\arcsec$. This would be enough to see the expected dust scattering clearly from the warm dust in $\epsilon$~Eri inferred from the $20~\mu$m infrared excess bump measured in SPITZER data \citep{Backman09}. However, we find that the contrast is limited by a weak systematic noise pattern to about $\Delta{S\!B}_{\rm p}\approx 15~{\rm mag/arcsec}^2$. Even this contrast is much deeper than previous measurements of the extended polarized emission in high-contrast imaging with SPHERE and GPI. Typical contrast limits for one hour observations are about $\Delta{S\!B}_{\rm p}\approx 8$ to $10~{\rm mag/arcsec}^2$ at $\rho\approx 0.4\arcsec$ for high inclination debris disks \citep[e.g. ][]{Engler17,Engler18,Esposito20}
or $\approx 10$ to $12~{\rm mag/arcsec}^2$ around $1\arcsec$ for protoplanetary disks \citep[e.g. ][]{Avenhaus18,Tschudi21}. We are not aware of deep searches for polarized circumstellar emission that combined long integrations of about 10~hours or more for a very bright star to reach much fainter extended sources. The presented $\epsilon$~Eri data show that much deeper limits $\Delta{S\!B}_{\rm p}\approx 15~{\rm mag/arcsec}^2$ can be achieved with high-contrast imaging polarimetry and the potential of such observations should be used more often for the future investigation of faint circumstellar dust around bright stars.

\begin{acknowledgements}
CT and HMS acknowledge the financial support by the Swiss National Science Foundation through grant 200020\_181983. SPHERE is an instrument designed and built by a consortium consisting of IPAG (Grenoble, France), MPIA (Heidelberg, Germany), LAM (Marseille, France), LESIA (Paris, France), Laboratoire Lagrange (Nice, France), INAF – Osservatorio di Padova (Italy), Observatoire de Gen`eve (Switzerland), ETH Zurich (Switzerland), NOVA (Netherlands), ONERA (France) and ASTRON (Netherlands), in collaboration with ESO. SPHERE was funded by ESO, with additional contributions from CNRS (France), MPIA (Germany), INAF (Italy), FINES (Switzerland) and NOVA (Netherlands). SPHERE also received funding from the European Commission Sixth and Seventh Framework Programmes as part of the Optical Infrared Coordination Network for Astronomy (OPTICON) under grant number RII3-Ct-2004-001566 for FP6 (2004–2008), grant number 226604 for FP7 (2009–2012) and grant number 312430 for FP7 (2013–2016). A.Z. acknowledges support from ANID -- Millennium Science Initiative Program -- Center Code NCN2021\_080. 
This research is based on observations obtained with ESO Telescopes at the La Silla Paranal Observatory under programme IDs: 
0104.C-0178(A), 0104.C-0178(B), 0104.C-0178(C) and 106.2144.001.
\end{acknowledgements}

\bibliographystyle{aa}
\bibliography{bibliography}

\begin{appendix} 

\section{Advanced improvements for the SPHERE/ZIMPOL data reduction}

This appendix describes improvements in the ZIMPOL data reduction for the very deep $\epsilon$~Eri observations presented in this work, which are new or go beyond the procedures described previously \citep{Hunziker20,Schmid18}. We had to improve our data analysis because this paper pushes the limits of the ZIMPOL performance to deeper limits.

\subsection{Camera 1 readout issue}\label{A1_cam1rd}
\begin{figure}[h]
    \includegraphics[trim=10 10 10 10, clip=true,width=0.46\textwidth]{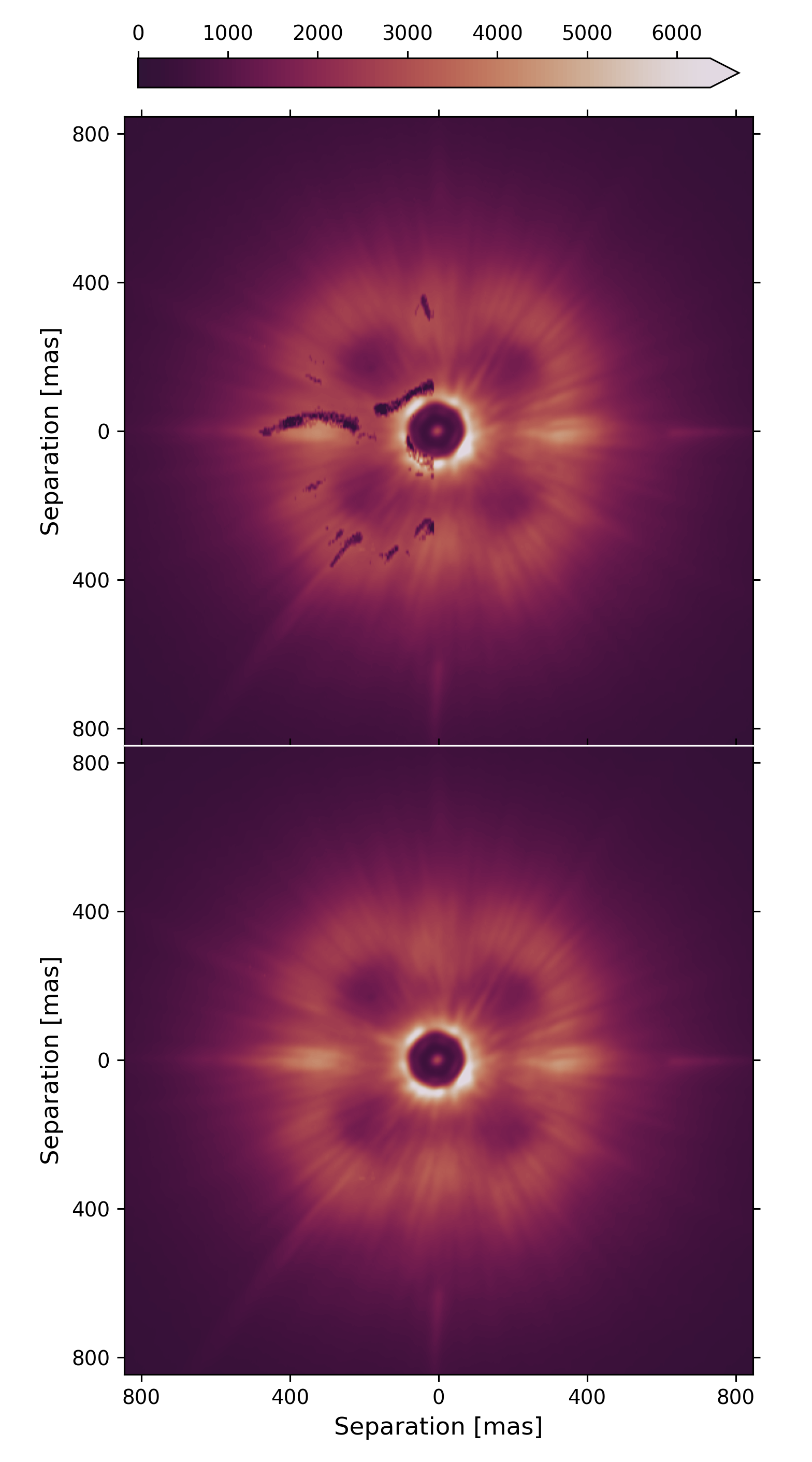}
    \centering
\caption{Illustration of the ZIMPOL camera 1 readout issue affecting the
left side of the image. Illustrated is one individual intensity image (two
5~s sub-integrations = 10~s) with readout problem before and after correction
of the affected pixels.} \label{cam1_readout}
\end{figure}

The detector of ZIMPOL camera 1 had issues with the electronics during the first eight nights of our observations. The analogue to digital converter of the left read out register produced wrong results for pixels with a count level of about 7000~ADUs as shown in Figure~\ref{cam1_readout}. The speckle pattern in the coronagraphic image varies as result of the atmospheric turbulence and therefore the affected pixels changed from frame to frame because different regions had exposures levels near 7000 ADUs. Typically, there are around 1000 affected pixels located near the bright speckle ring or near the 'coronagraph mask'. Fortunately, ZIMPOL takes data simultaneously with camera~1 and camera~2 and therefore the same image is taken twice with only a small scaling factor difference because the ZIMPOL beamsplitter sends a few percent more light to camera~2. This allowed a correction of the bad pixels in camera~1, identified with an outlier detection procedure, by replacing them with corresponding, scaled pixel values from camera~2. A corrected image is shown in Figure~\ref{cam1_readout}.

This reconstruction could save all the affected camera~1 data from the first eight nights without producing spurious effects for the applied post-processing procedures. Not correcting and not including the affected data in our analysis would have reduced the effective exposure time of our programme by about 28~\% (56~$\%$ of the observing time is affected, however only one camera, therefore 28~$\%$ of the 'photons'). Of course, the information of the affected $\approx 1000$~pixels per camera~1 frame is lost and only duplicated by camera~2 data, but this loss corresponds to only $\approx 0.1~\%$ of one camera~1 frame or only $\approx 0.05~\%$ of the total pixels for each integration registered with the two detectors. ESO has solved the issue before the $\rm 9^{\rm th}$ night by replacing read-out electronics boards and two broken cooling fans.

\subsection{Telescope polarization correction}\label{A2_telpol}
\begin{figure}[h]
    \includegraphics[trim=36 4 4 2, clip=true,width=0.46\textwidth]{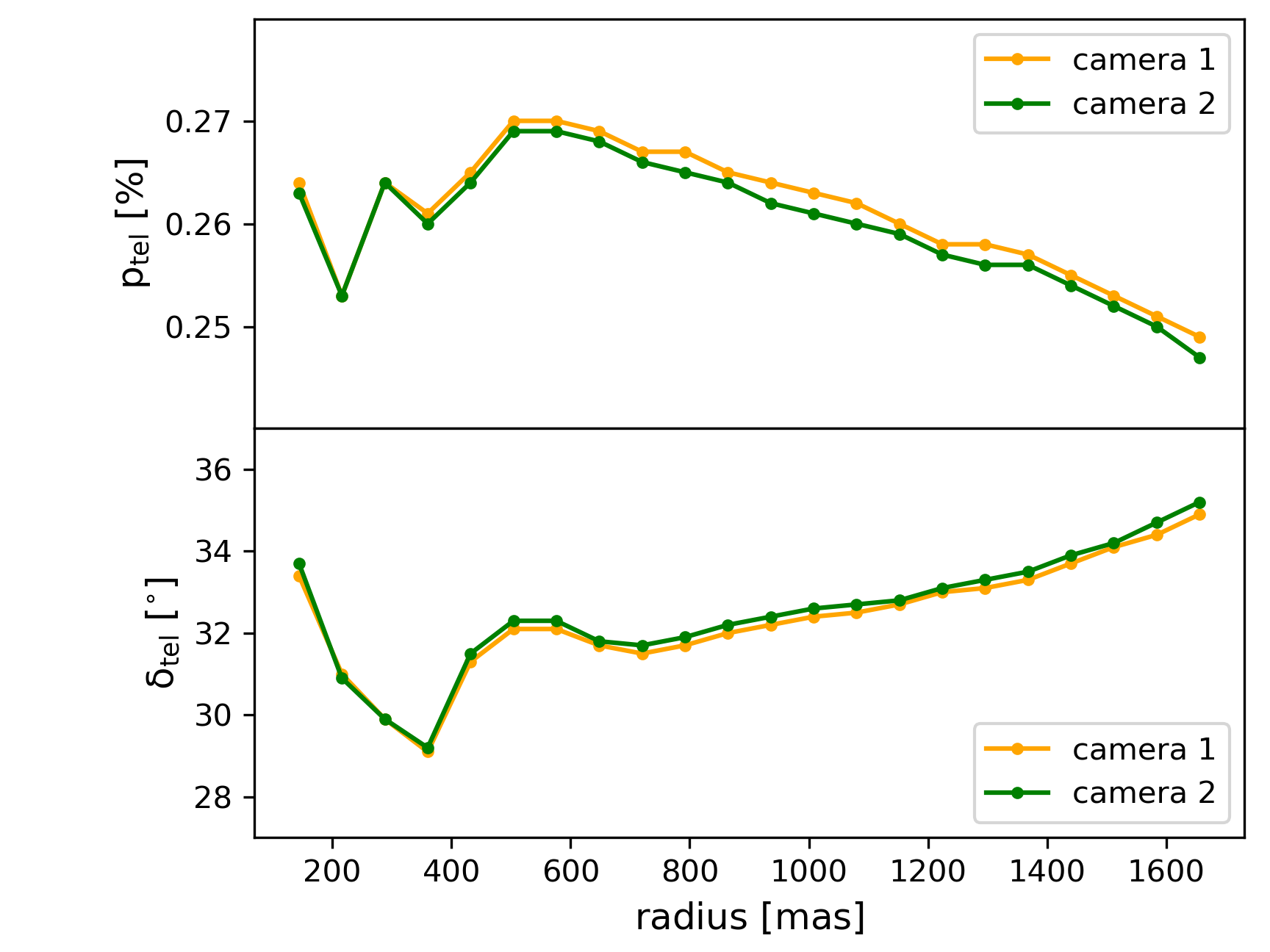}
    \centering
\caption{Radial dependence of the telescope polarization parameters $p_{\rm tel}$ and $\rm \delta_{\rm tel}$ for the Very Broad Band (VBB) filter. The plotted curves are the mean curve for all $\epsilon$~Eridani observations listed in Table~\ref{obsSHOW}.} \label{telpolfig}
\end{figure}

For deep polarimetric observations with SPHERE/ZIMPOL it is important to apply a correction for the residual telescope polarization as described in \citet{Schmid18}. A polarization of about 4~\% is introduced by the aluminium coated M3 mirror in the VLT, which is for ZIMPOL polarimetry compensated with a rotating half-wave plate and a 'crossed' M4 mirror. The compensation is not perfect, but the telescope effects are reduced to about $p_{\rm tel} \approx 0.5~\%$ or less. This residual polarization depends on the parallactic angle $\theta_{\rm para}$ of the telescope and the measured values lie in the Q/I-U/I plane on a circle with radius $p_{\rm tel}$ with position angle $\theta_{\rm tel}=\theta_{\rm para}+\delta_{\rm tel}$ \citep[see e.g.][]{Hunziker20,Tschudi21}. The centre ($q_{\rm m}, u_{\rm m}$) of the circle can be offset from the origin (0,0) due to interstellar or intrinsic polarization of the star. We measure for $\epsilon$~Eri with the VBB filter in 2019 and 2020 $p_{\rm tel} = 0.26 \pm 0.03$ and $\delta_{\rm tel} = 32.1 \pm 1.5 ^\circ$ and a centre of ($\rm q_{\rm m}=0.0032 \pm 0.006~\%, u_{\rm m}= -0.0013 \pm 0.004~\%$) with zero polarization as expected and in very good agreement with previous high precision measurements for this target $\epsilon$~Eri $q_{\rm m}: 0.00284 \pm 0.00056~\% $ and $u_{\rm m}: -0.00120 \pm 0.00057~\%$ and $p \approx 30 \cdot 10^{-6}$, $\theta = 168.5^\circ$ \citep{Cotton17}. 
The expected polarization from light scattering by the circumstellar dust around $\epsilon$~Eri is much lower, less than $10^{-4}~\%$ (see Section~\ref{SectDustModel}). For such objects without strong intrinsic polarization, a good first order correction for the telescope polarization is obtained by the normalizations $I_0=I_{90}$ and $I_{45}=I_{135}$ for each cycle. This is equivalent to setting the integrated polarization to zero: $Q=0$ and $U=0$. Not correcting for the telescope polarization would introduce an $I\rightarrow Q,U$ cross talk and intensity speckles would be visible as polarized features with a relative strength at the level of the telescope polarization $p_{\rm tel}$.

For $\epsilon$~Eri, we need to consider also second order effects of the telescope polarization. The two parameters describing the telescope polarization $p_{\rm tel}$ and $\delta_{\rm tel}$ depend on the wavelength \citep{Schmid18} and this is an issue for very deep polarimetry in the VBB filter with large bandwidth $590-890$~nm and therefore significantly different instrument polarization for short and long wavelengths. This is shown in \citet{Tschudi21} for the $R'$ and the $I'$ filter which cover roughly the short and long wavelength parts of the VBB filter, respectively. In addition, the SPHERE AO PSF for the VBB filter is a superposition of different radial profiles for the different wavelengths. This produces a 2-dimensional polarization effect which cannot be corrected with a single vector $(p_{\rm tel},\delta_{\rm tel})$ without leaving second order calibration errors. 

The dominant feature for the radial dependence of the instrument polarization is the wavelength dependent location of the strong PSF speckle ring, which is defined by the control radius ($20\, \lambda/D$) of the SPHERE AO system. This ring is for the $R$-band at a separation of $\rho\approx 0.35''$ and for the $I$-band at $\rho \approx 0.45''$ \citep[Figure~11 in][]{Schmid18} and therefore the $R$-band instrument polarization dominates at smaller separations while the $I$-band polarization contributes strongly for $\rho\geq 0.45''$. We could measure clearly a corresponding radial dependence of the telescope polarization $p_{\rm tel}(r)$ and $\delta_{\rm tel}(r)$ with amplitudes of $\pm 0.01~\%$ and $\pm 2^\circ$ for the mean radial curve as illustrated in Figure~\ref{telpolfig}. The curves $p_{\rm tel}(r)$ and $\delta_{\rm tel}(r)$ look similar in shape from night to night but show small variations $\Delta p_{\rm tel}< 0.02~\%$ and $\Delta \delta_{\rm tel}< 1^\circ$ because of PSF variations introduced by different observing conditions. We don't see a long term trend within the 13 months covered by our observations, but on longer timescales one should expect systematic changes because of the evolution of the coatings of the telescope mirror M3 and the first folding mirror M4 in SPHERE. 

For the general case with a ($q_{\rm m},u_{\rm m}$) offset from (0,0) we would determine the mean $p_{\rm tel}(r)$ and $\delta_{\rm tel}(r)$ curves of each night to correct the telescope polarization of that night. From these profiles we construct two interpolated 2d maps to correct each polarization image pixel-wise depending on the parallactic angle of that data. In the special case of $\epsilon$~Eri with no offset polarization $(q_{\rm m}, u_{\rm m}) = (0,0)$, we can measure Q/I, respective U/I in every image individually and correct it to 0. We do this for each radial annuli separately to fully account for the radial dependence as described above. The advantage of this method is that PSF variations, which can happen within minutes, and higher order telescope polarization effects are also corrected. 

This second order correction for the telescope polarization is not crucial for the search of a planet outside the speckle ring $\rho>0.6\arcsec$ where the dependence of $p_{\rm tel}(r)$ and $\delta_{\rm tel}(r)$ is smooth. At small separation ($<0.5 \arcsec$), the residual pattern of strong speckles is slightly reduced. The second order correction should however improve the search of an extended weak signal from dust scattering. For example, the effect is clearly seen in the $Q$ and $U$ images of the debris disk observations of HIP~79\,977 by \citet[][Figure~3]{Engler17} as over-corrected central area $r<0.5\arcsec$. At the time of that analysis the origin of this calibration problem was unknown.

\subsection{Beamshift correction}\label{A3_bs}

\begin{figure}[htp!]
    \includegraphics[trim=2 2 0 2, clip=true,width=0.48\textwidth]{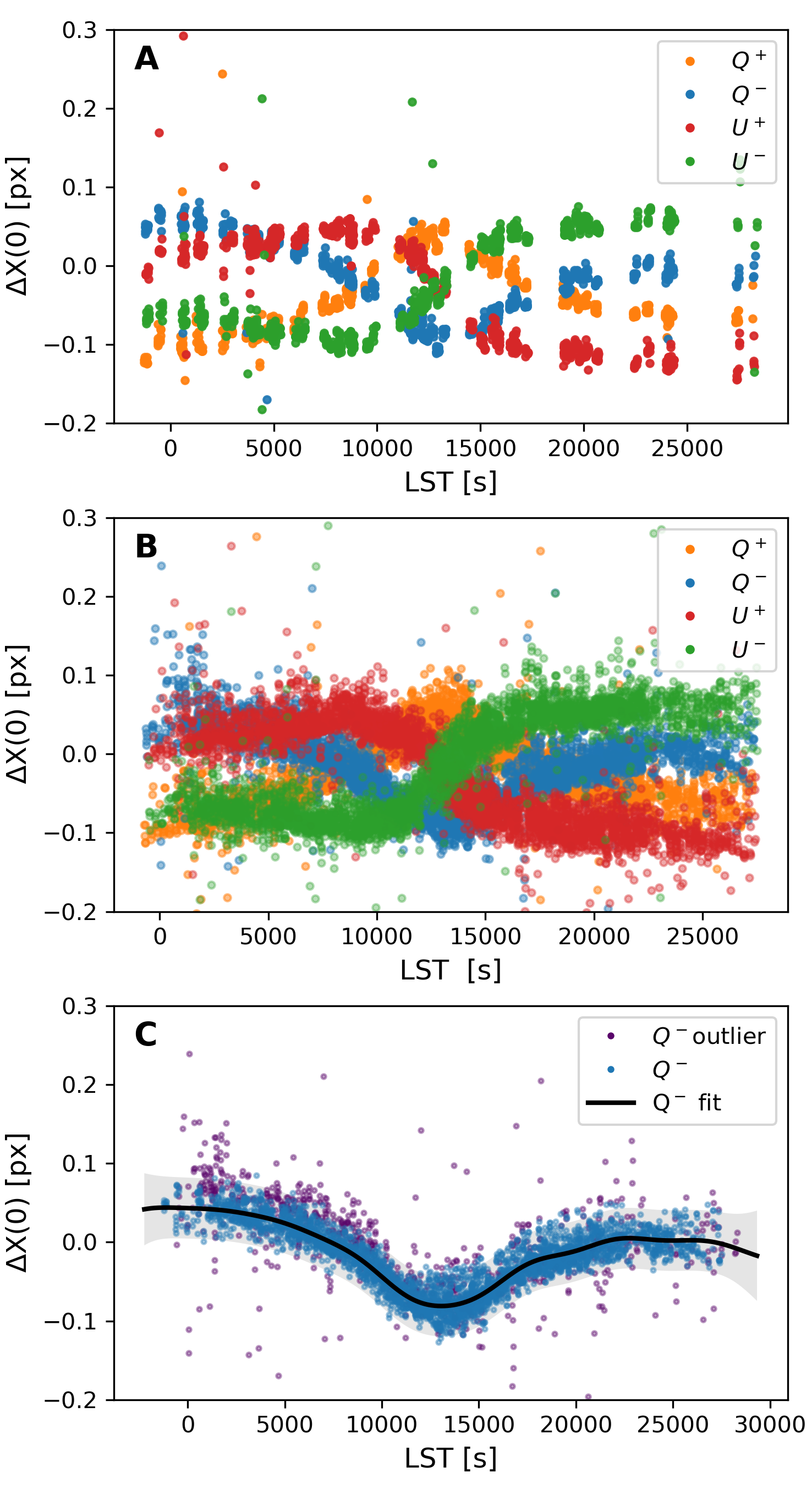}
    \centering
\caption{Beamshift correction parameters as a function of the pointing (local sidereal time). Illustrated as example is the shift in X direction for camera 2 and 0 phase images. (A): Beamshift as measured in the unsaturated (ND2) PSF images. Different colour represent the different polarization images $Q^+$, $Q^-$, $U^+$ and $U^-$. (B): Analogous figure as measured in the coronagraphic images. The measurement accuracy is reduced especially for bad observing conditions. (C): Example for the iterative process of Gaussian process fitting and outlier detection to calculate the fit of this parameter for $Q^-$.}
\label{bs_beamshift}
\end{figure}

It is important for polarimetric differential imaging (PDI) performance that the differential aberrations between the opposite polarization directions $I_\parallel$ and $I_\perp$ are very small so that unpolarized speckles and other PSF features cancel out by the subtraction of the two images and the polarized planet signal is easier to detect. The ZIMPOL design was optimized to reduce such differential aberrations and the two polarization states are for example recorded with the same detector pixels. However the inclined third mirror M3 of the telescope, the 45$^\circ$ pupil tip-tilt mirror and the three image derotator mirrors mainly introduce a, wavelength and telescope pointing dependant, differential beamshift of up to 0.3~pixels (or $\approx$ 1~mas) between $I_\parallel$ and $I_\perp$ \citet{Schmid18, Hunziker20}. For a given pointing of the telescope, meaning the same altitude and parallactic angle, the beamshift effect is the same and therefore we can sort the $\epsilon$~Eridani data and derive the beamshift parameter as a function of their local sideral time (LST). There are many beamshift parameters as the shifts are different in X and Y directions in the image, different for the polarization images $Q^+$, $Q^-$, $U^+$, $U^-$, different for camera 1 and camera 2 and different for the $0$ and $\pi$ phase images. To measure the beamshift accurately an unsaturated point source is required as available from the regularly taken non-coronagraphic observation using the ND2 filter. The used Lyot coronagraph (V$\_$CLC$\_$MT$\_$WF) has a slightly transparent mask and under good conditions it is possible to see the stellar PSF peak through the mask and to measure the beamshift also in the coronagraphic observations. In the top panel of Figure~\ref{bs_beamshift}(A) the non-coronagraphic measurements of the $\Delta X(\rm 0\ phase)$ beamshift parameters for $Q^+$, $Q^-$, $U^+$, $U^-$ are displayed. The same parameters measured for the coronagraphic data can be seen in the middle panel of Figure~\ref{bs_beamshift}(B). It is obvious that the dispersion is larger for the measurements with coronagraph, however there are many more images and no time gaps between the images, but sometimes the determination of the beamshift fails for the coronagraphic data (many of these points are far outside the showed y-axis in the middle panel of Figure~\ref{bs_beamshift}(B)). Unfortunately, there exists no model for SPHERE/ZIMPOL which could predict the beamshift effects. Therefore we use all the existing $\epsilon$~Eri data and derive the shift corrections from the best fit to the data. For this we calculate a Gaussian process fit, apply an outlier detection and removal method and iterate the procedure a few times (see bottom panel of Figure~\ref{bs_beamshift}(C)). The fitting procedure is particularly important for data taken under bad conditions (seeing > 1$\arcsec$). All the images are visually checked after applying the beam shift correction.
\\
\\
\\
\\
\\
\\
\\
\\
\\

\subsection{Chargetrap correction}\label{A4_ct}
\begin{figure*}
    \includegraphics[trim=1 1 1 1, clip=true,width=1\textwidth]{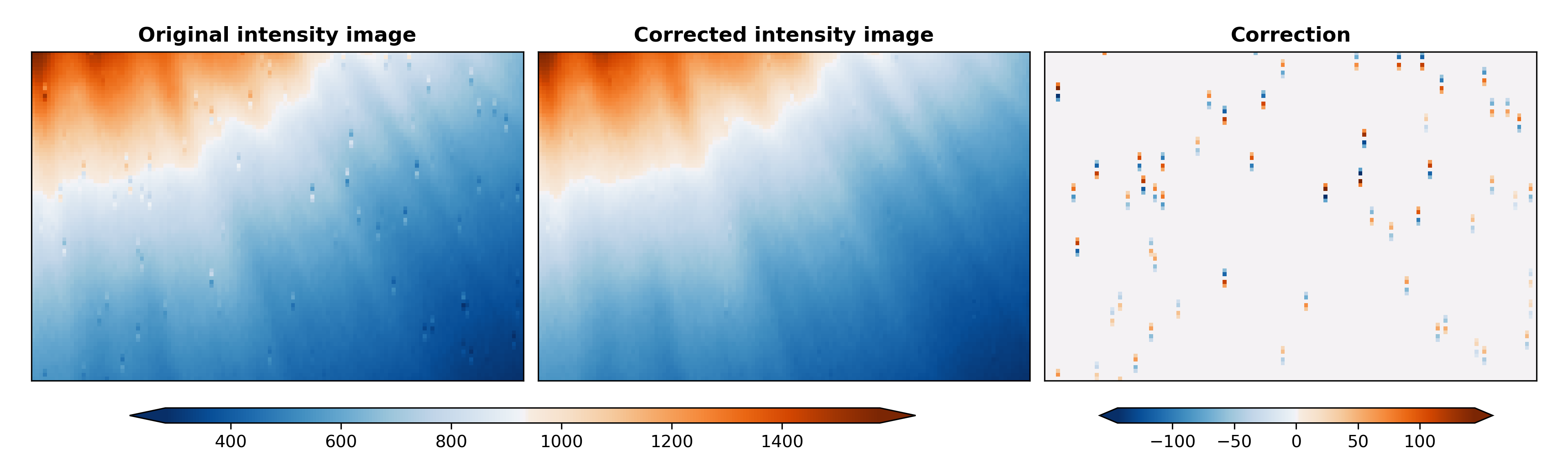}
    \centering
\caption{Illustration of the charge trap correction for ZIMPOL intensity images [$\rm ct \ DIT^{-1} \ px^{-1}$] obtained in polarimetric mode.} \label{chargetrap}
\end{figure*}

Both ZIMPOL detectors consist of alternating open and masked rows (512~rows each) with 1024~pixels. In the polarimetric ZIMPOL modes the charges created in the illuminated rows are shifted up and down in synchronization with the polarimetric modulation (frequency depends on fast or slow polarization mode) \citep{Schmid18}. The final detector frame consists of an “even rows” subframe with one polarization state $\rm I_\perp$, and an “odd rows” subframe for the opposite polarization state $\rm I_\parallel$. In column direction the subframes are then interpolated in a flux conserving manner to create 1024x1024 images. Although the charge transfer efficiency of the CCDs are better than 99.9995~$\%$ there exist pixels which do not shift electron charges perfectly. For example a pixel can block one electron which is not down shifted but in the following up-shift it is transferred. Because of the fast (de)modulation such a trap can cause a hole of many electrons in one polarization state and a corresponding spike in the other. To get rid of this effect in ZIMPOL polarimetric modes there are always an even number of subintegrations and in every second subintegration the charge shifting is reversed with respect to the polarization modulation. The images from zero to $\pi$ phase subintegration are then combined to create a double difference in which the charge trap effects are cancelled \citep{Schmid18,Gisler04,Schmid12}. Unfortunately, for the intensity frame $\rm I = I_\perp + I_\parallel$ derived from the polarimetric data the charge traps do not vanish with the combination of the subintegrations. The charge traps 
 produce a negative-positive pattern with a very specific appearance in column direction. First a pixel with too many counts, then an interpolated 'neutral' pixel and then a pixel with too few counts or the other way around. Sometimes secondary and tertiary pixels are also slightly affected. We search for this pattern in each image and correct it by shifting counts from a 'spike' to a corresponding 'hole' in a flux conserving manner as illustrated in Figure~\ref{chargetrap}). More than 99~$\%$ of the charge traps can be recognized and corrected with this method. This correction is especially helpful for polarimetric p2 mode when the field is fixed and one has only a few long exposures. 

\newpage

\section{Detection maps}

\begin{figure*}[bh!]
\begin{subfigure}{1\textwidth}
    \includegraphics[trim=1 1 1 1, clip=true, width=0.98\textwidth]{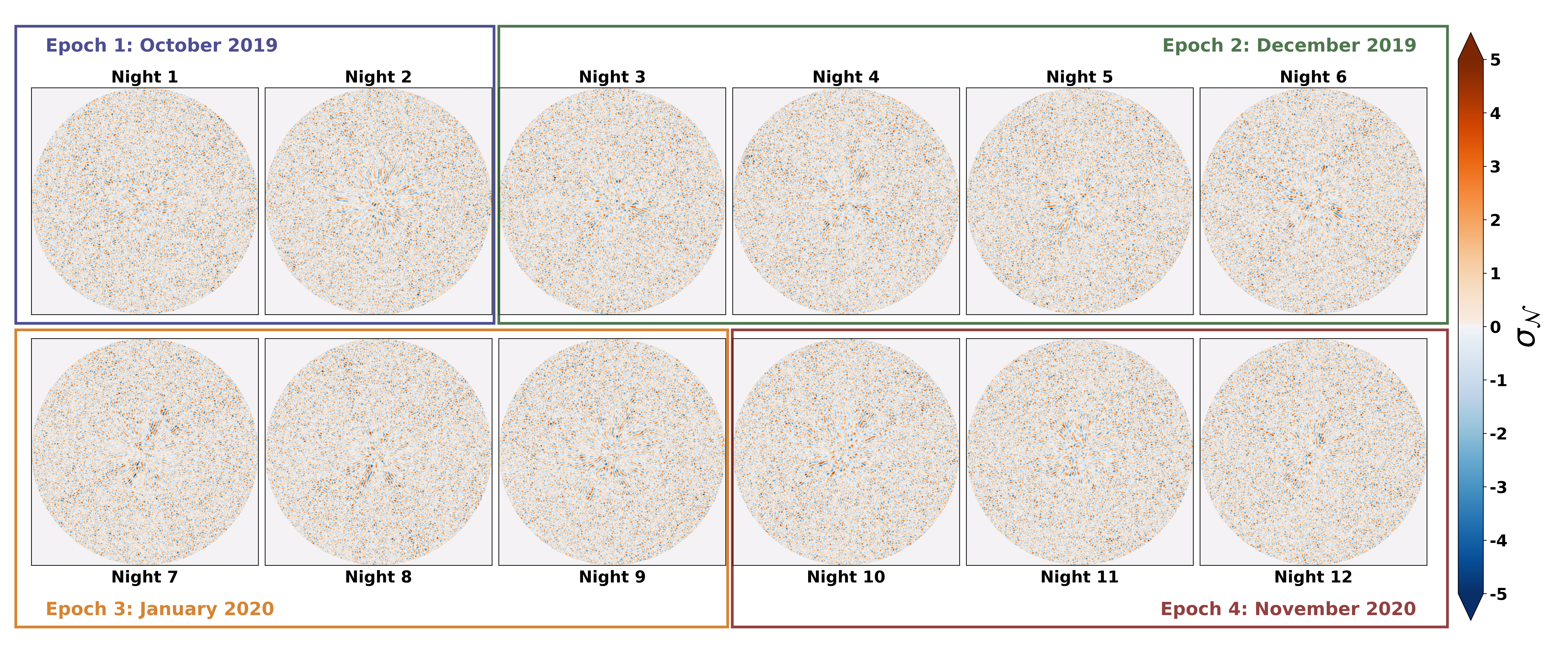}
     \centering
\end{subfigure}
\begin{subfigure}{1\textwidth}
   \centering
    \includegraphics[trim=1 1 1 1, clip=true, width=0.78\textwidth]{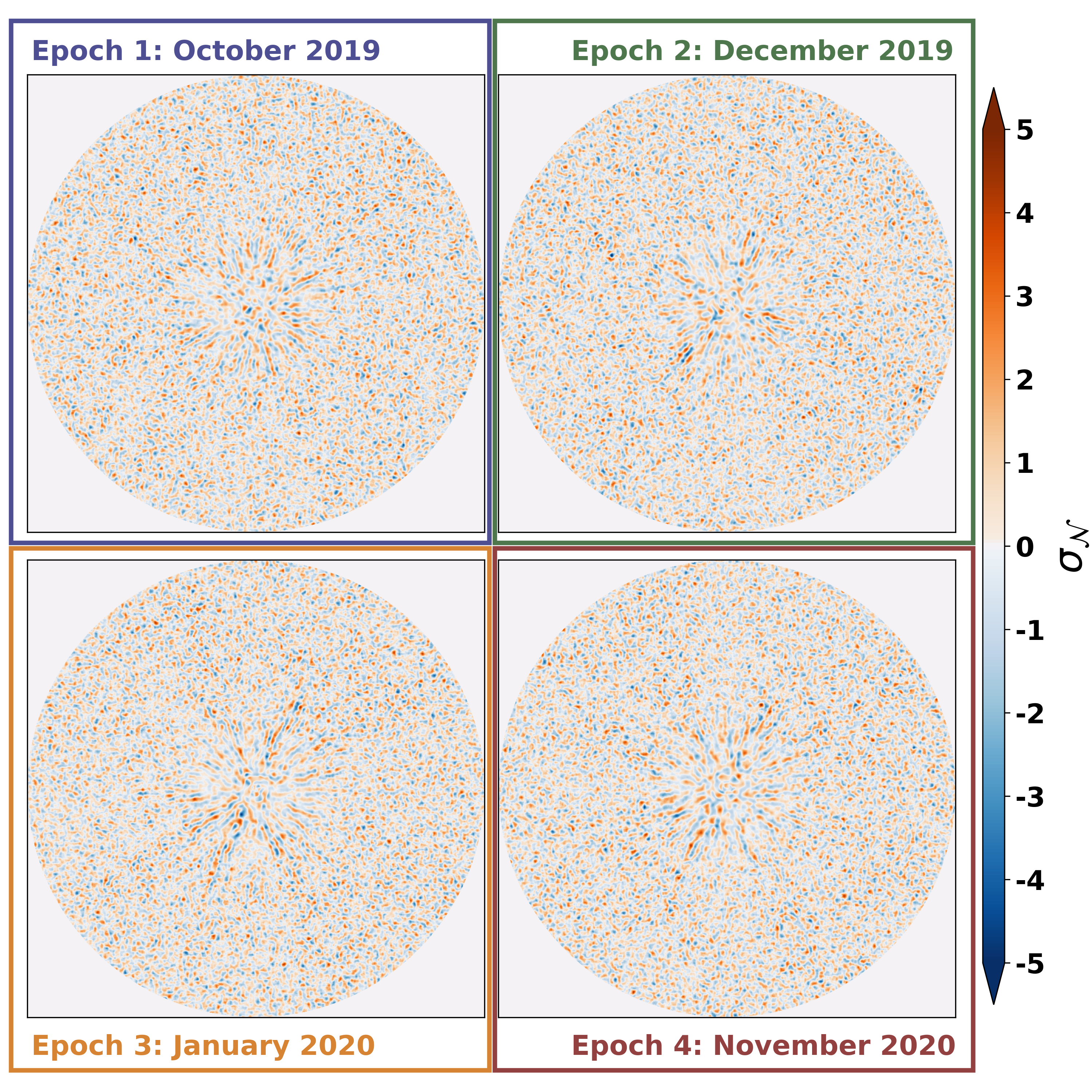}
     \centering
\end{subfigure}
\caption{Polarized light $Q_{\phi}$ detection maps showing the Gaussian significance of the individual pixels. North is up, east is to the left, and the image field of view is the same as in Figure~\ref{psfsub_09}. Top: Detection maps of all the individual nights. Bottom: Detection maps of the four epochs.}
\label{det_maps}
\end{figure*}

\end{appendix}

\end{document}